\pdfoutput=1
\documentclass[a4paper,10pt]{article}
\usepackage[ansinew]{inputenc} % ASCII (Western Windows CP1252)
\usepackage{graphicx}
\usepackage{color}
\usepackage[colorlinks]{hyperref}
\begin{document}
\small\rm
\begin{center}
{\large\bf DANSS: Detector of the reactor AntiNeutrino\\ based on Solid Scintillator}\\[2mm]
{\em
{I.~Alekseev~$^{a,b,c}$},
{V.~Belov~$^{d}$},
{V.~Brudanin~$^{d}$},
{M.~Danilov~$^{b,c,e}$},
{V.~Egorov~$^{d,f}$\footnote{Corresponding author; e-mail:  egorov@jinr.ru}},
{D.~Filosofov~$^{d}$},
{M.~Fomina~$^{d}$},
{Z.~Hons~$^{d,g}$},
{S.~Kazartsev~$^{d,f}$},
{A.~Kobyakin~$^{a,c}$},
{A.~Kuznetsov~$^{d}$},
{I.~Machikhiliyan~$^{a}$},
{D.~Medvedev~$^{d}$},
{V.~Nesterov~$^{a}$},
{A.~Olshevsky~$^{d}$},
{D.~Ponomarev~$^{d}$},
{I.~Rozova~$^{d}$},
{N.~Rumyantseva~$^{d}$},
{V.~Rusinov~$^{a}$},
{A.~Salamatin~$^{d}$},
{Ye.~Shevchik~$^{d}$},
{M.~Shirchenko~$^{d}$},
{Yu.~Shitov~$^{d,h}$},
{N.~Skrobova~$^{a,c}$},
{A.~Starostin~$^{a}$},
{D.~Svirida~$^{a}$},
{E.~Tarkovsky~$^{a}$},
{I.~Tikhomirov~$^{a}$},
{J.~Vl\'{a}\v{s}ek~$^{d,i}$},
{I.~Zhitnikov~$^{d}$},
{D.~Zinatulina~$^{d}$}
}\\[2mm]
$^a$ {\scriptsize\rm ITEP -- State Scientific Center, Institute for
 Theoretical and Experimental Physics, Moscow, Russia}\\
$^b${\scriptsize\rm MEPhI -- National Research Nuclear University
MEPhI, Moscow, Russia}\\
$^c$ {\scriptsize\rm MIPT -- Moscow Institute of Physics and Technology, Moscow Region, Dolgoprudny, Russia}\\
$^d$ {\scriptsize\rm JINR -- Joint Institute for Nuclear Research, Moscow Region, Dubna, Russia}\\
$^e${\scriptsize\rm LPI RAS -- Lebedev Physical Institute of the Russian Academy of Sciences, Moscow, Russia}\\
$^f$ {\scriptsize\rm DSU -- Dubna State University, Moscow Region, Dubna, Russia}\\
$^g$ {\scriptsize\rm NPI -- Nuclear Physics Institute, \v{R}e\v{z}, Czechia}\\
$^h$ {\scriptsize\rm ICL -- Imperial College London, SW7 2AZ, London, United Kingdom}\\
$^i$ {\scriptsize\rm CTU -- Czech Technical University in Prague, Czechia}\\
\end{center}

\begin{flushleft}
%PACS: 13.15.+g; 23.40.Bw; 14.60.St; 25.30.Pt
\end{flushleft}
\noindent

\begin{abstract}
The DANSS project is aimed at creating a relatively compact neutrino spectrometer which does not contain any flammable or other dangerous liquids and may therefore be located very close to the core of an industrial power reactor. As a result, it is expected that high neutrino flux would provide about 15,000 IBD interactions per day in the detector with a sensitive volume of 1 m$^3$. High segmentation of the plastic scintillator will allow to suppress a background down to a $\sim$1\% level. Numerous tests performed with a simplified pilot prototype DANSSino under a 3~GW$_{\rm th}$ reactor of the Kalinin NPP have demonstrated operability of the chosen design.

The DANSS detector surrounded with a composite shield is movable by means of a special lifting gear, varying the distance to the reactor core in a range from 10~m to 12~m. Due to this feature, it could be used not only for the reactor monitoring, but also for fundamental research including short-range neutrino oscillations to the sterile state. Supposing one-year measurement, the sensitivity to the oscillation parameters is expected to reach a level of $\sin^2(2\theta_{\rm new})\sim5\times10^{-3}$ with $\Delta m^2 \subset (0.02-5.0)$~eV$^2$.
\end{abstract}

\section{Introduction}
Neutrino is probably one of the most enigmatic and at the same time the most wide-spread particles in the Universe \cite{Neutrinos}. Due to its very weak interaction with matter, a target would have to be light-years thick before efficiently stopping a neutrino. Therefore, direct investigation of the neutrino properties requires intensive neutrino source and low background detector with a sensitive volume of at least cubic metre scale.

The most intensive laboratory neutrino source is provided by nuclear fission -- for instance, a typical Russian 3~GW$_{\rm th}$ industrial reactor WWER-1000 produces about $10^{21}$ antineutrinos per second. As the particle flux falls down very fast with distance, it is desirable to install the detector as close to the reactor core as possible. On the other hand, security rules do not allow to use big amount of inflammable, caustic, toxic or other dangerous liquids in a reactor building. That is why conventional liquid scintillator (LS) becomes ``persona non grata'' at nuclear power plant (NPP), and detectors of other type are needed.

If such detector exists -- it could be efficiently used for many applied and fundamental goals based on the precise measurement of the neutrino energy spectrum: on-line monitoring of the reactor power, fuel composition, burning space pattern (up to tomography), etc.
If made movable, the detector would be probably the best suited for testing the hypothesis of short-range neutrino oscillation to a sterile state~\cite{WhitePaper}.

The aim of the {\bf DANSS} ({\bf D}etector of {\bf A}nti{\bf N}eutrino based on {\bf S}olid {\bf S}cintillator) project~\cite{DANSS_TAUP2011,DANSS_AAP2011,DANSS_ICHEP2012} is to develop and create a relatively compact detector of the reactor antineutrinos which consists of highly segmented plastic scintillator with a total volume of 1 m$^3$, has appropriate Signal-to-Background (S/B) ratio and can be moved within few metres from the reactor core. Being installed in any available room close to an industrial reactor, the detector could register about $10^4$ neutrinos per day and measure their energy spectrum. Varying the core-detector distance in some range, detector could confirm or disprove ``sterile'' explanation of the reactor neutrino anomaly~\cite{Mention2011} within few weeks of data taking.

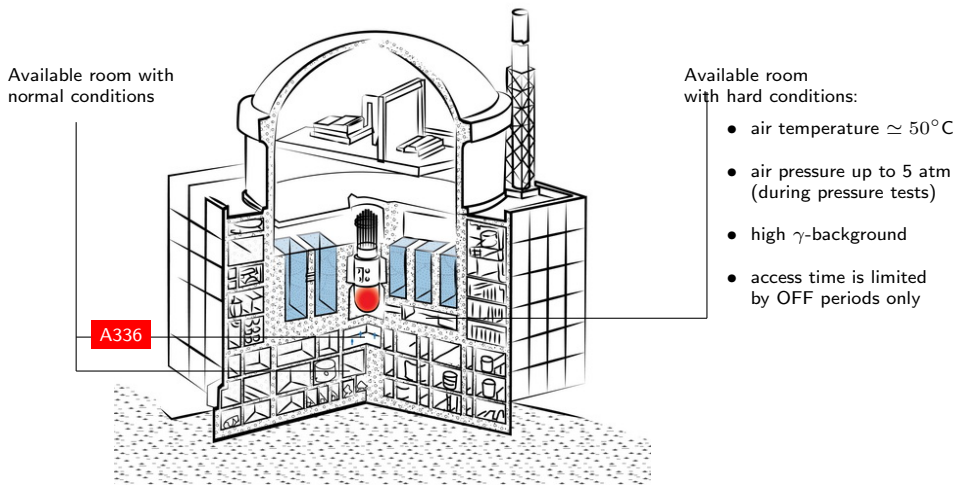
\begin{figure}[htb]
 \setlength{\unitlength}{1mm}
 \centering{
 \begin{picture}(130,70.0)(0,0)
  \put(15.0,0.0){\includegraphics[height=70mm]{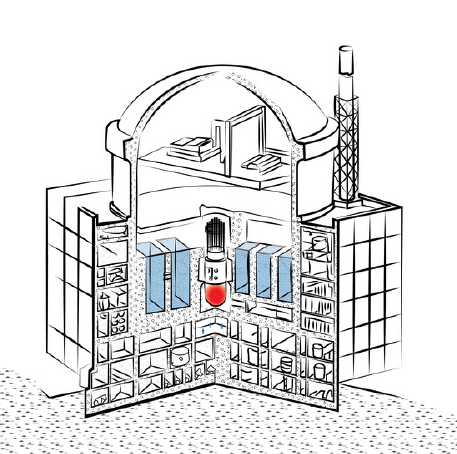}}
  %\put(0,0){\framebox(130,70.0)[b]{}}
  \put (10.0,19.5){\line(1,0){38}}
  \put (10.0,15.0){\line(1,0){36}}
  \put (10.0,15.0){\line(0,1){33}}
  \put(1,52.0){\parbox{30mm}{\scriptsize\sf Available room with\\ normal conditions}}
  \put(12.0,19.0){\colorbox{red}{\scriptsize\sf \color{white}A336}}
  \put (93.0,22.0){\line(-1,0){37}}
  \put (93.0,22.0){\line(0,1){30}}
  \put(90,37.0){\parbox{43mm}{\scriptsize\sf Available room\\ with hard conditions: \begin{itemize}
   \item air temperature $\simeq50^\circ$C
   \item air pressure up to 5~atm\\ (during pressure tests)
   \item high $\gamma$-background
   \item access time is limited\\ by OFF periods only
  \end{itemize} }}
 \end{picture}
 }
 \caption{Available rooms in a typical WWER-1000 reactor building (``B-320'' project).}
 \label{Fig.WWER1000_Building}
\end{figure}

One of the most appropriate sites in Russia for our neutrino experiments is Kalinin Nuclear Power Plant (KNPP) located between Moscow and Saint Petersburg and exploiting four pressurized water reactors WWER-1000~\cite{WWER-1000}. The first stage of KNPP includes 30 years old reactor units \#1 and \#2 of so-called ``B-338'' project, and the second stage -- younger units \#3 and \#4 of the ``B-320'' project.

Depending on the project, there are several potentially available places in the reactor building (Fig.~\ref{Fig.WWER1000_Building}), but only few of them could provide appropriate conditions for an experimental setup. For the ``B-338'' project, it was the A107 room located just under the reactor cauldron at 14 meters from the core. This site was used in our GEMMA experiment \cite{GEMMA}. The more recent ``B-320'' project has an even better location: room A336. In addition to extremely high neutrino flux ($\sim5\times10^{13}\; \bar\nu_e /{\rm cm}^2/{\rm s}$ at a distance of 11~m) such location provides very good shielding ($\simeq$50~m of water equivalent) against cosmic rays.

Indeed, in addition to numerous equipment, thick walls of heavy concrete and the reactor body with more than 70 tons of uranium, there are several huge reservoirs with technological liquids placed above the room -- repository of the primary coolant and boric acid for the reactivity adjustment, as well as a cooling pond for the spent fuel. These hydrogen-containing materials completely remove fast cosmic neutrons which are the main source of background in such measurements (see Section~\ref{Sect.DANSSino} below). The muon component is suppressed by a factor of $\simeq$6 also.

Two similar rooms 3-A336 and 4-A336 located under the twin reactor units \#3 and \#4 have been considered as an experiment site. Finally the latter one was chosen, because it provides enough space for big dimension equipment.

\subsection{Gamma background}\label{Section.Gamma_BG}

\begin{figure}[bht]
 \setlength{\unitlength}{1mm}
 \begin{picture}(150,40.0)(0,0)
  %\put(0,0){\framebox(150,40.0)[b]{}}
  \put(  0.0,0.0){\includegraphics[width=150mm]{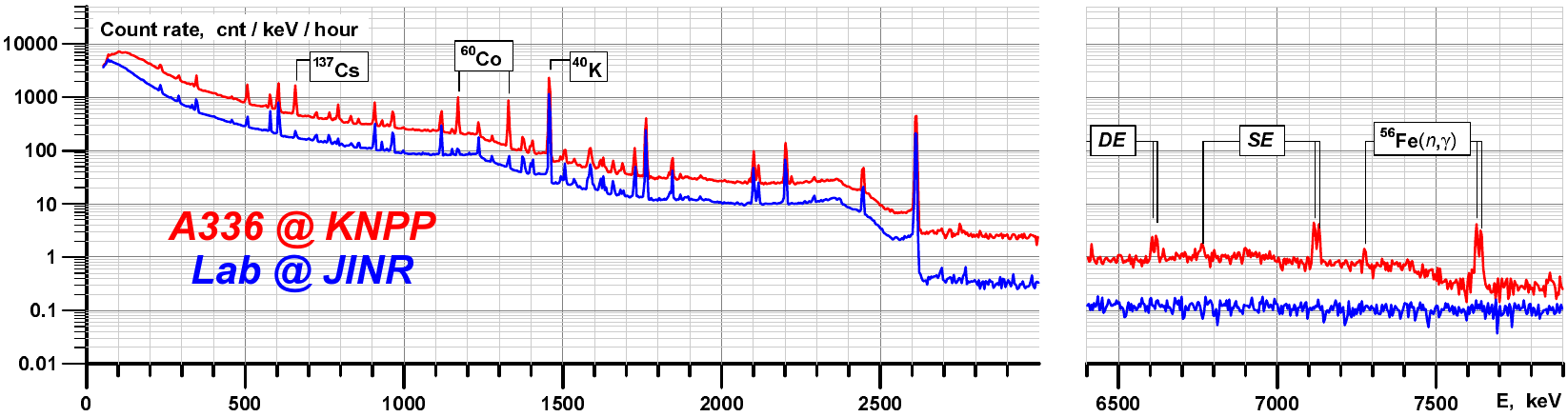}}%[width=0.4\textwidth]
 \end{picture}
 \caption{Gamma-background measured with 1.3 kg HPGe detector at the DANSS position in the KNPP room A336 and in the JINR laboratory.}
 \label{Fig.Gamma-BG}
\end{figure}

Gamma background in the center of the room is only twice higher than in an average laboratory building. A reason of the excess is big amount of heavy concrete in the surrounding walls. It contains natural $^{40}$K which is radioactive and emits $\gamma$-rays with $E$=1.461~MeV. Measurement performed with HPGe detector (Fig.~\ref{Fig.Gamma-BG}) does not indicate any deviation of the background structure from a natural one, except of some $^{60}$Co and $^{137}$Cs pollution typical for the reactor surroundings.

High energy part of the spectrum does not indicate any Gd $\gamma$-rays (5.903 or 6.750 MeV) from the fuel rods\footnote{One of the ways to increase efficiency of the fuel burning and prolong the reactor life-time is using more efficient fuel of higher $^{235}$U enrichment together with slowing down the fission reaction. That is why some of the fuel rods are doped with gadolinium which captures neutrons and partially dumps the fuel burning.}. It means that the direct $\gamma$-rays from the core surely do not penetrate into the room through the reactor shielding. Instead, the spectrum demonstrates several high energy lines corresponding to the neutron capture by Cr, Fe and Ni nuclei which are contained in numerous steel construction elements. The most visible of them is the 7.631+7.645 MeV doublet following the $^{56}{\rm Fe}(n,\gamma)$ reaction. Being integrally rather weak, these radiation nevertheless should be taken into account in the data analysis.

\subsection{Neutron background}\label{Section.Neutron_BG}
In our GEMMA experiment the neutron flux in room 2-A107 was found to be 20--30 times lower than in a usual laboratory. Unfortunately, this is not the case for the second KNPP stage with reactors of the ``B-320'' project. Here 27 steel tubes $\oslash$100 mm are immured vertically in the concrete walls around the reactor body. These tubes commence at the ceiling of the room A336 and are used to monitor the reactor with movable ion chambers. Low energy~\footnote{As the tubes are {\sl tangent} to the reactor cauldron, initial MeV-range neutrons from the reactor core cannot come to the room directly, but only after numerous scattering in water and concrete, loosing their energy each time.} neutrons enter the room through these tubes, thus increasing the neutron background.

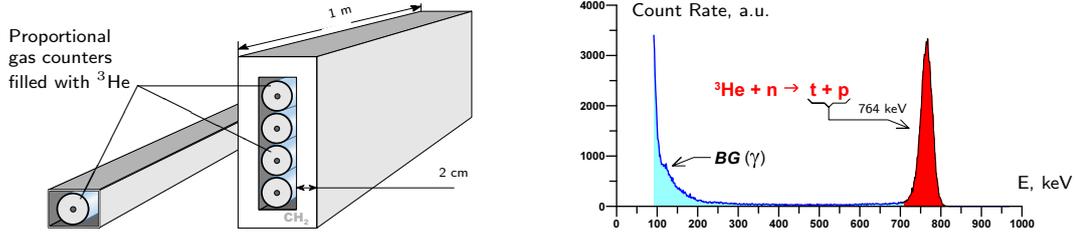
\begin{figure}[bht]
 \setlength{\unitlength}{1mm}
 \begin{picture}(150,28.0)(0,2)
  %\put(0,0){\framebox(150,30.0)[b]{}}
  \put(10.0,0.0){\includegraphics[height=19mm]{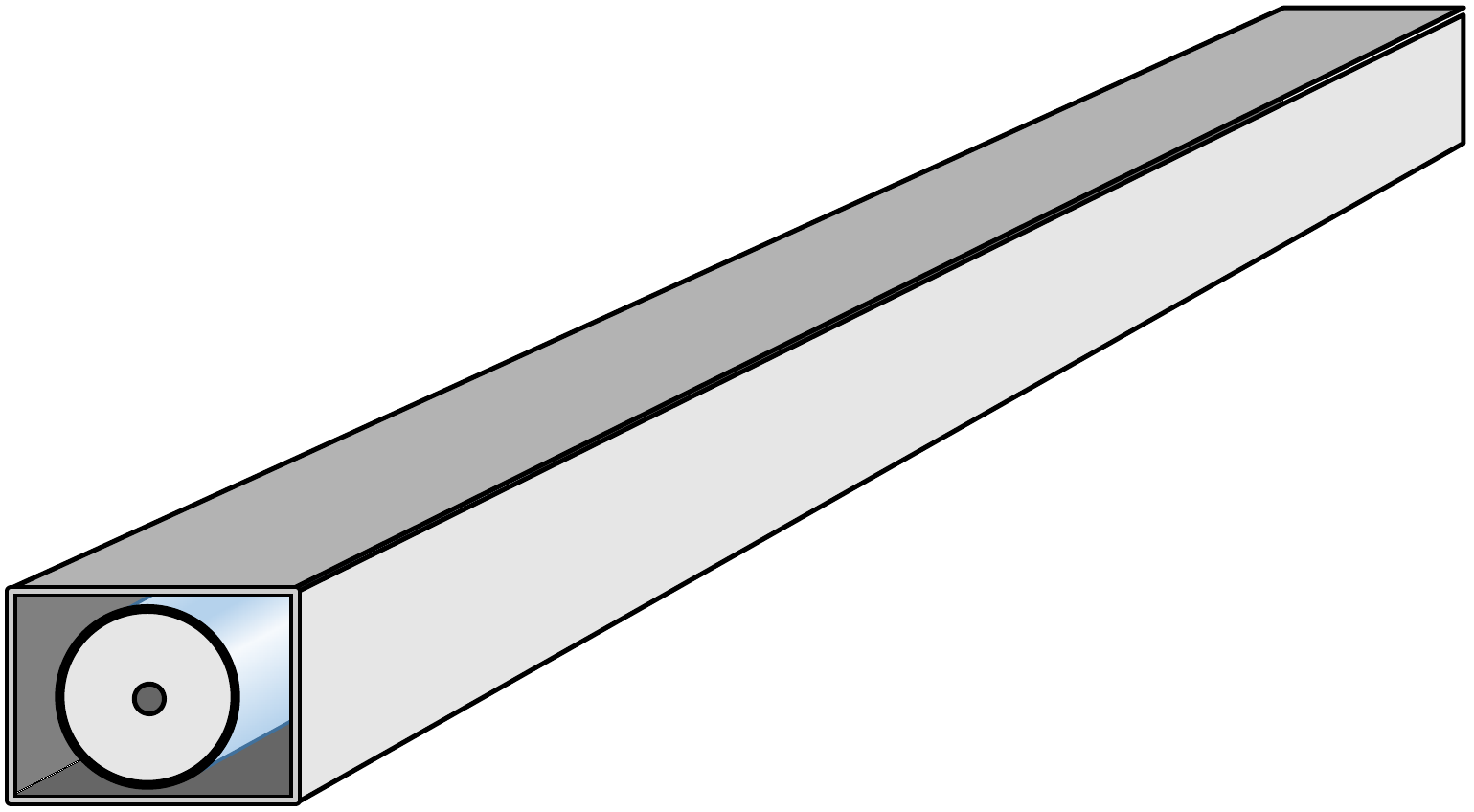}}
\put(35.0,0.0){\includegraphics[height=30mm]{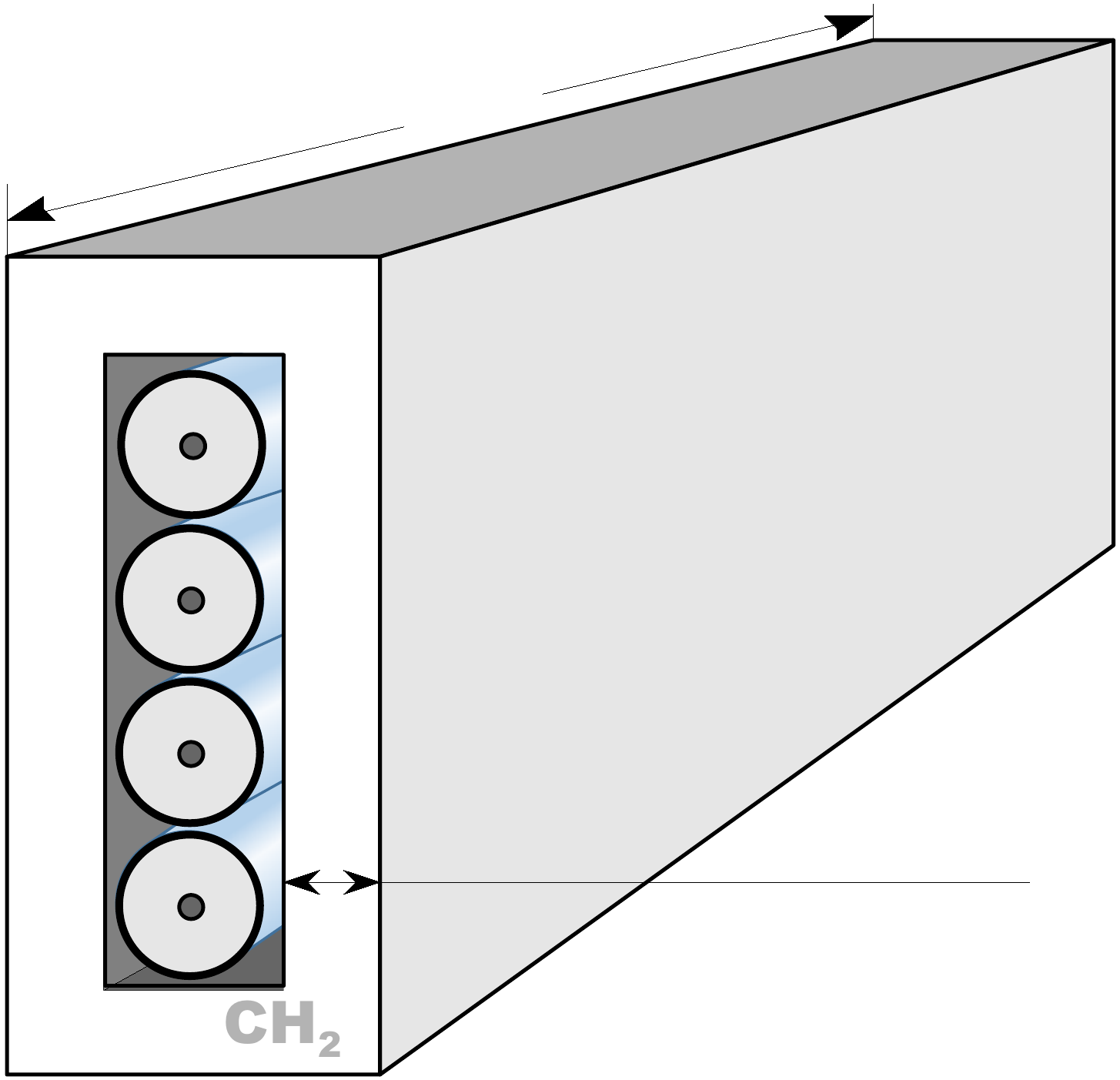}}
\put(47.0,28.0){\rotatebox{15}{\tiny\sf 1 m}}

\put(5.0,20.0){\parbox{20mm}{\begin{flushleft}\scriptsize\sf
    Proportional\\gas counters\\[0.5mm]filled with $^3$He\end{flushleft}}}
   \put(22.0,19.0){\line(1,0){17.5}}
   \put(22.0,19.0){\line(-1,-2){7.5}}
   \put(22.0,19.0){\line(2,-1){17.5}}
   \put(65.5,7.5){\makebox(0,0)[r]{\tiny\sf 2 cm}}
  \put(80.0,0.0){\includegraphics[width=60mm]{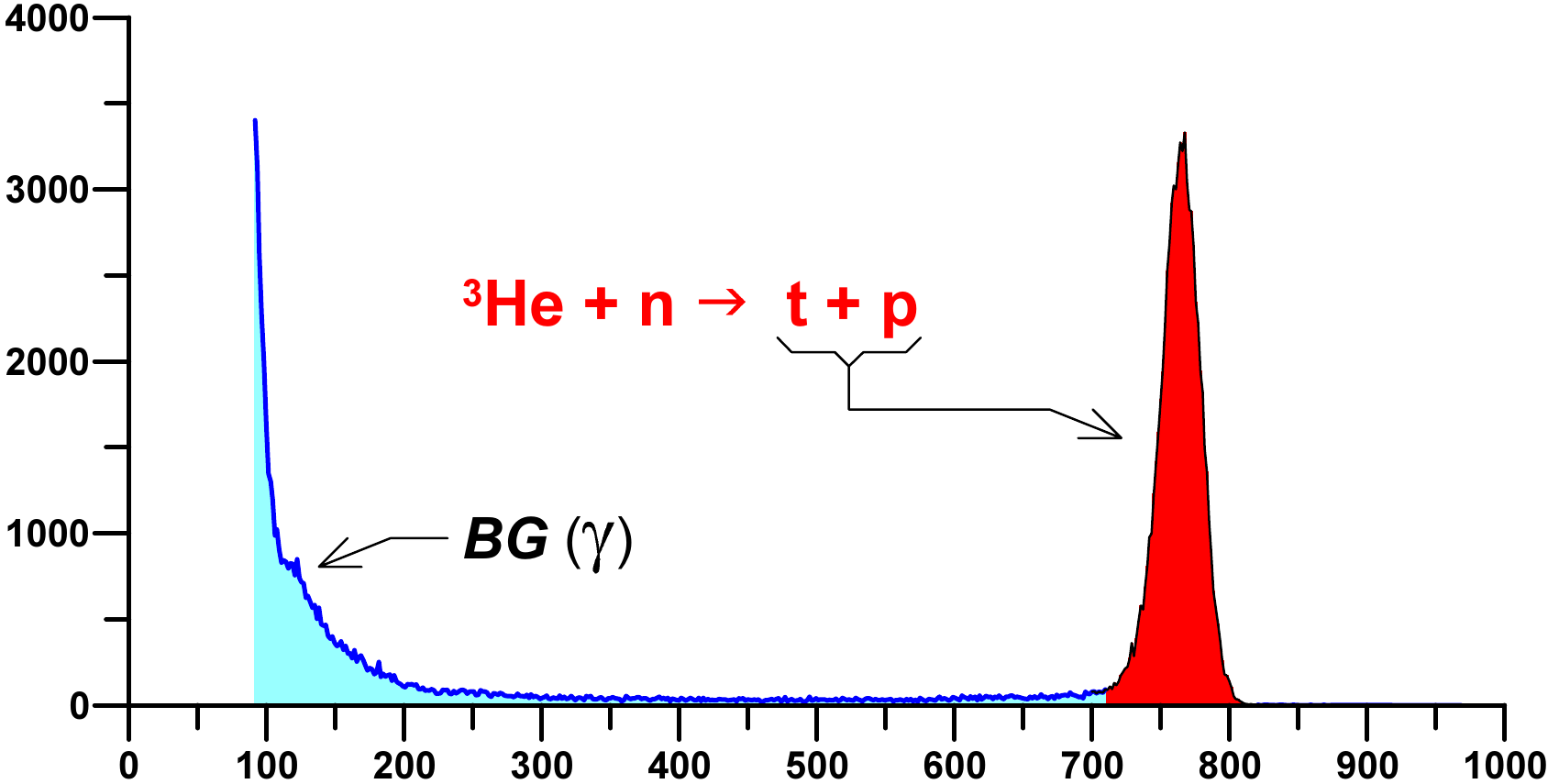}}
   \put(87.0,30.0){\makebox(0,0)[lt]{\scriptsize\sf Count Rate, a.u.}}
   \put(145.0,5.0){\makebox(0,0)[rb]{\scriptsize\sf E, keV}}
   \put(120.0,16.0){\makebox(0,0)[c]{\tiny\sf 764 keV}}
 \end{picture}
 \caption{Neutron detectors based on the $^3$He gas counters and an energy spectrum measured with it.}
 \label{Fig.He3}
\end{figure}

Two simple spectrometers with proportional $^3$He gas counters (Fig.~\ref{Fig.He3}) were used to detect these neutrons. Neutron capture with high cross-section (5333 barn) is followed by disintegration of the nucleus to a pair of charged particles -- proton and triton. Their total kinetic energy (764~keV) is detected in the gas and can be used to identify the process and thus improve the detector selectivity.
It should be stressed that only \emph {thermal} neutrons are captured in $^3$He. As for a neutron of keV energy range -- it can be captured only after moderation in any hydrogen-containing substance.
As the bigger (four-fold) detector includes relatively thin polyethylene moderator, it is therefore sensitive to both thermal and \emph{epithermal} neutrons (but not to the MeV-neutrons). Results of the measurements performed under different conditions are shown in Table~\ref{Tab.Neutron_Flux}.

\begin{table}[hbt]
{\footnotesize
\begin{tabular*}{150mm}{@{\extracolsep{\fill}}|l||c||c||c|c|c|c|c|c||c|c|c|}\hline
 Site                  & Outdoor & 2-A107 & \multicolumn{6}{c||}{3-A336} & \multicolumn{3}{c|}{4-A336 \rule[-1.0mm]{0mm}{5mm}}\\ \hline
 $P_r$ [GW$_{\rm th}$] & --  &  3.0 & \multicolumn{2}{c|}{ 0 }  & \multicolumn{3}{c|}{ $\simeq$1.6 } & {\color{red}\bf 3.1} & 0 & 3.1 & {\color{red}\bf 3.1}  \rule[-1.0mm]{0mm}{5mm}\\ \hline
 CHB [cm]       & 0 & 0 & 0 & 8 & 0 & 8 & 16 & {\color{red} 0} & 0 & 0 & {\color{red} 0}  \rule[-1.0mm]{0mm}{5mm}\\ \hline \hline
 $\Phi_n$ [n/m$^2$/s] & 11$_{\it 2}$ &  0.87$_{\it 1}$   &  0.77$_{\it 4}$   &  0.09$_{\it 1}$  &  222$_{\it 9}$ & 5$_{\it 1}$ &  0.25$_{\it 1}$ & {\color{red}\bf 140$_{\it 2}$} & 1.2$_{\it 1}$ & 1240$_{\it 5}$ & {\color{red}\bf 590$_{\it 3}$} \rule[-1.5mm]{0mm}{6mm}\\ \hline
\end{tabular*}
}
\caption{Flux of the thermal and epithermal neutrons ($\Phi_n$) measured with the four-fold detector under different conditions -- the reactor thermal power ($P_r$) and the detector shielding with borated polyethylene (CHB). Red colour corresponds to the final values after modification of the neutron stoppers.}
\label{Tab.Neutron_Flux}
\end{table}

It is seen that two layers of borated polyethylene made of standard C3 modules are quite enough to suppress the total flux of low energy neutrons by three orders of magnitude. Dedicated works have been done with some of 27 steel tubes in both rooms 3-A336 and 4-A336, increasing their built-in neutron stoppers and thus reducing the background by a factor of four.
As a result, the flux of (thermal + epithermal) neutrons in the rooms is now about 140 and 590 n/m$^2$/s respectively. Distinction between the background in two rooms is caused probably by slightly different equipment installed in the tubes.

In order to get more detailed information about neutron energy spectrum some additional tests were performed with a single $^3$He detector covered with a polyethylene moderator of different thickness and cadmium neutron absorber (0.5~mm thick Cd foil) outside the PE moderator (Table~\ref{Tab.Neutron_Moderator_Cd}).

\begin{table}[hbt]
{\small
\begin{tabular*}{150mm}{@{\extracolsep{\fill}}|l||r|r|r|r|r|r|r|r|}\hline
 Moderator thickness [mm] & 0 & 2 & 4 & 6 & 8 & 10 & 20 & 30 \rule[-1.0mm]{0mm}{5mm}\\ \hline
 CR [1/s] without shielding & 9.66 & 9.84 & 10.46 & 11.42 & 12.53 & 13.42 & 14.32 & 12.77
 \rule[-1.0mm]{0mm}{5mm}\\ \hline
 CR [1/s] with ext. Cd shield & 1.12 &   &   &   &   & 6.37 & 10.44 & 9.95
 \rule[-1.0mm]{0mm}{5mm}\\ \hline
\end{tabular*}
}
\caption{Count rate detected with a single $^3$He tube covered with PE moderator and external absorber of thermal neutrons (0.5 mm cadmium foil). Statistical accuracy is within 1\%.}
\label{Tab.Neutron_Moderator_Cd}
\end{table}

From the Tab.~\ref{Tab.Neutron_Moderator_Cd} it is seen that neutron spectrum consists mainly of approximately comparable number of thermal and epithermal neutrons. Indeed, cadmium foil cuts thermal neutrons by order of magnitude, but being placed outside the moderator reduces the count rate by a quarter only. It means that only this rejected quarter corresponds to incoming thermal neutrons. More detailed MC simulation gives the following estimation of the spectrum:
\begin{equation}
\left(E_n\!\!<\!0.025\,{\rm eV}\right)\!:\!
\left(0.025\,{\rm eV}\!\!<\!E_n\!\!<\!1\,{\rm eV}\right)\!:\!
\left(1\,{\rm eV}\!\!<\!E_n\!\!<\!1\,{\rm keV}\right)\!:\!
\left(1\,{\rm keV}\!\!<\!E_n\right)
\simeq 1\!\!:\!\!100\!:\!\!18\!:\!40.
\end{equation}

\subsection{Muon background}\label{Section.Muon_BG}
\begin{figure}[ht]
 \setlength{\unitlength}{1mm}
\begin{minipage}[t]{45mm}
\centering
 \begin{picture}(45,42)(0,0)
  %\put(0,0){\framebox(45,42)[b]{}}
  \put(17,0){\includegraphics[height=23mm]{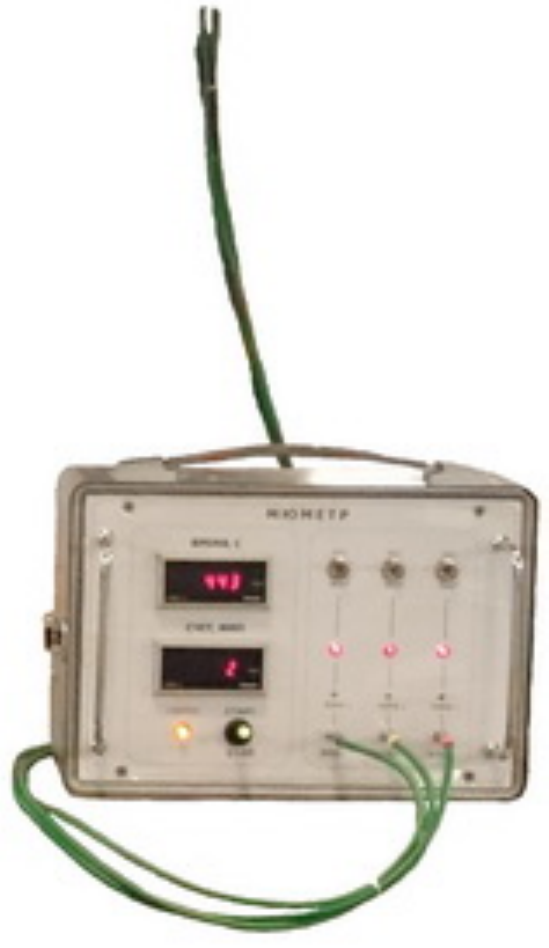}}
  \put(0,0){\includegraphics[height=42mm]{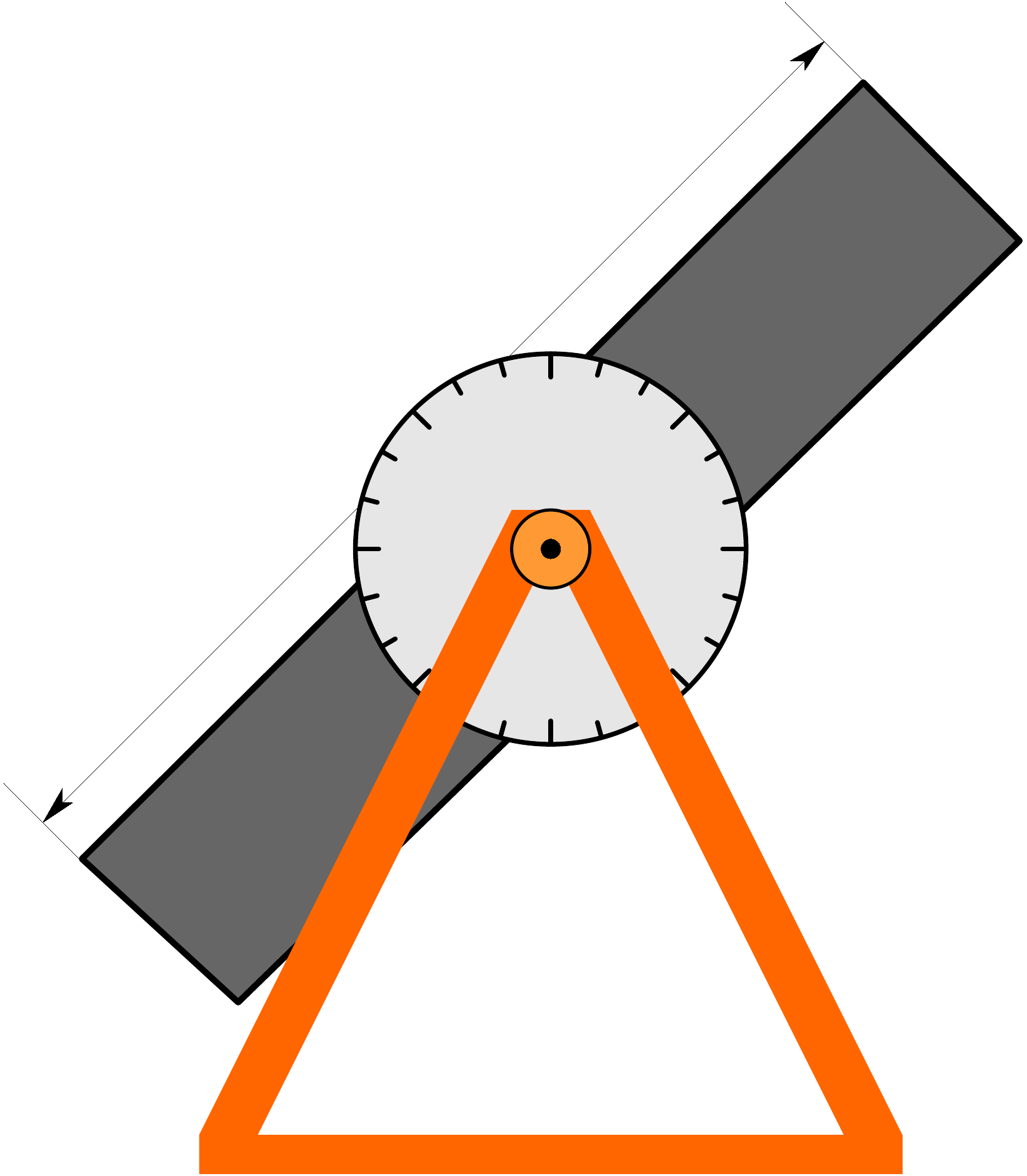}}
  \put(11,26){\rotatebox{45}{\scriptsize\sf 1 m}}
  \put(10.0,40.0){\makebox(0,0)[c]{\normalsize\sl ``MuMeter''}}
 \end{picture}
 \caption{Telescopic detector of cosmic muons placed on a turret.}
 \label{Fig.MuMeter}
\end{minipage}\hfill{ }
\begin{minipage}[t]{55mm}
\centering
\begin{picture}(55,42)(0,0)
  %\put(0,0){\framebox(55,42)[b]{}}
  \put(0,0){\includegraphics[height=43mm]{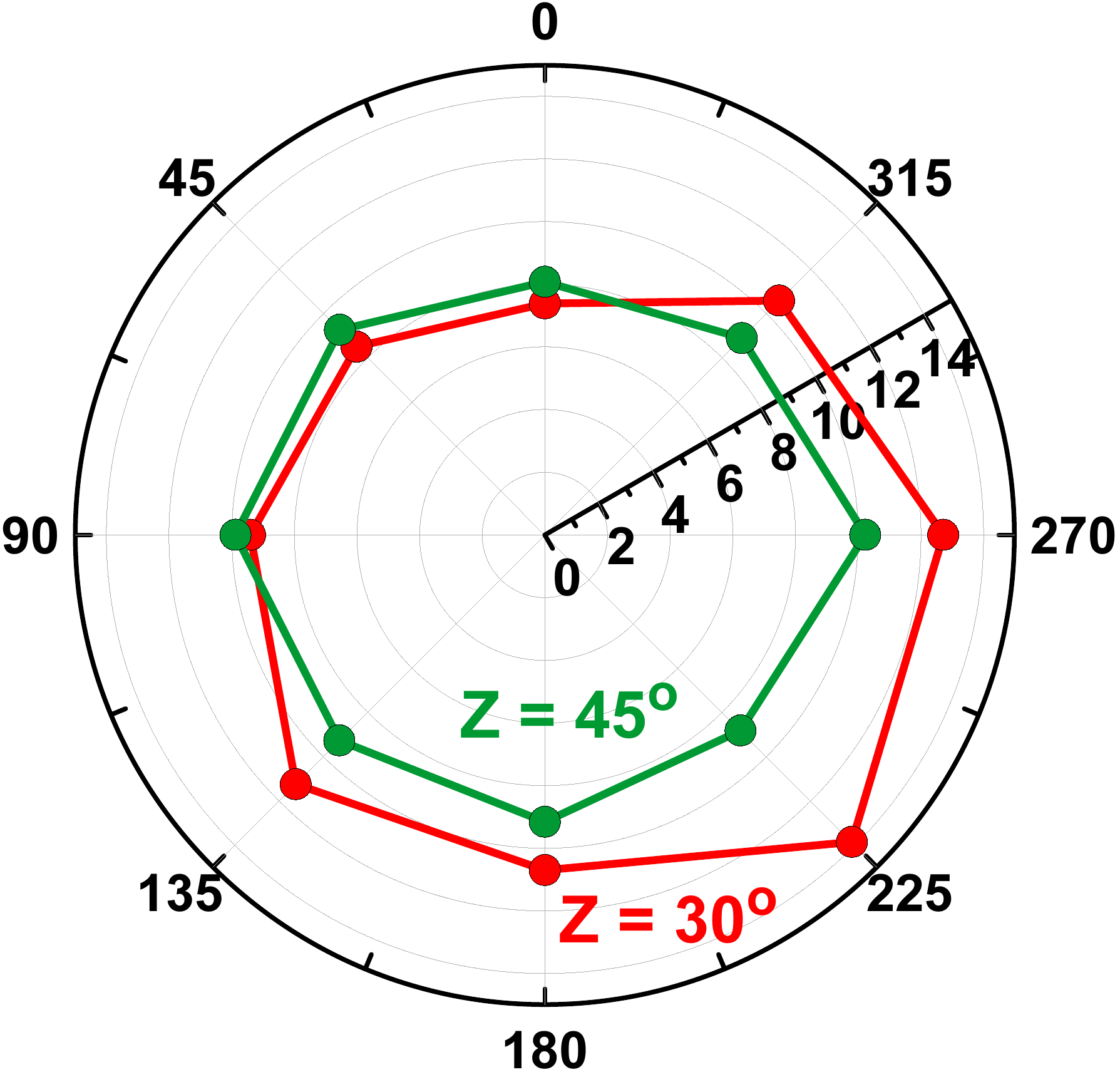}}
  \put(40.0,31.0){\makebox(0,0)[l]{\small\sf $\Phi_\mu$, \scriptsize\sf $\mu$/m$^2$/sr/s}}
\end{picture}
 \caption{An absolute flux $\Phi_\mu$ vs azimuth angle A at the DANSS site.}
 \label{Fig.Mu(A)}
\end{minipage}\hfill{ }
\begin{minipage}[t]{43mm}
\centering
\begin{picture}(43,42)(0,0)
  %\put(0,0){\framebox(43,42)[b]{}}
  \put(0,0){\includegraphics[height=42mm]{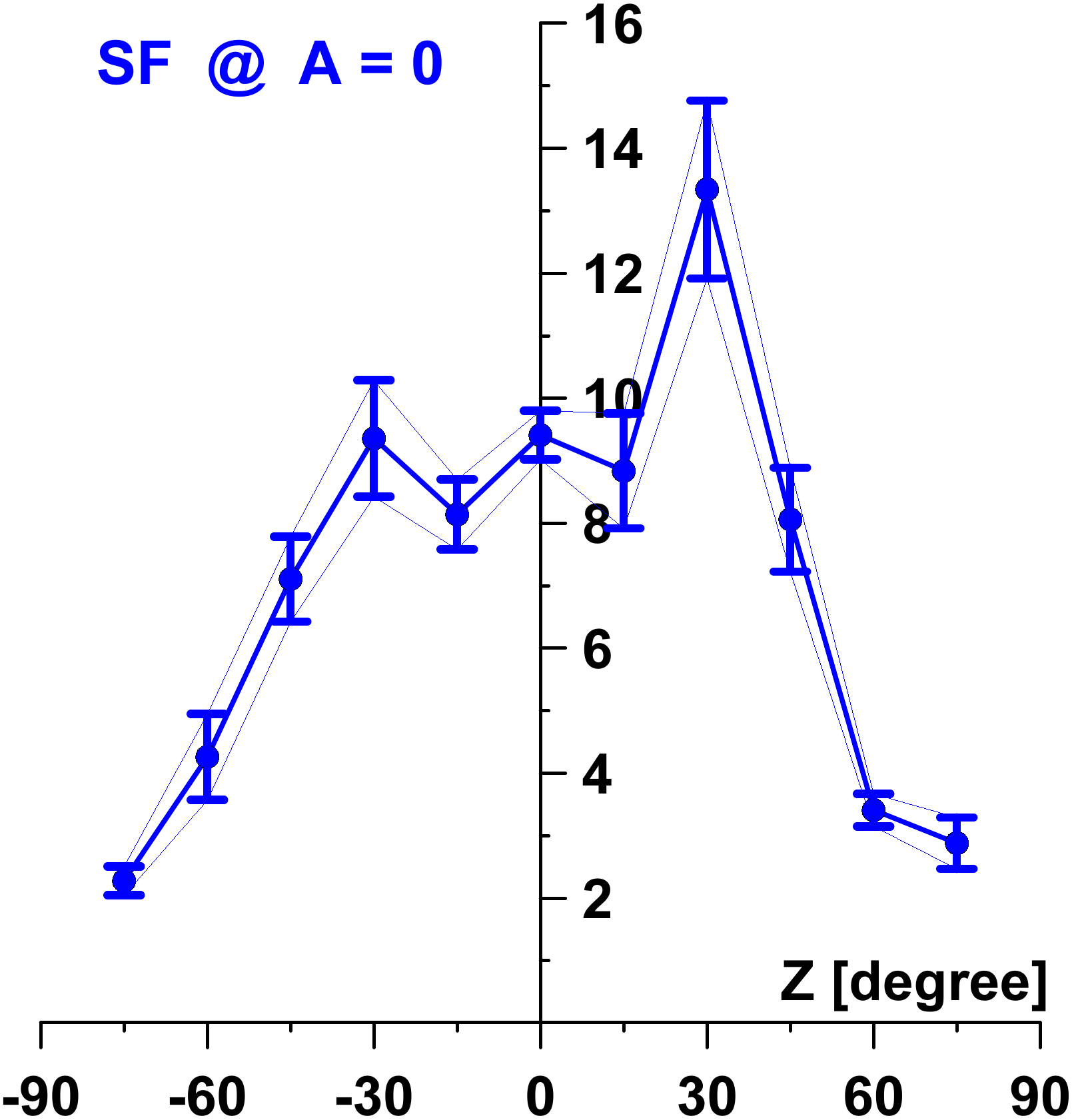}}
 \end{picture}
 \caption{Suppression factor SF vs zenith angle Z.}
 \label{Fig.Att(Z)}
 \end{minipage}
\end{figure}

Absolute value of the flux and angular distribution of cosmic rays in the room A336 differ from typical ones \cite{Heusser} because of the above building structure.
In order to investigate charged component of the cosmic background (i.e., muons) a simple ``MuMeter'' detector (Fig.~\ref{Fig.MuMeter}) was used. It includes three plastic scintillator disks ($\oslash$12.7 cm $\times$ 3 cm) coupled to five-inch PMTs and operating in coincident mode\footnote{Triple coincidences between three aligned counters reduce background count rate down to negligible level.}.
As a result, the telescope has an aperture of 0.013~m$^2$ and solid angle of 0.014~sr.

As an example, Fig.~\ref{Fig.Mu(A)} shows an azimuth angular distribution of the muon flux $\Phi_\mu$ at zenith angles of Z=30$^\circ$ and Z=45$^\circ$. Due to complicated structure of the reactor building this distribution is rather far from isotropic.

Comparing measurements performed at the DANSS site and at {\sl plein air}, one can deduce a suppression factor SF for differend directions. Fig.~\ref{Fig.Att(Z)} shows zenith angle dependence of this factor.

\section{The detection idea}\label{Section.Idea}

The most appropriate reaction which is generally used to detect reactor antineutrino was mentioned first by H.~Bethe and R.~Peierls more than 80 years ago~\cite{IBD1934}. It is Inverse Beta-Decay (IBD):
\begin{equation}
 \overline{\nu}_e+p\rightarrow e^++n\;.
 \label{Eq.IBD}
\end{equation}
In case the proton is not bound in a compound nucleus, the IBD cross-section can be expressed~\cite{Llewellyn,Vogel99,Mikaelyan} as
\begin{equation}
\sigma(E)\simeq 9.556\times10^{-44}\cdot\frac{886}{\tau_n}\cdot(E-\Delta)\cdot\sqrt{\left(E-\Delta\right)^2-m_e^2}\;,
\label{Eq.IBD_Section}
\end{equation}
where neutron lifetime $\tau_n$ is given in seconds and $\Delta$=1.293~MeV. More precise formula includes some important corrections (see, e.g., \cite{Kopeikin-Mikaelyan-Sinev} or \cite{Mention2011}) which could change slightly the absolute value of $\sigma$ but not its general character. It is easily seen that energy threshold of the IBD reaction is
\begin{equation}
E^{\rm thr} = \Delta + m_e = 1.804\;{\rm MeV}\;,
\label{Eq.IBD_Threshold}
\end{equation}
and the detection efficiency depends very much on the neutrino energy.

An absolute energy spectrum of antineutrino emitted by nuclear reactor is a subject of numerous theoretical and experimental investigations. Unfortunately, it cannot be extracted directly from experimentally measured overall cumulative $\beta$-spectrum emitted by the reactor fuel, but instead requires precise knowledge of intensities and spectra of {\sl each} decay branch of {\sl each} fission product {\sl separately}. This information is not available at the present time -- only few tens $\beta$-transitions from several thousands have been ever measured with enough precision. So, the reactor neutrino spectrum could be evaluated merely as a result of very imprecise and model-dependent theoretical estimations started 35 years ago \cite{Kopeikin80,Vogel81,Schrekenbach82} and continued up to now \cite{FrenchNeutrinoSpectrum,Kopeikin12}.

Additional uncertainty is caused by variation of the fuel composition within the reactor life-time: initial active $^{235}$U isotope burns out intensively, whereas so-called ``weapon'' $^{239}$Pu is produced from a ballast $^{238}$U which represents about 88\% of the fuel mass \footnote{It is not the case of small research reactors which use almost pure $^{235}$U.}.
The spent fuel is kept in a coolant pond nearby the reactor for 3--4 years before evacuation and $\beta$-decay of its long-lived nuclei contributes to the neutrino spectrum as well, so that it depends on many factors such as reactor type and actual status, previous fuel history, construction of the reactor building, etc.

According to the work \cite{Kopeikin12}, typical Russian reactor WWER1000 of the PWR type\footnote{Pressurized Water Reactors (PWRs) use water under high pressure as a coolant and neutron moderator at the same time. They constitute the large majority of all nuclear power plants.} operating in the middle of the campaign undergoes $0.915\times10^{20}$~fissions per second, emitting $\sim6.7$~antineutrinos per fission.

Significant portion of these neutrinos cannot be detected because of the reaction threshold (\ref{Eq.IBD_Threshold}), especially in case of inertial long-lived part of the fission products. Indeed, as the beta-decay probability depends on the decay energy roughly as $Q^5$, only short-lived nuclei are those which can emit above-threshold neutrinos (exception could be forbidden transitions or so-called ``generator pairs'', but they are not numerous).
It means that an IBD-based neutrino detector is useless for monitoring of the spent fuel or radioactive waste. On the other hand, it makes the detector even more sensitive to the fission process itself -- to be exact, to numerous extremely short-lived fission products which reflect the reactor status.

The IBD detection (\ref{Eq.IBD}) proceeds in two steps: the first one applies to the positron and the second to the neutron. After subtraction of the threshold value most of the remaining neutrino energy is transferred to the positron and only few keV to the neutron\footnote{Unfortunately, angular correlation between momenta of the neutrino and positron is negligible and only neutron ``remembers'' the neutrino direction, but multiple neutron scattering during thermalization makes direction sensitivity of the IBD process very weak.}. The positron deposits its energy within a short range of few cm and then annihilates emitting two 511~keV photons at 180$^\circ$. As a result, the first (Prompt) energy deposit is distributed in space in a very specific way.

The second (Delayed) step is the detection of the neutron. After moderation in 1-3 cm of the plastic scintillator it is captured by $^{157}$Gd or $^{155}$Gd with a very high cross-section. In both cases a cascade of $\gamma$-rays is emitted with the total energy of about 8~MeV. Because of high multiplicity and deep penetration in plastic these $\gamma$-rays produce a flash which is spread widely within a sphere with a diameter of about 30-40~cm, so that a number of strips in several X and Y modules are usually fired. Distribution of time between the Prompt and Delayed signals is described by a combination of two exponents
 \begin{equation} \label{Eq.f(t)}
 f_1(t)=\frac1{\tau_c-\tau_m}\left(e^{-t/\tau_c}-e^{-t/\tau_m}\right)\;,
 \end{equation}
where the characteristic times $\tau_m$ and $\tau_c$ correspond to the neutron moderation and capture respectively and depend on the detector structure.

Though the IBD event has a very specific signature, it occurs under intense external and internal $\gamma$, $n$ and $\mu$ background. Therefore, adequate selection rules including {\sl hardware trigger} (see section~\ref{Section.ACQ}) should be strong enough to allow extraction pairs of Prompt (P) and Delayed (D) signals originating from the same position and with characteristic space pattern and energy.

\section{The basic scintillator element}\label{Section.Strips}
The basic element of DANSS is a polystyrene-based extruded scintillator strip ($1\times4\times100$~ cm$^3$) similar to one used in the MINOS detectors \cite{MINOS_scintillator}, but with gadolinium additive.

\subsection{Gadolinium contamination}

\begin{figure}[bht]
 \setlength{\unitlength}{1mm}
\begin{minipage}[t]{65mm}
\centering
 \begin{picture}(65,39)(0,1)
  %\put(0,0){\framebox(65,40)[b]{}}
  \put(10,0.5){\includegraphics[height=36mm]{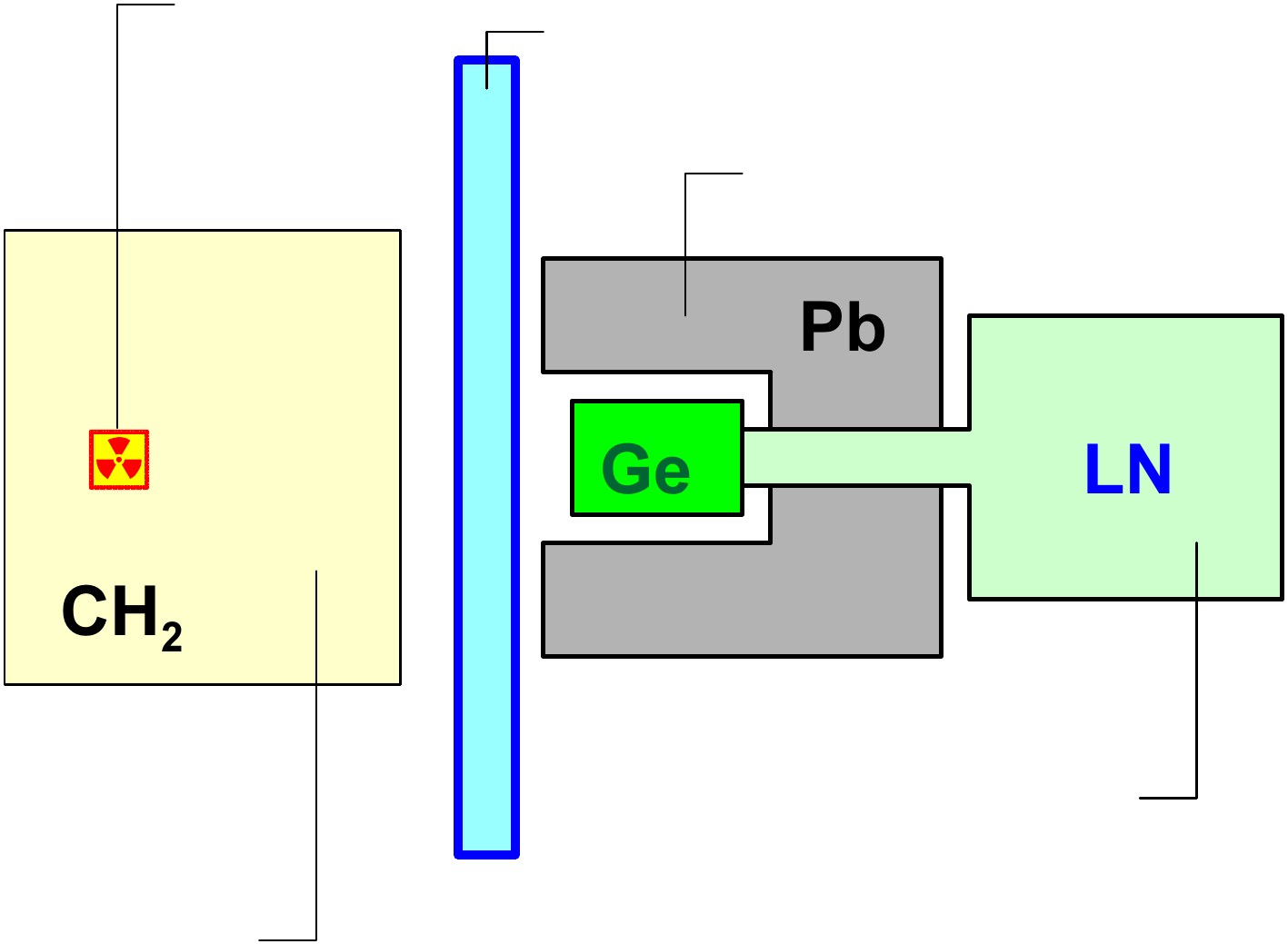}}
  \put(13.0,34.0){\parbox{16mm}{\begin{center}\scriptsize\sf Neutron\\[-0.9mm] Source \end{center}}}
  \put(5.0,1.5){\parbox{16mm}{\begin{center}\scriptsize\sf Neutron\\[-0.9mm] Moderator \end{center}}}
  \put(39.0,30.0){\makebox(0,0)[l]{\scriptsize\sf Lead shield}}
  \put(31.5,35.0){\parbox{36mm}{\scriptsize\sf Strip with Gd-coating\\[-0.5mm] being measured}}
  \put(43.0,4.0){\parbox{26mm}{\scriptsize\sf Cryostat\\[-0.5mm] with Ge-detector}}
 \end{picture}
 \caption{A simplified scheme of the test bench with a neutron source and Ge $\gamma$-detector.}
 \label{Fig.Gd_Scheme}
 \end{minipage}\hfill{ }
\begin{minipage}[t]{80mm}
\centering
\begin{picture}(80,39)(0,1)
  %\put(0,0){\framebox(80,40)[b]{}}
  \put(10,0){\includegraphics[height=40mm]{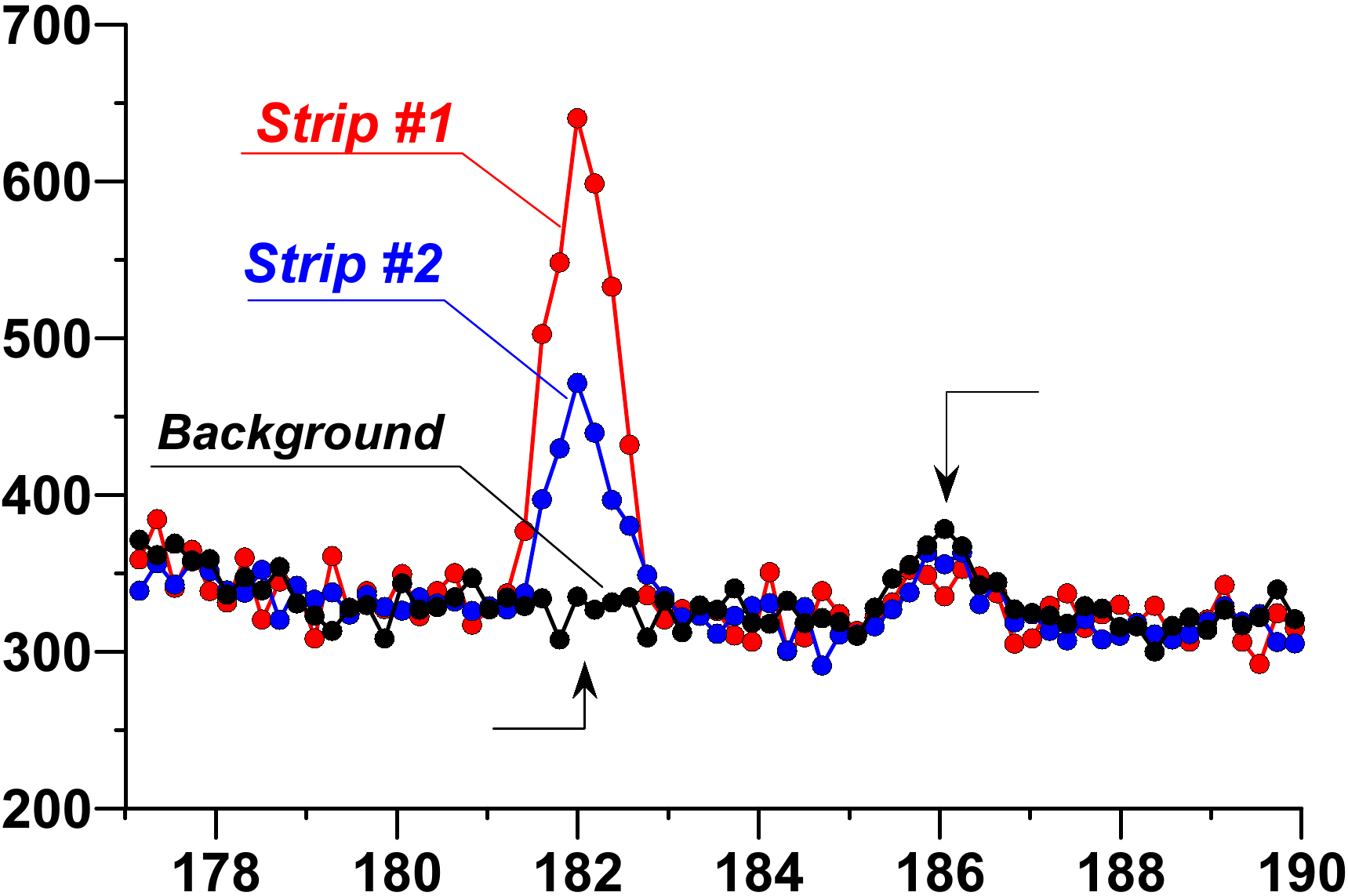}}
    \put(17.0,39.5){\makebox(0,0)[lt]{\scriptsize\sf Counts / keV / hour}}
    \put(70.0,5.0){\makebox(0,0)[rb]{\scriptsize\sf Gamma Energy, keV}}
    \put(55.0,25.0){\makebox(0,0)[l]{\footnotesize\sf $^{226}$Ra}}
    \put(31.5,7.5){\makebox(0,0)[r]{\footnotesize\sf $^{157}$Gd($n,\gamma$)}}
 \end{picture}
 \caption{Fragments of $\gamma$-spectra measured for two strips with different Gd percentage in the coating.}
 \label{Fig.Gd_Spectra}
 \end{minipage}
\end{figure}

The plastic is doped with 1\% PPO plus 0.03\% POPOP fluors and co-extruded with a white layer (about 0.1--0.2~mm) to contain scintillation light. In order to provide ($n,\gamma$)-conversion, chemical composition of this layer was changed. The optimized coating consists of polystyrene with 18\% admixture of rutile and 6\% of gadolinium oxide, so that the final Gd density is about 1.6~mg/cm$^2$, which corresponds to $\sim$0.35\%$_{\rm wt}$ with respect to the whole detector body. Verification of Gd percentage in the final strips is presented in  Figs.~\ref{Fig.Gd_Scheme} and \ref{Fig.Gd_Spectra}.

\subsection{Signal yield}\label{Section.Strips.Yield}
\begin{figure}[ht]
 \setlength{\unitlength}{1mm}
\begin{minipage}[t]{75mm}
\centering
 \begin{picture}(75,36)(0,1)
  %\put(0,0){\framebox(75,31)[b]{}}
  \put(0,0){\includegraphics[width=75mm]{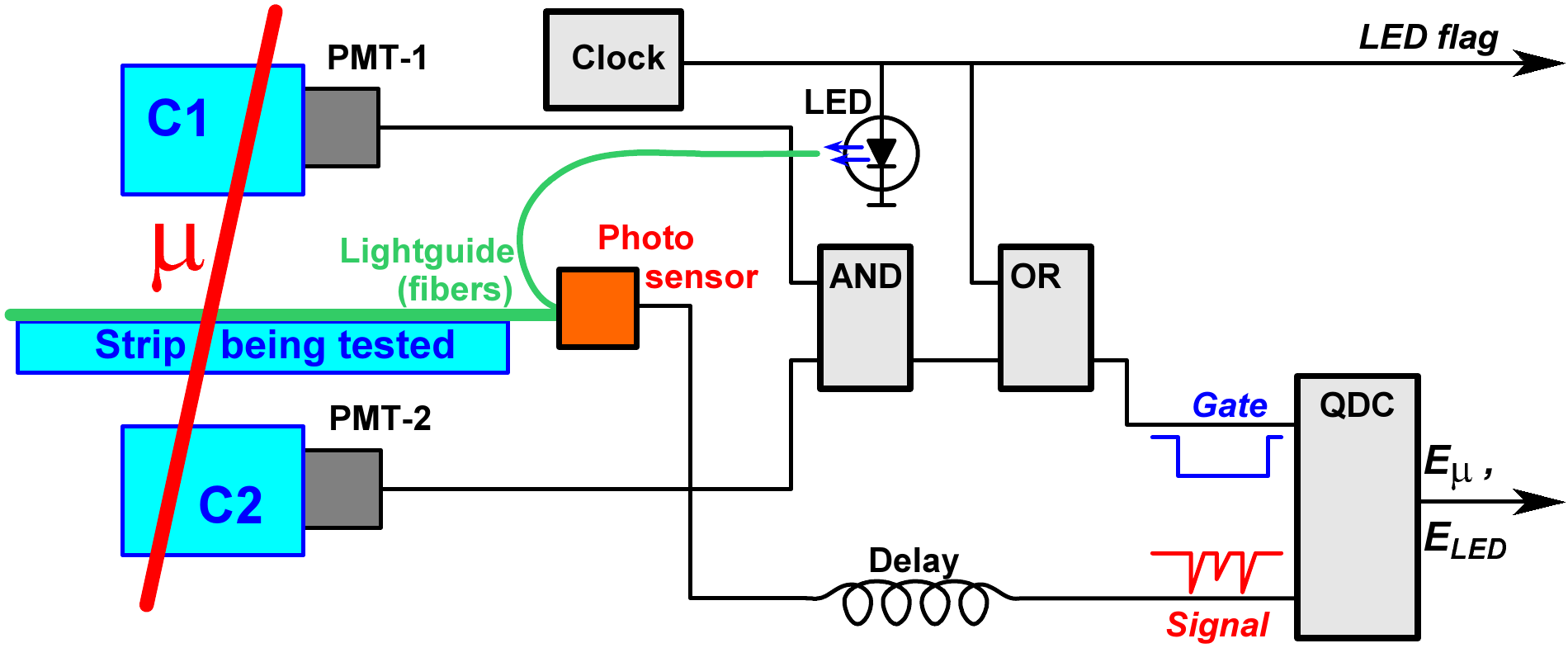}}
    \put(55.0,23.0){\makebox(0,0)[l]{\scriptsize $\lambda=\overline{n\cdot E_{\rm LED}}$}}
 \end{picture}
 \caption{A simplified scheme of the test bench with a telescope of cosmic muons (C1,C2) and LED producing $\lambda$ photo electrons per pulse on the average.}
 \label{Fig.Test_Bench}
 \end{minipage}\hfill{ }
\begin{minipage}[t]{65mm}
\centering
\begin{picture}(65,36)(0,1)
  %\put(0,0){\framebox(65,31)[b]{}}
  \put(0,3.0){\includegraphics[width=50mm]{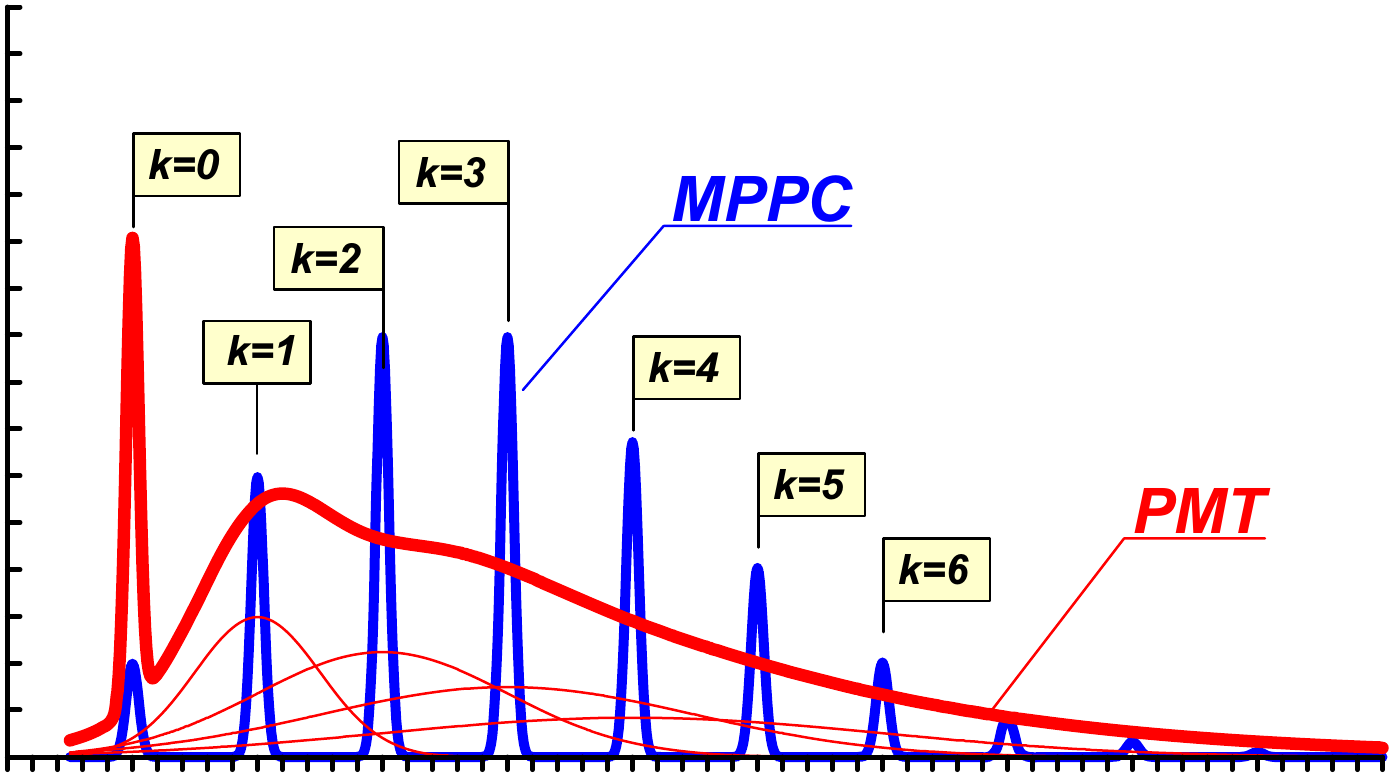}}
    \put(2.0,31.0){\makebox(0,0)[tl]{\scriptsize\sf Relative Count Rate}}
    \put(64.0,0.0){\makebox(0,0)[rb]{\scriptsize\sf Hypothetical QDC channels}}
    \put(54.0,30.0){\makebox(0,0)[t]{\scriptsize $\mathcal{P}(k)$ = \normalsize \begin{math}\frac{e^{-\lambda}\,\cdot\, \lambda^k}{k\,!}\end{math}}}
    \put(54.0,23.0){\makebox(0,0)[t]{\scriptsize $I_0/I_\Sigma\;=\;\mathcal{P}(0)$}}
    \put(54.0,18.0){\makebox(0,0)[t]{\scriptsize $\lambda\;=\;-\ln\left(I_0/I_\Sigma\right)$}}
 \end{picture}
 \caption{Two spectra of the LED weak pulses detected with different photo sensors -- PMT and MPPC.}
 \label{Fig.Poisson_MC_Spectra}
 \end{minipage}
\end{figure}

One of the most important characteristics of any detector is its energy resolution. In our case it is determined mainly by a tiny signal produced at an input of a photo sensor -- photomultiplier tube (PMT) or multipixel photon counter (MPPC). Later, this signal is amplified by 5-6 orders of magnitude with the PMT dynode multiplying or the MPPC Geiger avalanche, but the final uncertainty still corresponds to the statistical dispersion of a number of initial photoelectrons at the PMT cathode (or a number of the MPPC pixels fired). A norm of this number $n$ corresponding to a 1~MeV energy deposit in the detector depends on the scintillator quality, light collection efficiency and quantum yield of the photo sensor. In order to optimize these parameters and increase the $n$ value, a simple test-bench shown in Fig.~\ref{Fig.Test_Bench} was used. It allows to measure a real signal produced in the detector by cosmic muons and weak light pulses produced by a light-emitting diode (LED).

Energy spectrum measured with MPPC consists of a number of narrow peaks (Fig.~\ref{Fig.Poisson_MC_Spectra}). Each peak corresponds to a number of pixels fired ($k=0,1,2,\ldots$), so that it is very easy to calibrate the spectrometric channel in terms of photoelectrons.

\begin{figure}[htb]
\setlength{\unitlength}{1mm}
\begin{minipage}[t]{70mm}
\centering
\begin{picture}(70,30)(0,0)
  %\put(0,0){\framebox(70,30)[b]{}}
\put(0,0){\includegraphics[width=69mm]{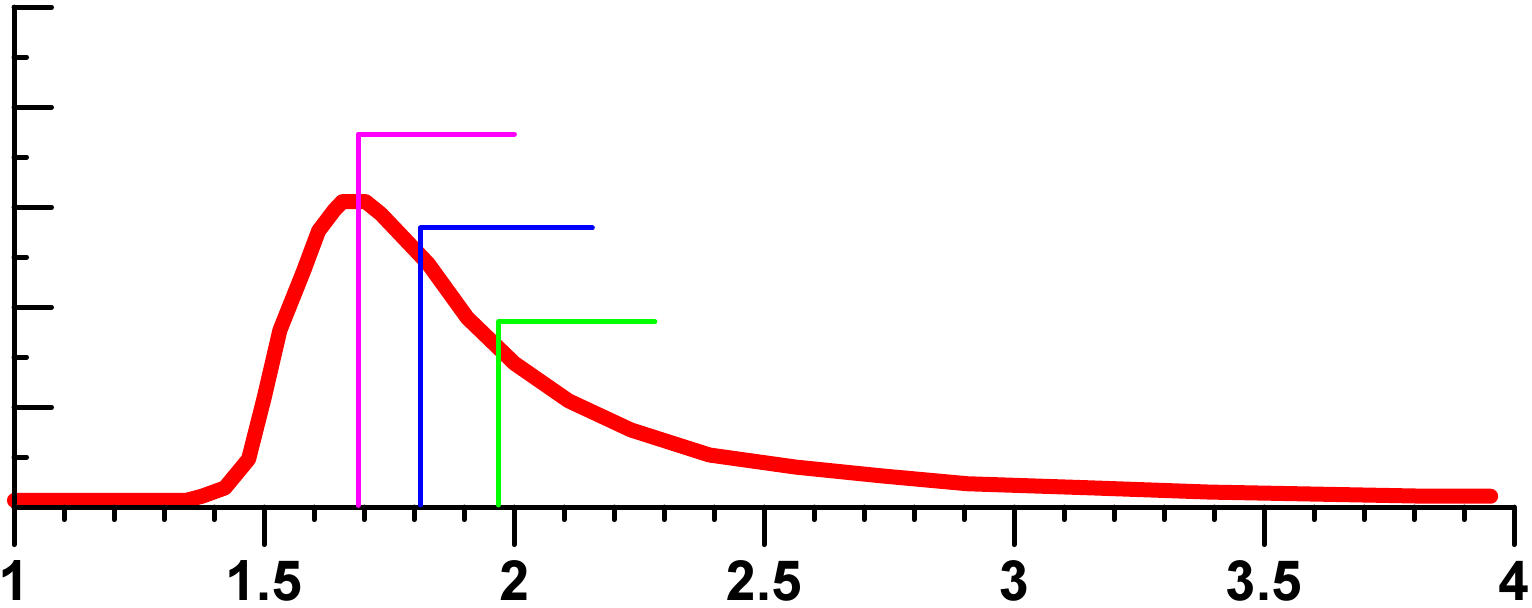}}
  \put(3.0,30.0){\makebox(0,0)[lt]{\scriptsize\sf Relative Intensity}}
  \put(23.0,21.0){\makebox(0,0)[l]{\scriptsize\sf\color{magenta} Mode = 1.67 MeV}}
  \put(27.5,18.0){\makebox(0,0)[l]{\scriptsize\sf\color{blue} Median = 1.82 MeV}}
  \put(30.0,12.3){\makebox(0,0)[l]{\scriptsize\sf\color{green} Mean = 1.97 MeV}}
  \put(68.0,7.0){\makebox(0,0)[rb]{\scriptsize\sf Energy,  MeV}}
  \put(55.0,20.0){\color{red}\framebox(15,10)[c]{\Large\bf MC}}
\end{picture}
\caption{MC-simulated energy deposit produced by cosmic muons in the DANSS strip.}
\label{Fig.Cosmic_MC}
\end{minipage}\hfill{ }
\begin{minipage}[t]{70mm}
\centering
\begin{picture}(70,30)(0,0)
  %\put(0,0){\framebox(70,30)[b]{}}
\put(0,0){\includegraphics[width=70mm]{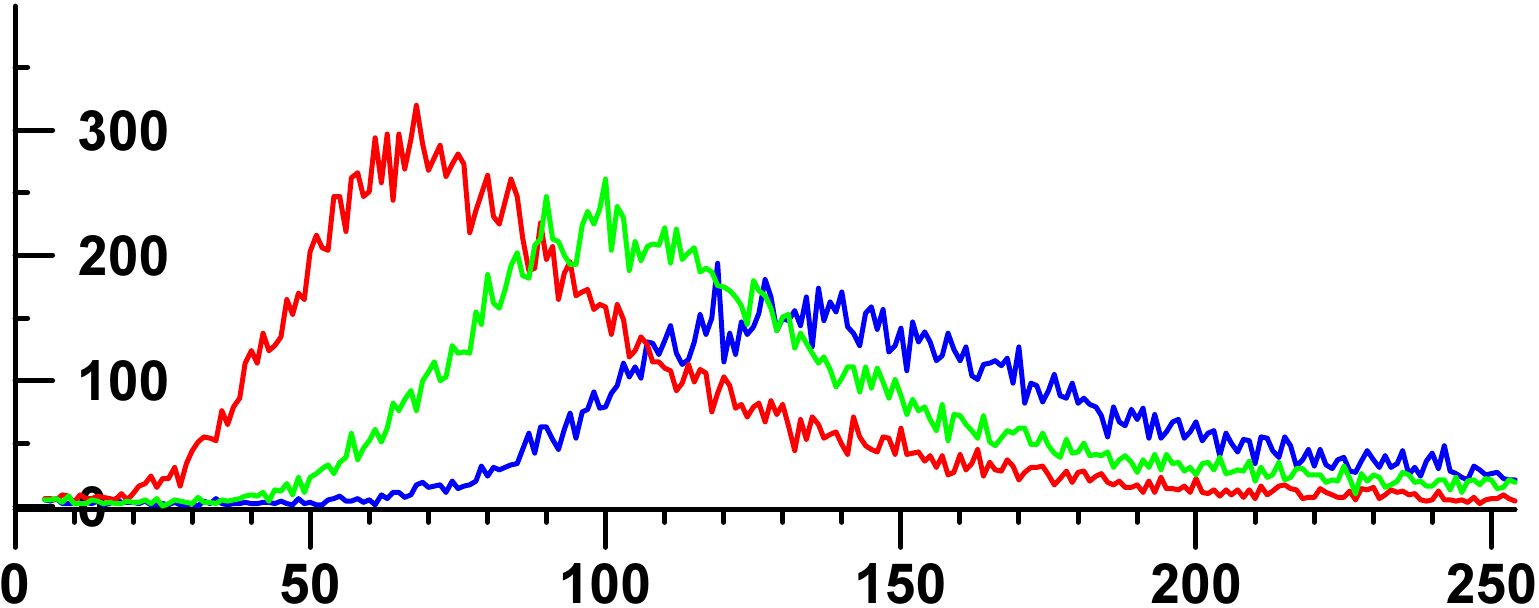}}
  \put( 3.0,30.0){\makebox(0,0)[lt]{\scriptsize\sf Counts per channel}}
  \put(69.0,8.5){\makebox(0,0)[rb]{\scriptsize\sf QDC chnl.}}
  \put(48,22){\parbox{25mm}{\footnotesize\sl Energy deposit\\ by cosmic muons\\ in a strip}}
\end{picture}
\caption{Examples of energy spectra detected with PMT under different conditions.}
\label{Fig.Cosmic_Measured}
\end{minipage}
\end{figure}

In case of PMT the spectrum is more complicated because of a random nature of its gain.  All the peaks here are proportionally broadened and cannot be resolved. The only exception is the first peak corresponding to $k=0$ when the signal is absent and the gain is not applied at all. Analysing relative intensity of this peak with a method developed in~\cite{PhotoElectrons}, a mean number of photoelectrons ($\overline{k}=\lambda$) can be evaluated. Matching this $\lambda$ value with a centroid of the whole LED spectrum, the QDC scale can be graduated in terms of photoelectrons.

To do the same in MeV units, a spectrum of the energy deposited in the strip by cosmic muons is used. It depends on many factors and has a specific shape (the Landau distribution convoluted with an energy spectrum and angular distribution of initial cosmic muons). The spectrum simulated for the test-bench geometry with GEANT4 is shown in Fig.~\ref{Fig.Cosmic_MC}. A real response function of the detection system broadens the spectrum (Fig.~\ref{Fig.Cosmic_Measured}) but does not change significantly an energy of some specific points: the peak point (the mode), the equal-areas point (the median) and the balance point (the mean). In our analysis we use the median energy (1.82~MeV) because it can be extracted with a good precision even for relatively low statistics and requires only a number of overflows instead of detailed measurement of a long high-energy tail of the spectrum.

\begin{figure}[htb]
\setlength{\unitlength}{1mm}
\begin{minipage}[t]{50mm}
\centering
\begin{picture}(50,23)(0,1)
  %\put(0,0){\framebox(50,24)[b]{}}
  \put(10,0){\includegraphics[height=20mm]{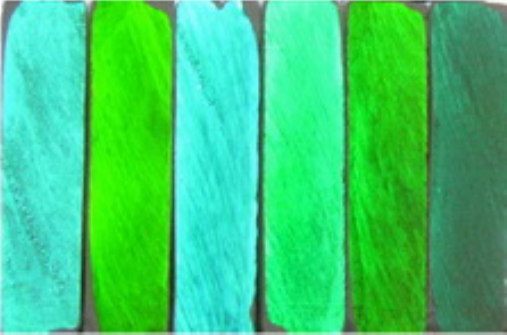}}
  \put(10.0,17.0){\includegraphics[width=6.0mm]{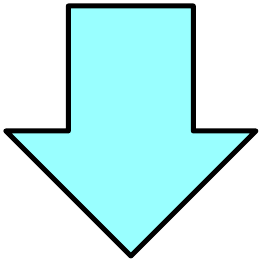}}
  \put(12.8,21.0){\makebox(0,0)[c]{\tiny\bf 16}}
  \put(15.0,18.0){\includegraphics[width=6.0mm]{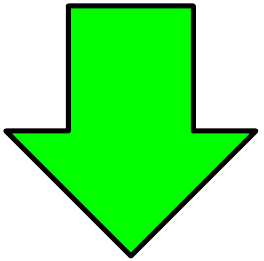}}
  \put(17.8,22.0){\makebox(0,0)[c]{\tiny\bf 10}}
  \put(20.0,17.0){\includegraphics[width=6.0mm]{arrow_blue.pdf}}
  \put(22.8,21.0){\makebox(0,0)[c]{\tiny\bf 17}}
  \put(25.0,18.0){\includegraphics[width=6.0mm]{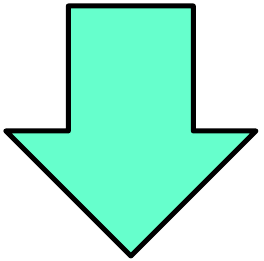}}
  \put(27.8,22.0){\makebox(0,0)[c]{\tiny\bf 13}}
  \put(30.0,17.0){\includegraphics[width=6.0mm]{arrow_green.pdf}}
  \put(32.8,21.0){\makebox(0,0)[c]{\tiny\bf 10}}
  \put(35.0,18.0){\includegraphics[width=6.0mm]{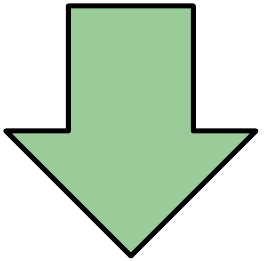}}
  \put(37.8,22.0){\makebox(0,0)[c]{\tiny\bf 12}}
\end{picture}
\caption{Colour of raw strips when UV-illuminated and their light yield in terms of p.e./MeV. }
\label{Fig.Colors}
\end{minipage}\hfill{ }
\begin{minipage}[t]{90mm}
\centering
\begin{picture}(90,23)(0,1)
  %\put(0,0){\framebox(90,24)[b]{}}
  \put(0,0){\includegraphics[width=90mm]{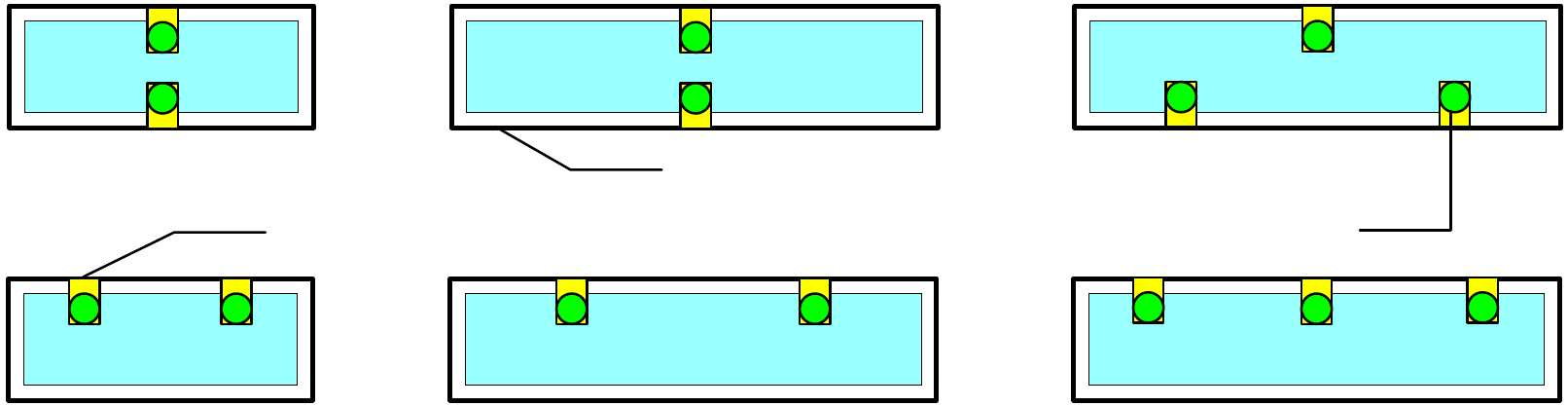}}
  \put(39.0,13.5){\makebox(0,0)[l]{\footnotesize\sf Light reflecting coat}}
  \put(77.5,10.5){\makebox(0,0)[r]{\footnotesize\sf WLS fiber}}
  \put(16.0,10.0){\makebox(0,0)[l]{\footnotesize\sf Transparent glue}}
 \end{picture}
 \caption{Different geometry of light collection from 2.5~cm and 4~cm wide strips with WLS fibers of different type and different producers.}
 \label{Fig.Strips}
\end{minipage}
\end{figure}
Parameters of the low-cost scintillator strips depend very much on the production technology and are widely scattered. For instance, if the after ends of the strips be UV-illuminated, one can see difference in their colour (Fig.~\ref{Fig.Colors}) normally invisible. To increase the signal and optimize the detector construction, different strips and photo sensors, as well as several schemes of light collection (Fig.~\ref{Fig.Strips}) have been tested with the above test-bench.

\subsection{Final construction}
\begin{figure}[htb]
\setlength{\unitlength}{1mm}
 \begin{picture}(150,34)(0,1)
  %\put(0,0){\framebox(150,35)[b]{}}
  \put( 20,0.5){\includegraphics[width=60mm]{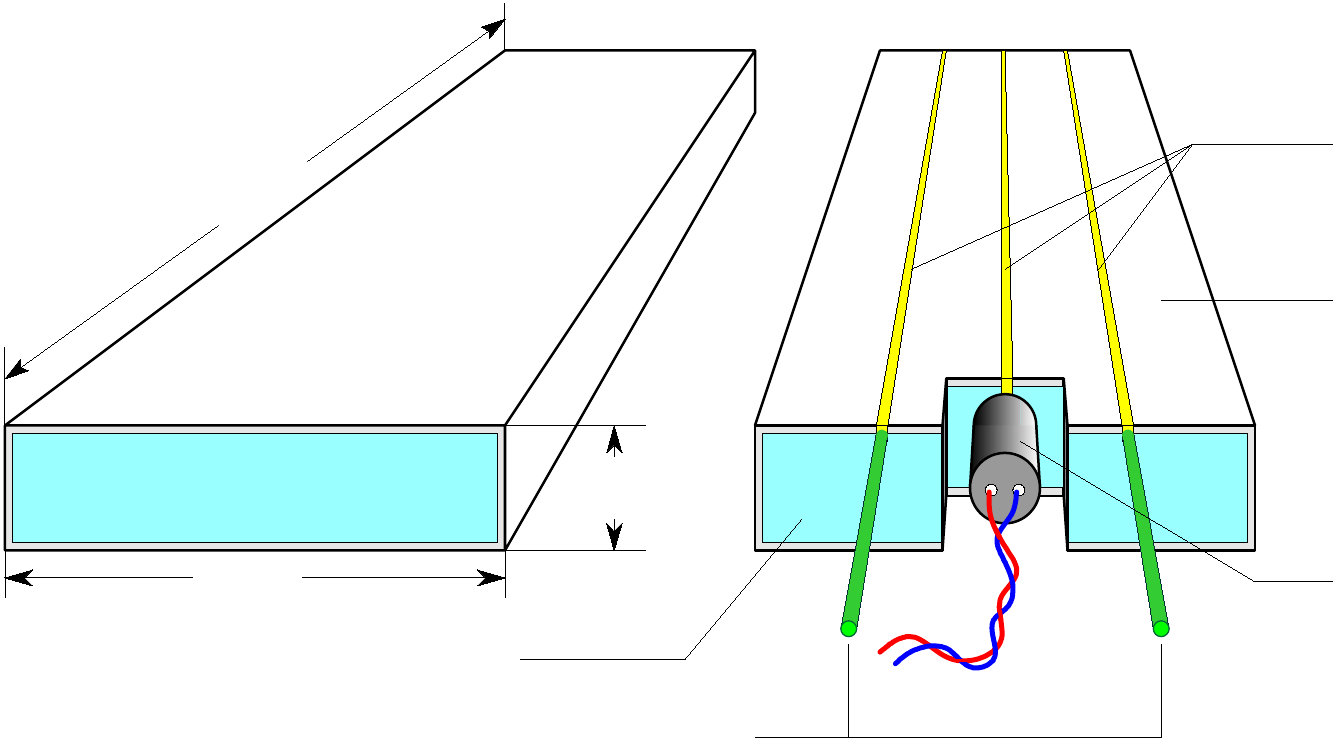}}
  \put(118,3.5){\includegraphics[width=20mm]{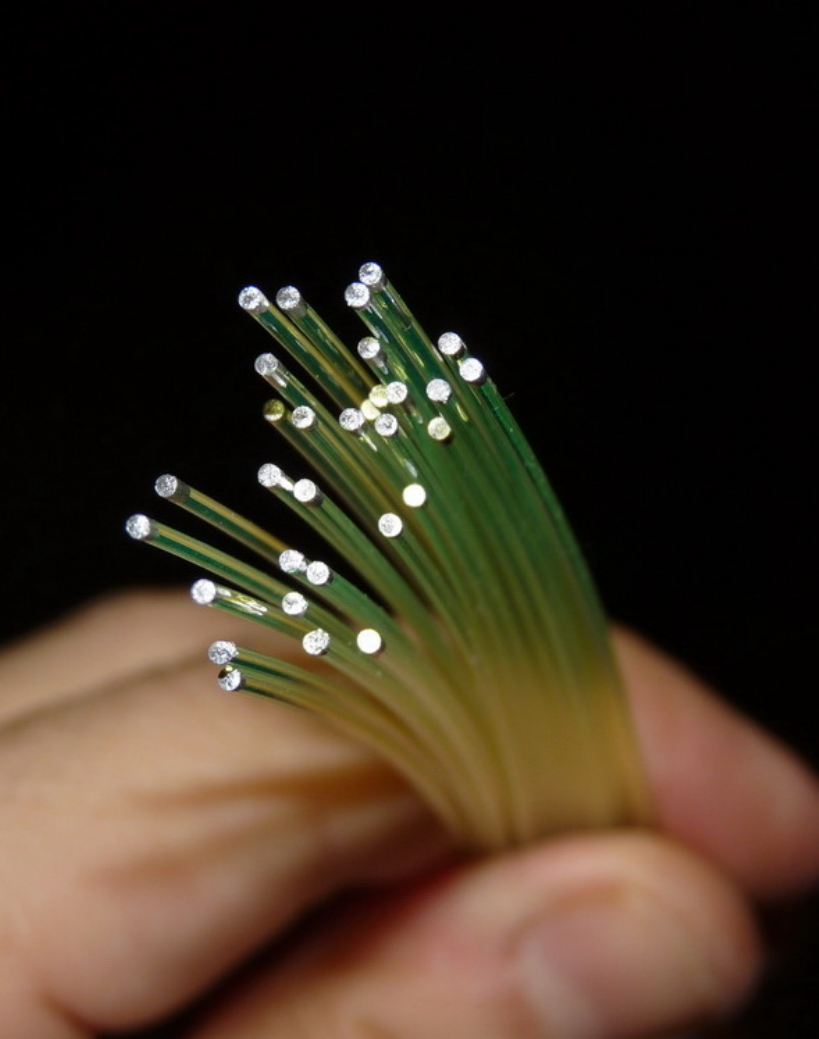}}
  \put(31.0,7.0){\makebox(0,0)[b]{\tiny\sf 4 cm}}
  \put(48.0,12.0){\makebox(0,0)[c]{\tiny\sf 1 cm}}
  \put(30.0,23.9){\rotatebox{35}{\tiny\sf 1 m}}
\put(79.0,15.0){\parbox{25mm}{\begin{center}\scriptsize\sf Gd-containing coat\\ (1.6 mg/cm$^2$)\end{center}}}
  \put(43.0, 4.0){\makebox(0,0)[r]{\scriptsize\sf Polystyrene-based scintillator}}
  \put(80.4,26.5){\makebox(0,0)[bl]{\scriptsize\sf Grooves}}
  \put(80.4,8.0){\makebox(0,0)[l]{\scriptsize\sf MPPC}}
  \put(53.0,0.0){\makebox(0,0)[rb]{\scriptsize\sf WLS fibers}}
  \put(100.0,35.0){\makebox(0,0)[t]{\scriptsize\sf Rear mirror ends of the fibers}}
  {
  \linethickness{0.001in}
  \color{white}
  \put(125,32.8){\line(0,-1){10}}
  \color{black}
  \linethickness{0.001in}
  \put(60.5,32.8){\line(0,-1){1.0}}
  \put(63.2,32.8){\line(0,-1){1.0}}
  \put(65.9,32.8){\line(0,-1){1.0}}
  \put(125,32.8){\line(-1,0){64.5}}
  \put(125,32.8){\line(0,-1){4.0}}
  }
\end{picture}
\caption{The basic element -- scintillator strip. }
\label{Fig.Strip}
\end{figure}
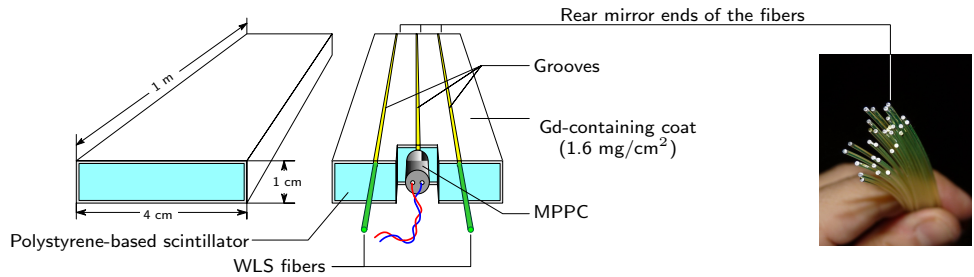

Finally, the light collection from the strip (Fig.~\ref{Fig.Strip}) is done via three wavelength shifting (WLS) Kuraray fibers Y-11, $\oslash$~1.2~mm, glued into grooves along all the strip. An opposite (blind) end of each fiber is polished and covered with a mirror paint, which decreases a total lengthwise attenuation of a light signal down to $\sim$15~\%/m.

\begin{figure}[ht]
\setlength{\unitlength}{1mm}
\begin{picture}(150,33)(0,2)
  %\put(0,0){\framebox(150,35)[b]{}}
  \put(0.0,0){\includegraphics[height=35mm]{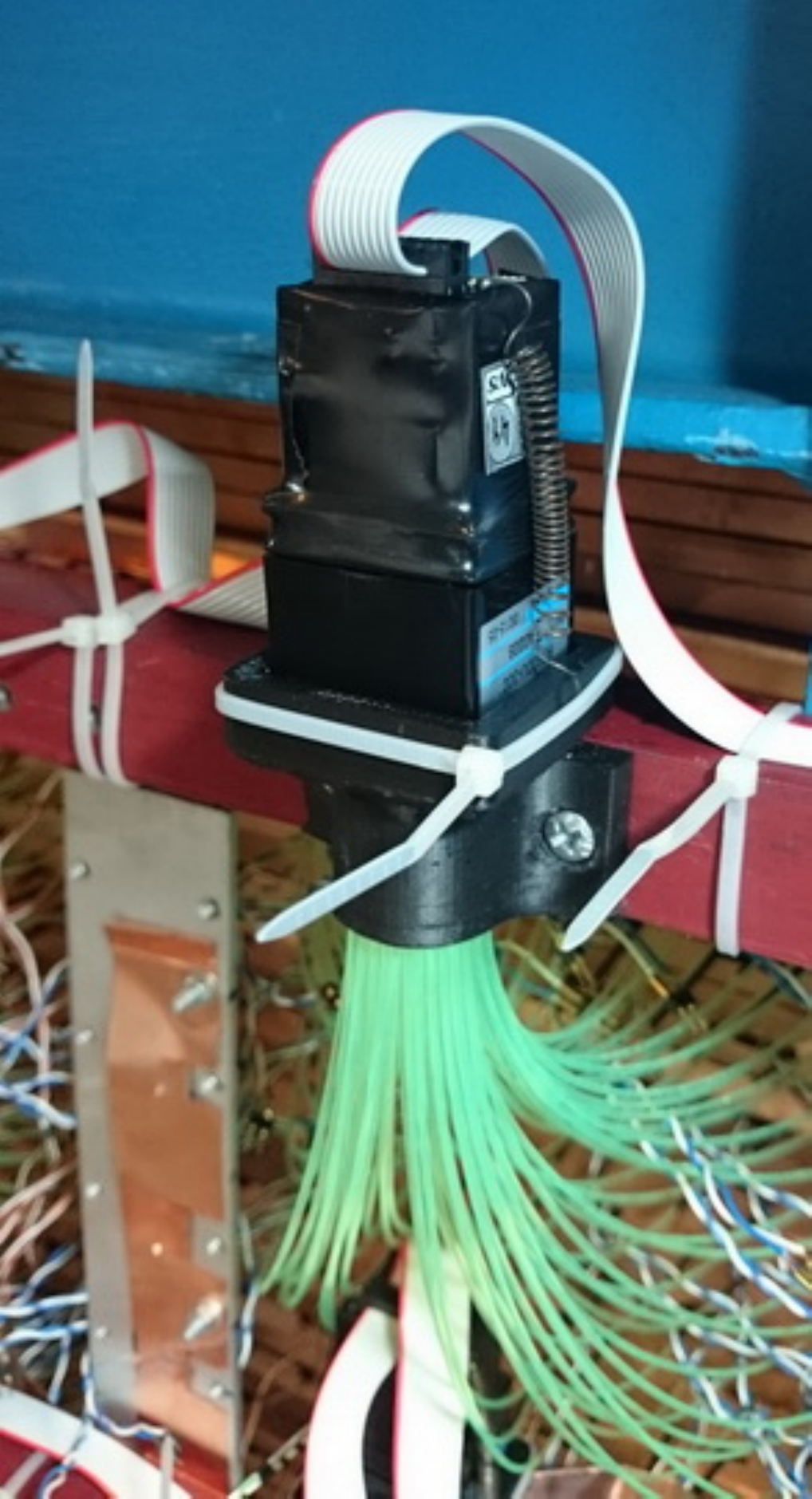}}
  {
  \linethickness{0.001in}
  \color{white}
  \put(17,26.0){\line(-1,0){7}}
  \put(17,20.0){\line(-1,0){7}}
  \put(17,14.0){\line(-1,0){8}}
  \put(17, 8.0){\line(-1,0){8}}
  \color{black}
  \linethickness{0.001in}
  \put(17,26.0){\line(1,0){10}}
  \put(17,20.0){\line(1,0){10}}
  \put(79,16.8){\vector(1,0){15}}
  \put(17,14.0){\line(1,0){10}}
  \put(17, 8.0){\line(1,0){10}}
  }
  \put(27.5,26.0){\makebox(0,0)[l]{\scriptsize\sf Front-end electronics}}
  \put(27.5,20.0){\makebox(0,0)[l]{\scriptsize\sf PMT R7600U-300 with}}
  \put(30.0,17.0){\makebox(0,0)[l]{\scriptsize\sf a photocathode of 18$\times$18 mm$^2$ active area}}
  \put(27.5,14.0){\makebox(0,0)[l]{\scriptsize\sf 3D-printed plastic adapter}}
  \put(27.5, 8.0){\makebox(0,0)[l]{\scriptsize\sf Bundle of 100 WLS fibers}}

  \put(107.0,0){\includegraphics[height=35mm]{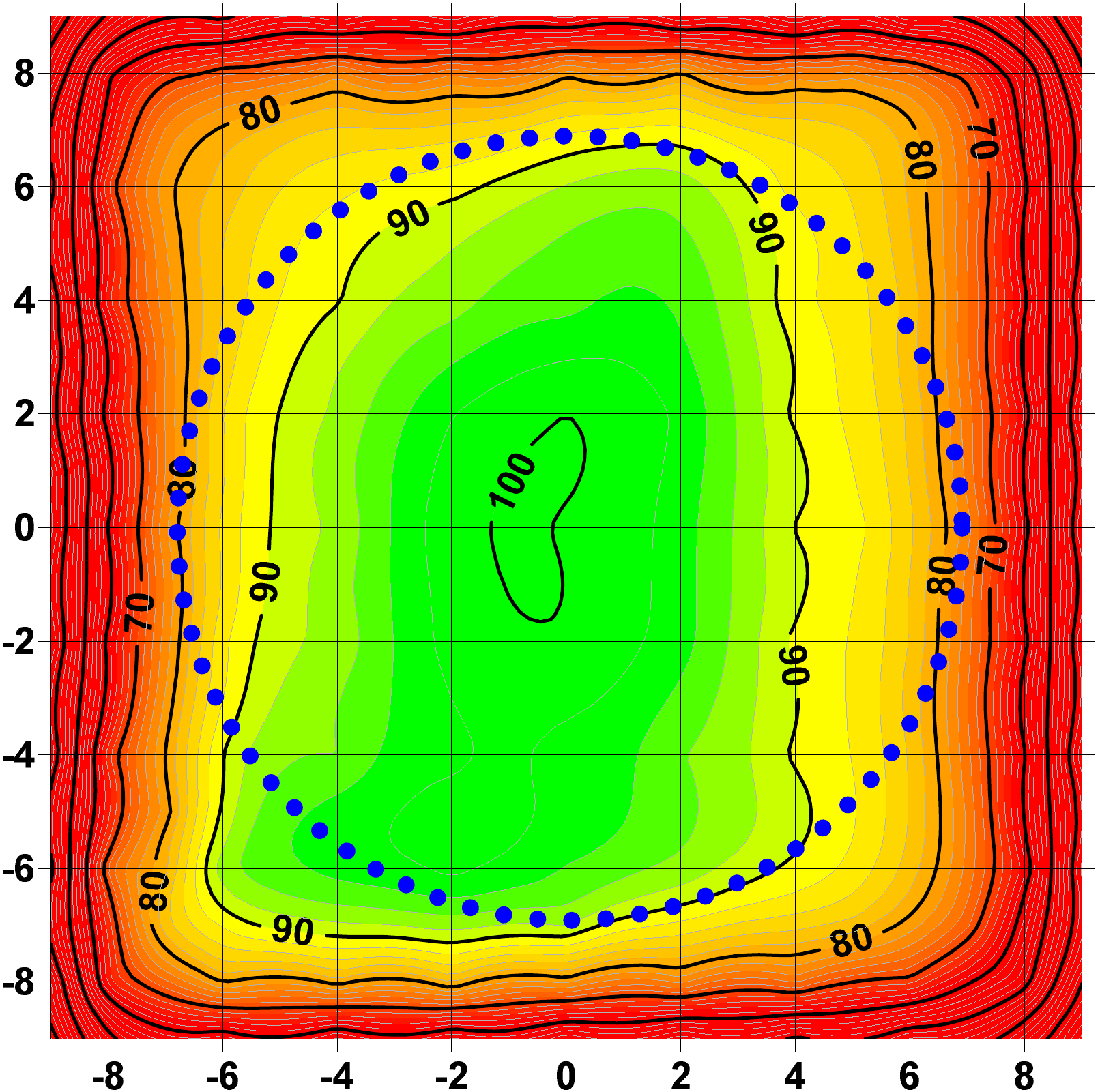}}
  \put(80.0,23.0){\parbox{25mm}{\scriptsize\sf Color represents\\ the relative amplitude\\of a test light signal}}
  \put(65,6.0){\parbox{45mm}{\scriptsize\sf Central part of the photocathode\\[-0.5mm] corresponds to the bundle diameter}}
  \put(140.0,0.0){\parbox{9mm}{\begin{center}\scriptsize\sf X\\ mm\end{center}}}
  \put(99.0,30.0){\parbox{10mm}{\begin{center}\scriptsize\sf Y\\ mm\end{center}}}
  {
  \linethickness{0.001in}
  \color{blue}
  \put(106.7,8.5){\line(1,0){9}}
  \color{black}
  \linethickness{0.001in}
  \put(106.7,8.5){\line(-1,0){7}}
  }
\end{picture}
\caption{Coupling of the fiber bundle to the photocathode of a compact PMT.}
\label{Fig.Bundle}
%\end{minipage}
\end{figure}
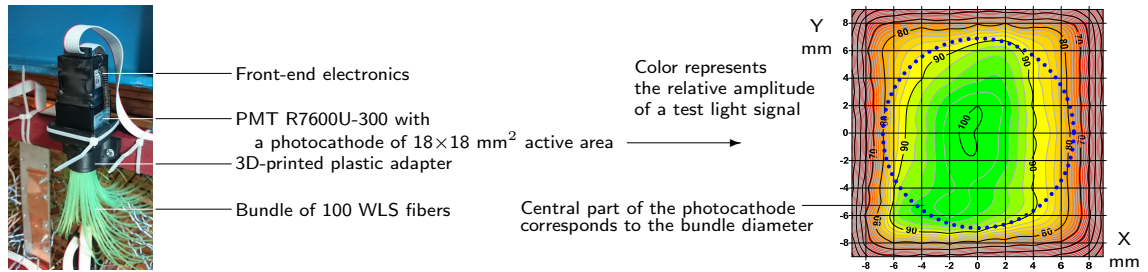

The central fiber is coupled to a multi-pixel photon counter\footnote{MPPC or SiPM, depending on the producer. We use Hamamatsu MPPC product S12825-050C(X) with a sensitive area of 1.3$\times$1.3~mm$^2$ and reduced dark current.} operating in Geiger mode. This sensor is individual for each strip and is connected to front-end electronics via twisted pair.
The rest two fibers are longer by 15-30 cm. They are coupled to a compact photomultiplier tube (PMT) which is common for a whole module -- a group of 50 parallel strips. As the emission spectrum of the fibers has a maximum in a green region (480-520~nm), the PMT photocathode should be also green-sensitive. The best suited for that are Hamamatsu products with the photocathode suffix 20 or 300. Optical coupling of the fibers and photocathode is done with transparent polymer gel. In order to hold all hundred fibers in front of the PMT center (an area of the highest sensitivity) the whole assembly is fixed in space with a special 3d-printed adapter (Fig.~\ref{Fig.Bundle}).

\section{Tests with the DANSSino pilot detector}\label{Sect.DANSSino}
\begin{figure}[bht]
 \setlength{\unitlength}{1mm}
 \begin{picture}(150,35.0)(0,1)
 %\put(0,0){\framebox(150,38.5)[b]{}}
 \put(0,0){\includegraphics[height=36.5mm]{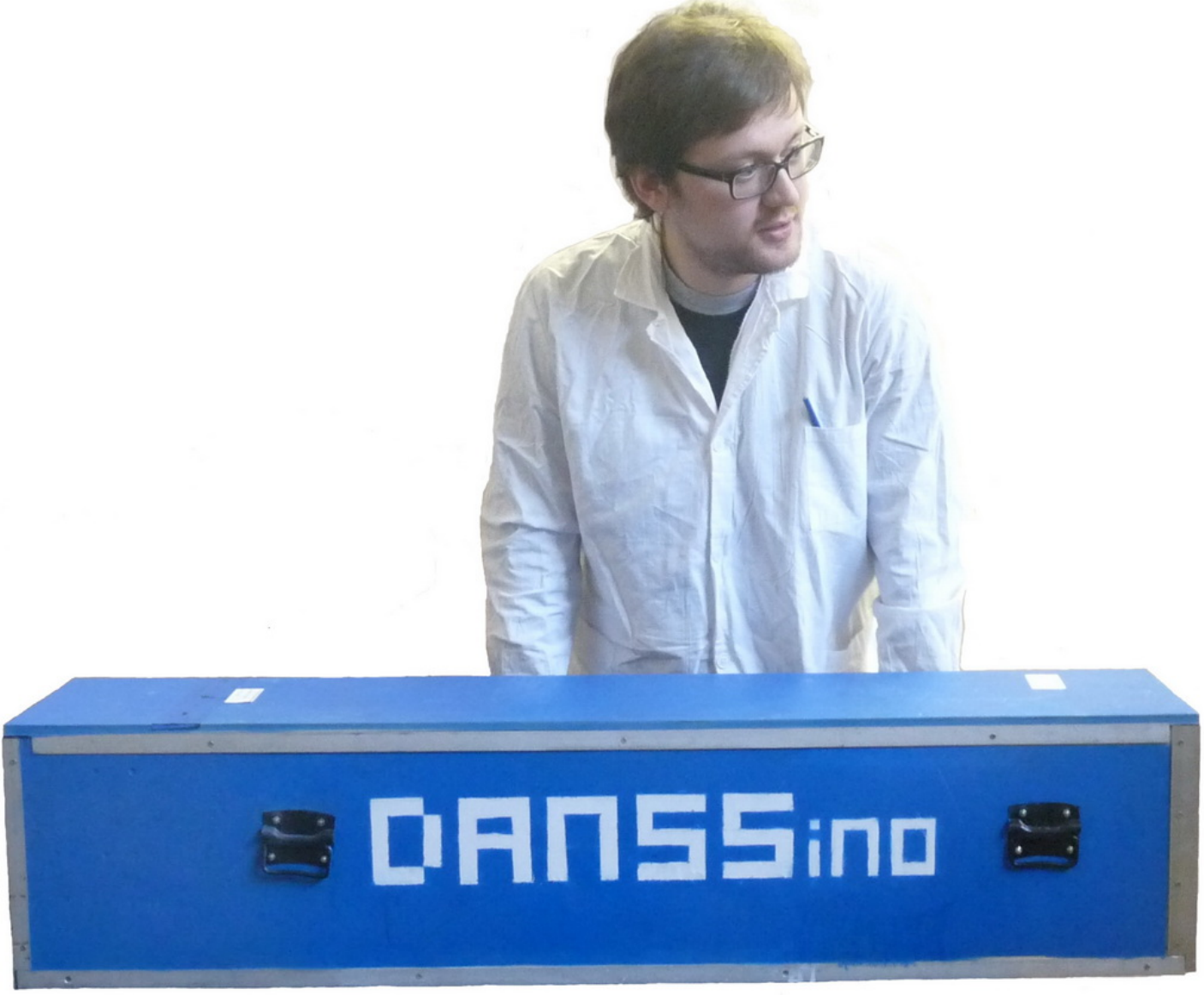}}
 \put(59.5,7){\includegraphics[width=90mm]{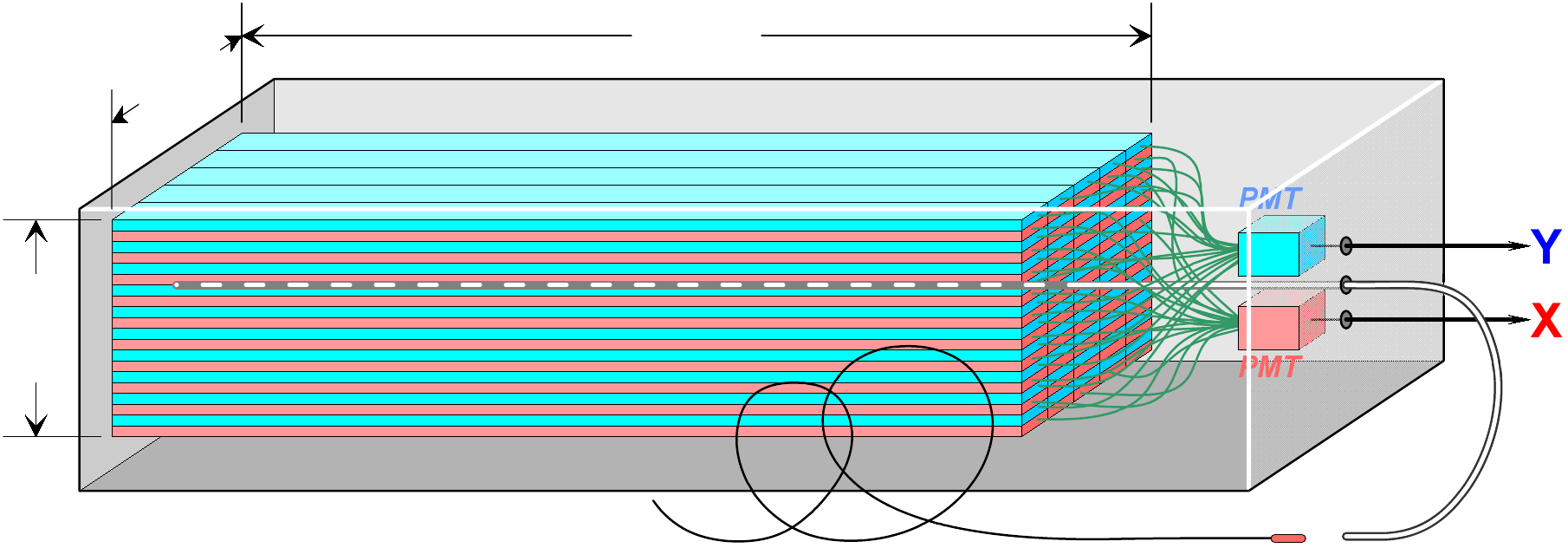}}
 \put(60.7,17.0){\rotatebox{90}{\tiny\sf 20 cm}}
   \put(67.2,32.2){\rotatebox{35}{\tiny\sf 20 cm}}
 \put(99.4,36.5){\makebox(0,0)[c]{\tiny\sf 1 m}}
 \put(133,4.5){\line(0,1){2}}
 \put(133,4.5){\line(-1,0){15}}
 \put(76.0,0.0){\parbox{40mm}{\begin{flushright}\scriptsize\sf Radioactive calibration source\\[-0.5mm] with a flexible string\end{flushright}}}
 \put(145,0.5){\line(0,1){10}}
 \put(145,0.5){\line(-1,0){5}}
 \put(138.0,0){\makebox(0,0)[rb]{\scriptsize\sf Teflon tube}}
\end{picture}
 \caption{The DANSSino detector.}
 \label{Fig.DANSSino}
\end{figure}
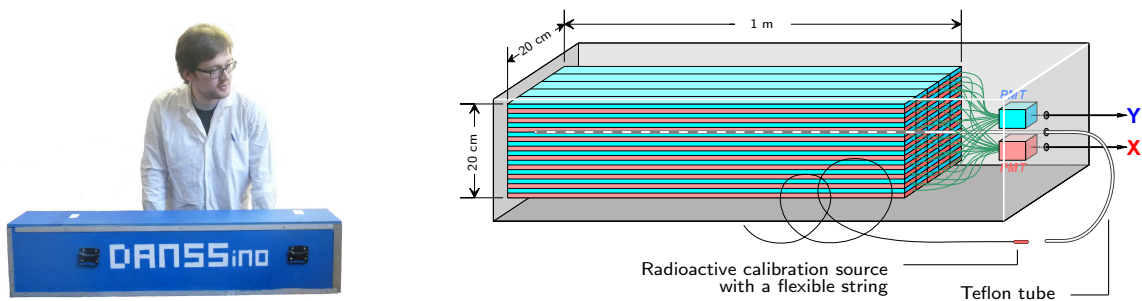

In order to check operability of the DANSS design, compare different acquisition schemes and reveal the main origins of the background, a simplified pilot version of the detector was created. Figure~\ref{Fig.DANSSino} shows this small prototype -- DANSSino\cite{DANSSino1,DANSSino2}, which is 1/25th part of the whole DANSS detector (2 modules of 50).

DANSSino consists of exactly the same basic elements as the main DANSS detector. One hundred strips of DANSSino (Fig.~\ref{Fig.DANSSino}) form a bar divided logically into two modules: the odd strip layers are coupled to the X-PMT and the even ones to the Y-PMT. Both modules are equipped with preamplifiers and placed into a light-tight box. Individual photodiodes are not used. Thin teflon tube allows to introduce a tiny radioactive source (see section~\ref{Section.Calibration} below) inside the detector body.

Initial tests performed with DANSSino in JINR laboratory have shown huge number of false neutrino-like events caused by cosmic-ray fast neutrons with energy up to GeV \cite{Hess}. These neutrons being scattered by the polystyrene protons produce the Prompt signal in a wide energy range and then, after moderation, are captured in Gd and give the regular Delayed signal, thus imitating the IBD.

As expected, operation in the room A336 under the KNPP reactor \#3 has demonstrated absence of such neutron-induced background. On the other hand, there still are some fast neutrons but with much lower energy than the cosmic ones. The events caused by these neutrons are definitely correlated with cosmic muons but uncorrelated with the reactor operation. Numerous tests with different combination of shielding materials have shown that the neutrons are produced in the surrounding lead and copper parts of the detector shielding. As a result, they have so-called ``evaporation spectrum'' with the main energy of $\sim$2~MeV, are not very dangerous\footnote{Due to a quenching factor of about 5, even the high-energy tail of the muon-induced neutron spectrum cannot produce a signal with $E_P> (1-1.5)$~MeV.} for our experiment and can be easily tagged with appropriate muon-veto system. Nevertheless, in order to suppress these muon-induced background, the initially designed DANSS shielding was reinforced with additional internal CHB$_{\rm int}$ layer (see Fig.~\ref{Fig.Shielding} in section~\ref{Section.Shielding} below).

 \begin{figure}[thb]
 \setlength{\unitlength}{1mm}
 \begin{minipage}[t]{90mm}
  \centering
  \begin{picture}(90,50)(0,1)
   %\put(0,0){\framebox(90,50)[b]{}}
   \put(0, 0){\includegraphics[height=50mm]{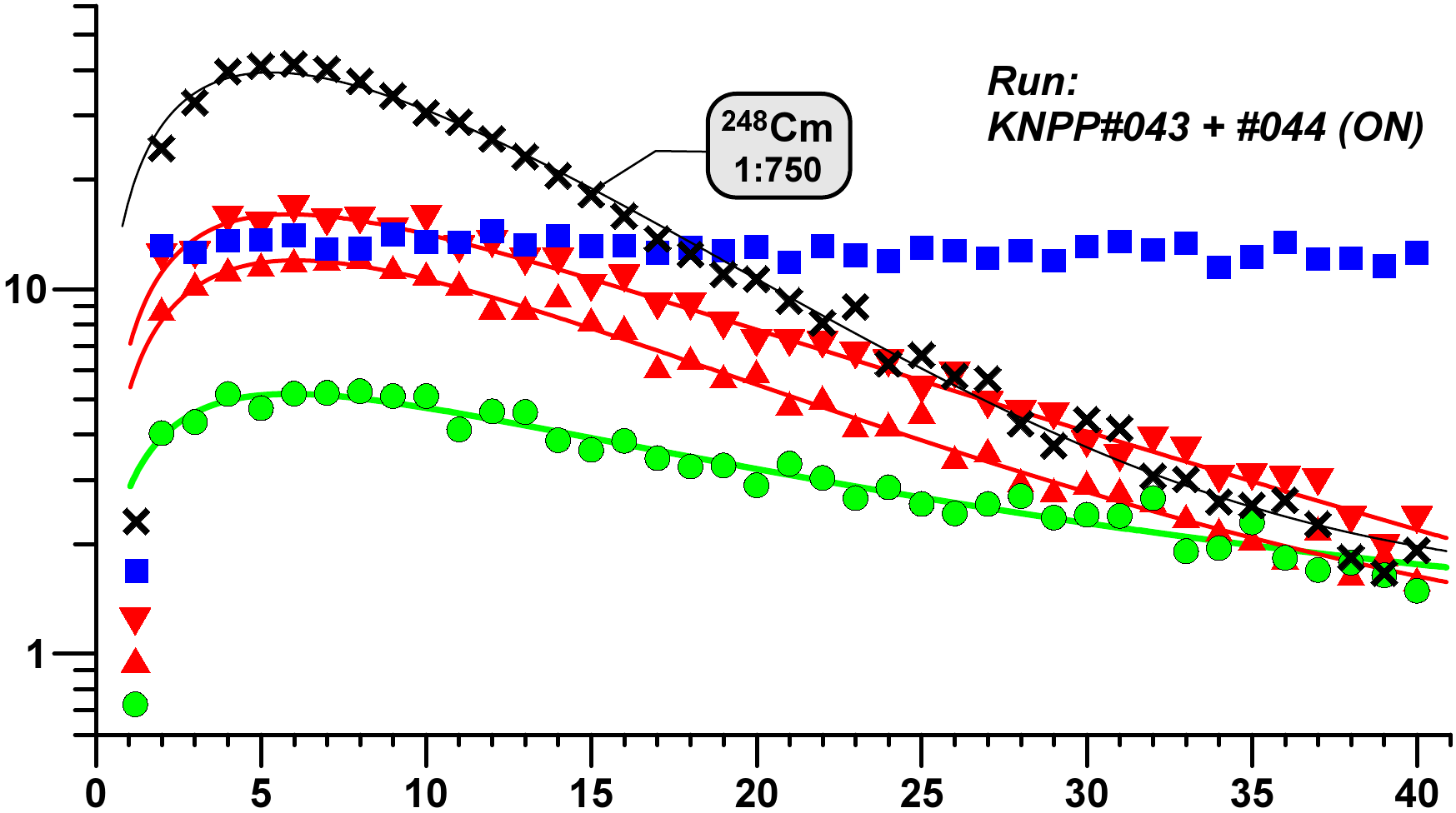}}
  \put(7.0,50.0){\makebox(0,0)[tl]{\scriptsize\sf $N$(ON), events / $\mu$s / day}}
   \put(89.0,7.5){\makebox(0,0)[r]{\scriptsize\sf $T_{PD}\;,\;\;\mu$s}}
  \put(11.0,18.8){\makebox(0,0)[l]
    {\tiny\sf = \scriptsize\sf $^{248}$\bf Cm neutron source}}
  \put(11.0,15.5){\makebox(0,0)[l]
    {\tiny\sf = $(X_P\!\oplus\! Y_P)\!\wedge\!\overline{\mu}\wedge\!(E_P\!<1.5$ MeV)}}
  \put(46.0,16.0){\makebox(0,0)[l]{\scriptsize\bf \color{blue} -- random BG}}
  \put(11.0,12.5){\makebox(0,0)[l]
    {\tiny\sf = $(X_P\!\oplus\! Y_P)\!\wedge\!\mu$}}
  \put(11.0,9.7){\makebox(0,0)[l]
    {\tiny\sf = $(X_P\!\wedge\! Y_P)\!\wedge\!\mu$}}
  \put(28.0,11.1){\makebox(0,0)[l]{\color{red}\scriptsize\bf $\left.\rule[-1.5mm]{0mm}{4.0mm}\right\}$ -- muon-induced fast neutrons}}
  \put(11.0,07.0){\makebox(0,0)[l]
    {\tiny\sf = $(X_P\!\wedge\! Y_P)\!\wedge\!\overline{\mu}$}}
  \put(31.0,07.2){\makebox(0,0)[l]{\colorbox{green}{\scriptsize\bf -- true IBD}}}
  \end{picture}
  \caption{The $T_{PD}$ time distribution measured with the $^{248}$Cm neutron source compared to the neutrino-like events detected under the operating reactor.}
  \label{Fig.DANSSino_T-spectrum}
 \end{minipage}\hfill{ }
 \begin{minipage}[t]{50mm}
  \centering
  \begin{picture}(50,50)(0,1)
   %\put(0,0){\framebox(50,50)[b]{}}
   \put(0, 0){\includegraphics[height=50mm]{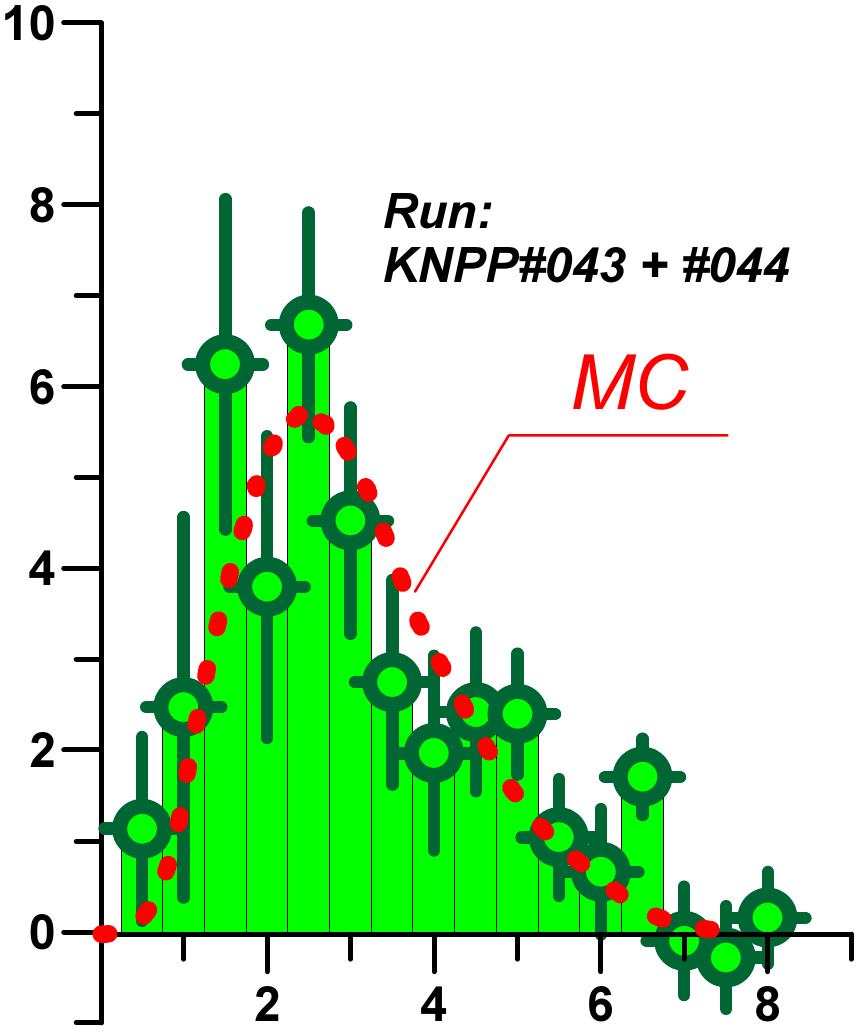}}
  \put(7.0,50.0){\makebox(0,0)[lt]{\scriptsize\sf $N$(ON) -- $N$(OFF),}}
  \put(7.0,47.0){\makebox(0,0)[lt]{\scriptsize\sf events / 0.5 MeV / day}}
  \put(48.5,5.0){\makebox(0,0)[r]{\scriptsize $E_P$,}}
  \put(49.5, 1.3){\makebox(0,0)[r]{\scriptsize\sf MeV}}
  \end{picture}
  \caption{Differential $E_P$ energy spectrum of the neutrino-like events measured with DANSSino.}
  \label{Fig.DANSSino_E-spectrum}
\end{minipage}
 \end{figure}

Fig.~\ref{Fig.DANSSino_T-spectrum} represents a number of the neutrino-like events versus time between the Prompt and Delayed signals ($T_{PD}$) detected with DANSSino under the operating reactor. The data were selected according to the expected IBD signature:
\begin{itemize}
\item the Delayed signal should correspond to the Gd($n$,$\gamma$) reaction, i.e., both the X and Y modules should be fired ($X_D\wedge Y_D$) with a reasonable\footnote{As the detector is small, significant part of the $\gamma$-cascade is not detected, and therefore the acceptable $E_D$ range is extended to the lower energy.} total energy: $E_{XD}+E_{YD}=E_D \in\left[1-8\right]$~MeV;
\item the Prompt energy must also be in a right range: $E_{XP}+E_{YP}$ =$E_P \in\left[1-7\right]$~MeV.
\end{itemize}
The Prompt signal of true IBD events, in addition to a positron itself, includes (at least, partial) detection of annihilation 511~keV photons. As a result, the ($X_P\wedge Y_P$) condition, as well as the absence of any preceding muons ($\overline{\mu}$) is required. Such events are represented in Fig.~\ref{Fig.DANSSino_T-spectrum} by green circles. Their distribution corresponds to the prediction (\ref{Eq.f(t)}) with \vspace{-2mm}
\begin{equation}\label{Eq.f(t)_DANSSino}
\tau_m=(3\pm1)\,{\rm \mu s};\hspace{5mm}\tau_c=(24\pm1)\,{\rm \mu s}\;.
\end{equation}

Other (non-IBD) events producing neutrons with multiplicity $k>1$ must give the distribution with much steeper slope corresponding to
\vspace{-3mm}
\begin{equation} \label{Eq.fk(t)}
f_k(t)=f_1(t)\cdot\!\left(1\!-\!\!\int\limits_0^t \!f_1(\tau)d\tau \right)^{\!k-1}\!\!\!\!\!\!=\;\;
\frac{e^{\frac{-t}{\tau_c}}-e^{\frac{-t}{\tau_m}}}{\left(\tau_c-\tau_m\right)^k}\cdot
\left(\tau_c\, e^{\frac{-t}{\tau_c}}-\tau_m\, e^{\frac{-t}{\tau_m}}\right)^{k-1},
\end{equation}
which reflects the fact than only \emph{the first} of $k$ neutrons is detected by the DANSSino acquisition system. Such is indeed the case of artificial $^{248}$Cm source emitting $k\simeq3.2$ neutrons per fission in average (black crosses) or muon-induced signals (red triangles) corresponding to production of $k\sim2$ neutrons in copper.

On the other hand, there could be neutrino-like non-IBD events which are not associated with muons but originate from random ($n-n$) or ($\gamma-n$) coincidences due to gamma- or thermal neutron background. The raw flux of thermal neutrons inside the detector was very low \footnote{Even with incomplete DANSSino shielding, it was not higher than few n/m$^2$/s.}, so that relative probability of random ($n-n$) coincidences is negligible. On the contrary, ($\gamma-n$) coincidences rally take place, but due to low efficiency of the plastic scintillator to gamma-rays they are registered only with low multiplicity and at low energy: $(X_P\!\oplus\! Y_P)\!\wedge\!(E_P\!<1.5$~MeV). Flat distribution of such events (blue squares in Fig.~\ref{Fig.DANSSino_T-spectrum}) confirm their random character.

An energy distribution of the neutrino-like events (Fig.~\ref{Fig.DANSSino_E-spectrum}) is in a good agreement with our MC simulations as well, thus confirming that the events observed are really the IBD ones.

\begin{figure}[hbt]
 \setlength{\unitlength}{1mm}
 \begin{picture}(150,23)(0,2)
  %\put(0,0){\framebox(150,25)[b]{}}
 \put(1,0){\includegraphics[width=150mm]{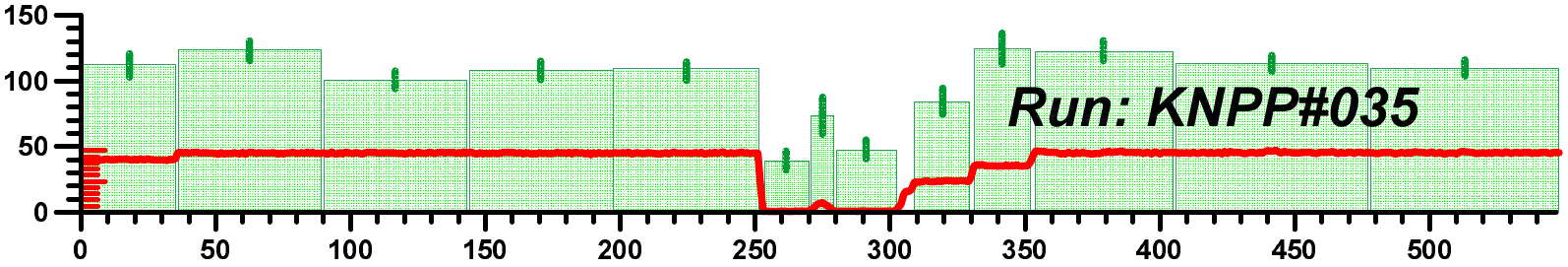}}
  \put(10.0,25.0){\makebox(0,0)[lt]{\scriptsize\sf Number of neutrino-like events per day}}
  {\color{red}
  \put(13.0,8.0){\makebox(0,0)[l]{\scriptsize\sf Relative reactor power}}
  }
  \put(150.0,0.3){\makebox(0,0)[rb]{\scriptsize\sf Time, hr}}
 \end{picture}
 \caption{Time plot of the reactor power (red curve) and the number of the neutrino-like events detected by DANSSino in one of the measurement periods.}
 \label{Fig.Run035}
 \end{figure}
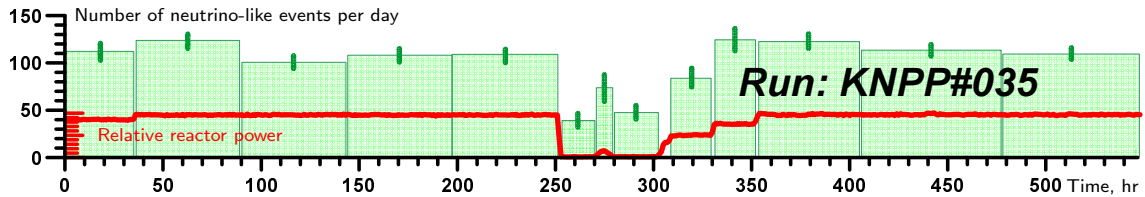

Because of its small size (0.04~m$^3$ only) DANSSino was not aimed really to detect reactor anti\-neu\-tri\-nos. Nevertheless, it turned out to be quite sensitive and able to detect about 70 IBD events per day with the signal-to-background ratio about unity. Figure~\ref{Fig.Run035} shows time evolution of the rate of neutrino-like events detected in one of the measurement runs including a two-days reactor stop. An interesting feature of this evolution is that the count rate within few hours of the reactor {\em starting} is somewhat higher than the ``cruising'' one. It means probably that while the fission number in these periods is more-or-less constant, the neutrino spectrum is of a bit higher energy (long-lived fission products with lower decay energy are not accumulated yet). The above conclusion is not strong enough because of very poor statistics and requires more detailed investigation.

As a result of numerous tests with DANSSino in JINR laboratory and under the industrial 3~GW$_{\rm th}$ reactor of the Kalinin Nuclear Power Plant at a distance of 11 m from the reactor core, the following conclusions were drawn:\\
--- The most important background under the WWER-1000 reactor originates from fast neutrons produced by cosmic muons in high-Z surroundings. Therefore, one should not place heavy materials inside the neutron moderator.\\% According to that, the second (internal) layer of CHB was introduced to the DANSS shielding.\\
--- Efficiency of the muon-veto system should not be less than 95-97\%. \\ %To reach this level, initially planned single layer of scintillator plates was replaced by double layer operating in coincidence mode with lower thresholds.\\
--- Operation of such detectors at a shallow depth with overburden less than 10-20 m w.e. seems to be questionable, as the neutron component of cosmic rays cannot be tagged by any veto system and produces a signature very similar to the IBD but outnumbers it by orders of magnitude.

\section{Detector description}

\subsection{Modular structure of the detector }\label{Section.Modular}
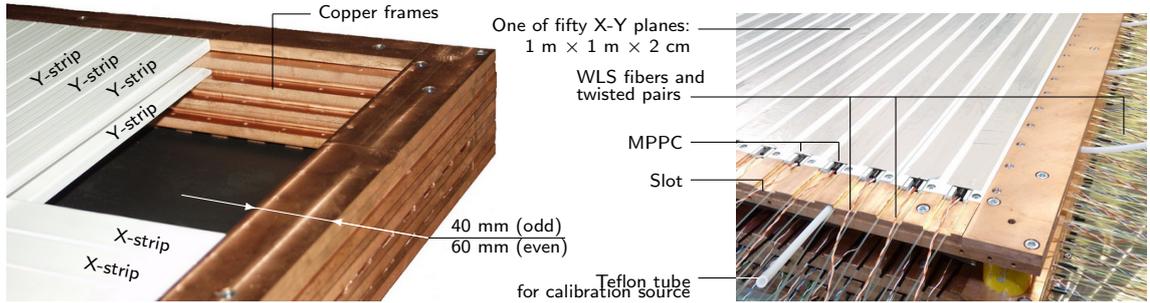
\begin{figure}[bht]
 \setlength{\unitlength}{1mm}
\begin{picture}(150,38)(0,2)
  %\put(0,0){\framebox(150,40)[b]{}}
  \put(0,0){\includegraphics[height=40mm]{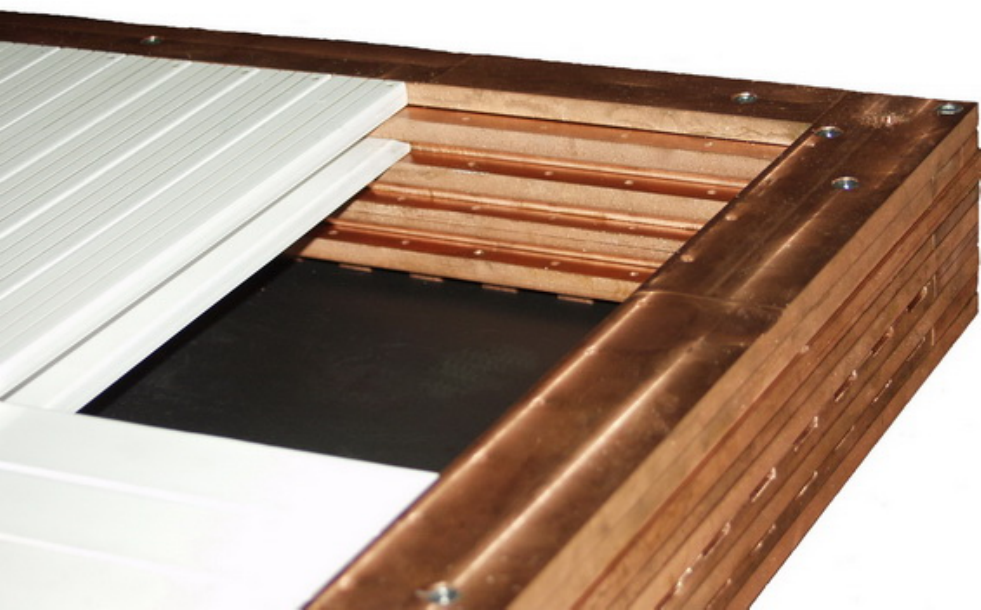}}
  \put( 3.0,28.0){\rotatebox{37}{\scriptsize\sf Y-strip}}
  \put( 8.0,27.3){\rotatebox{37}{\scriptsize\sf Y-strip}}
  \put(13.0,26.6){\rotatebox{37}{\scriptsize\sf Y-strip}}
  \put(13.0,20.9){\rotatebox{37}{\scriptsize\sf Y-strip}}
  \put(14.0,8.0){\rotatebox{-12}{\scriptsize\sf X-strip}}
  \put(10.0,4.3){\rotatebox{-12}{\scriptsize\sf X-strip}}
  {\color{white}
   \put(23.0,14.0){\rotatebox{-10}{\vector(1,0){10}}}
   \put(55.5,8.3){\rotatebox{-10}{\line(-1,0){23}}}
   \put(55.5,8.3){\rotatebox{-10}{\vector(-1,0){13}}}
   }
   \put(55.5, 8.3){\line(1,0){18}}
  \put(58.5,8.5){\makebox(0,0)[lb]{\scriptsize\sf 40 mm (odd)}}
  \put(58.5,8.2){\makebox(0,0)[lt]{\scriptsize\sf 60 mm (even)}}
  \put(41.0,38.0){\makebox(0,0)[l]{\scriptsize\sf Copper frames}}
  \put(35.0,38.0){\line(1,0){5}}
  \put(35.0,38.0){\line(0,-1){10}}

  \put(96,0){\includegraphics[height=38mm]{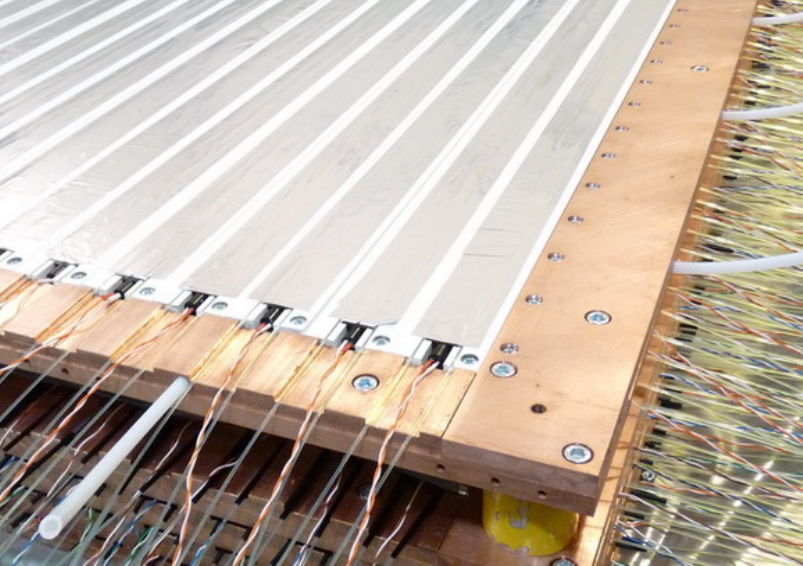}}
  \put(55.0,32.4){\parbox{35mm}{\begin{flushright}\scriptsize\sf One of fifty X-Y planes:\\1 m $\times$ 1 m $\times$ 2 cm \end{flushright}}}
  \put(91.0,36.0){\line(1,0){20}}

  \put(75,27.5){\parbox{20mm}{\scriptsize\sf WLS fibers and\\[-0.5mm] twisted pairs}}
  \put(90,27.0){\line(1,0){57}}
  \put(111,27.0){\line(0,-1){15}}
  \put(117,27.0){\line(0,-1){16}}
  \put(147,27.0){\line(0,-1){5}}

  \put(89,21.0){\makebox(0,0)[r]{\scriptsize\sf MPPC}}
  \put(90,21.0){\line(1,0){19.5}}
  \put(104.5,21.0){\line(0,-1){1.2}}
  \put(109.5,21.0){\line(0,-1){2.5}}

  \put(89,16.0){\makebox(0,0)[r]{\scriptsize\sf Slot}}
  \put(90,16.0){\line(1,0){9.5}}
  \put(99.5,16.0){\line(0,-1){1.5}}

  \put(65.0, 1.0){\parbox{25mm}{\begin{flushright}\scriptsize\sf Teflon tube \\[-1.5mm]for calibration source \end{flushright}}}
  \put(91.0,3.0){\line(1,0){5}}
  {\color{white}
   \put(96,3){\line(1,0){3}}
  }

\end{picture}
\caption{Scintillator strips mounted in copper frames.}
\label{Fig.Strips_in_Frames}
%\end{minipage}
\end{figure}

Sensitive volume of the DANSS detector (1~m$^3$) consists of scintillator strips laid in two perpendicular directions -- odd layers are parallel to X and even layers to Y axis (Fig.~\ref{Fig.Strips_in_Frames}). Frames made of copper flat bars are used to hold the strip layers, thus forming rigid X-Y planes. The frames not only fix position of the strips, but play also two additional roles -- they act as an internal part of passive shielding and as a neutron reflector.

Two WLS fibers and one twisted pair from each strip go out through a shallow slot in the copper bar. Some of X-Y planes are equipped with thin teflon tubes which allow introducing a tiny radioactive source for tests and calibration (see Sect.~\ref{Section.Calibration} below).

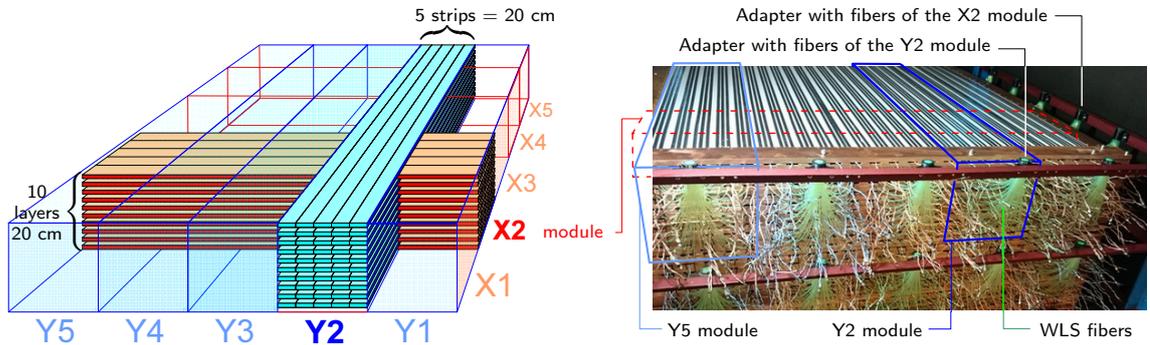
\begin{figure}[ht]
 \setlength{\unitlength}{1mm}
\begin{picture}(150,43)(0,2)
  %\put(0,0){\framebox(150,45)[b]{}}
  \put(0,0){\includegraphics[height=40mm]{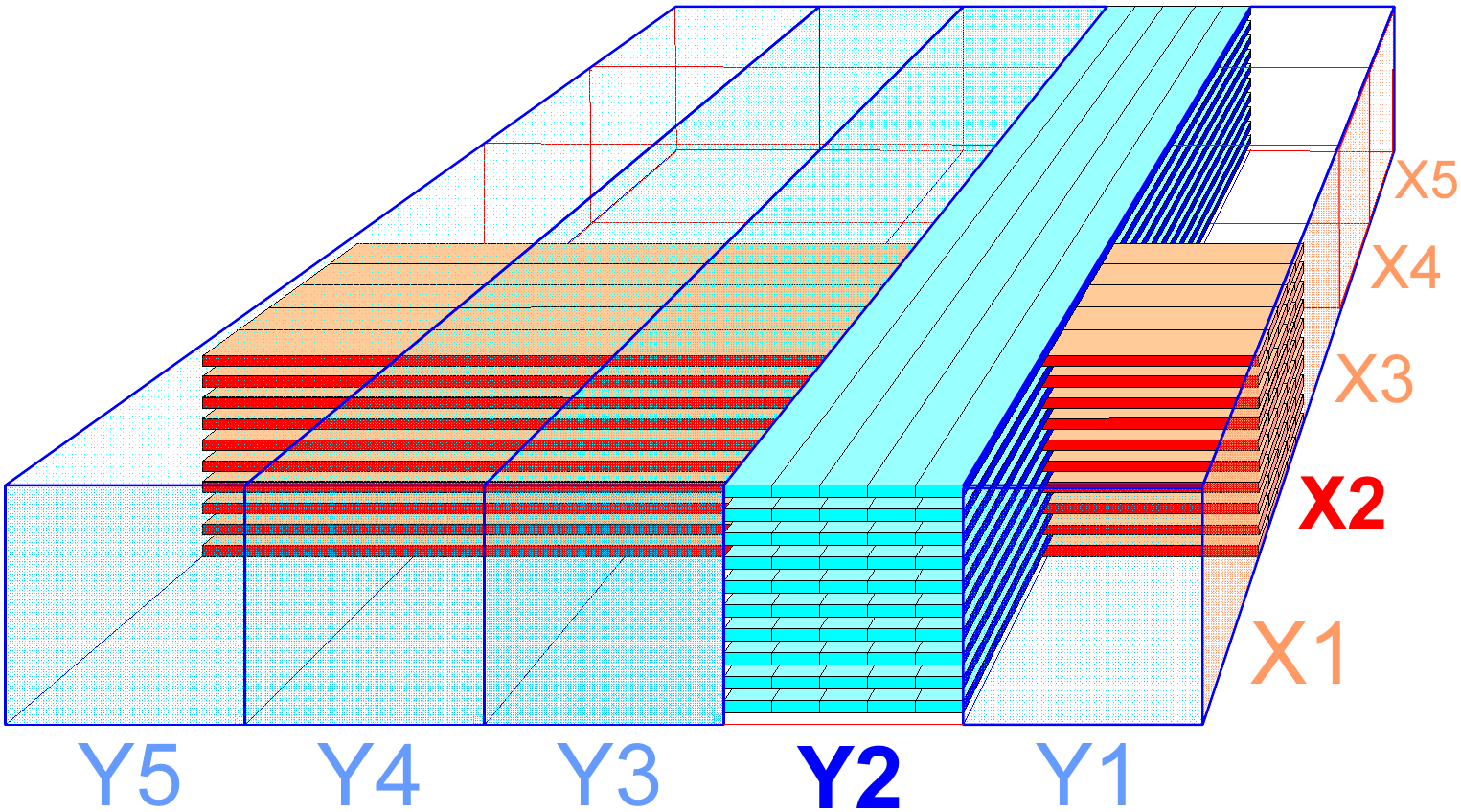}}

  \put(7.5,17.5){\makebox(0,0.5)[l]{\tiny\sf $\left\{\rule{0mm}{6.0mm}\right.$}}
  \put(0.5,14.8){\parbox{7mm}
     {\begin{center}\scriptsize\sf 10 layers\\ 20 cm\end{center}}}

  \put(63.0,43.5){\makebox(0,0)[c]{\scriptsize\sf 5 strips = 20 cm}}
  \put(58.0,40.0){\makebox(0,0)[b]{\tiny\sf $\overbrace{\rule{7mm}{0mm}}$}}

  \put(85,4.5){\includegraphics[width=65mm]{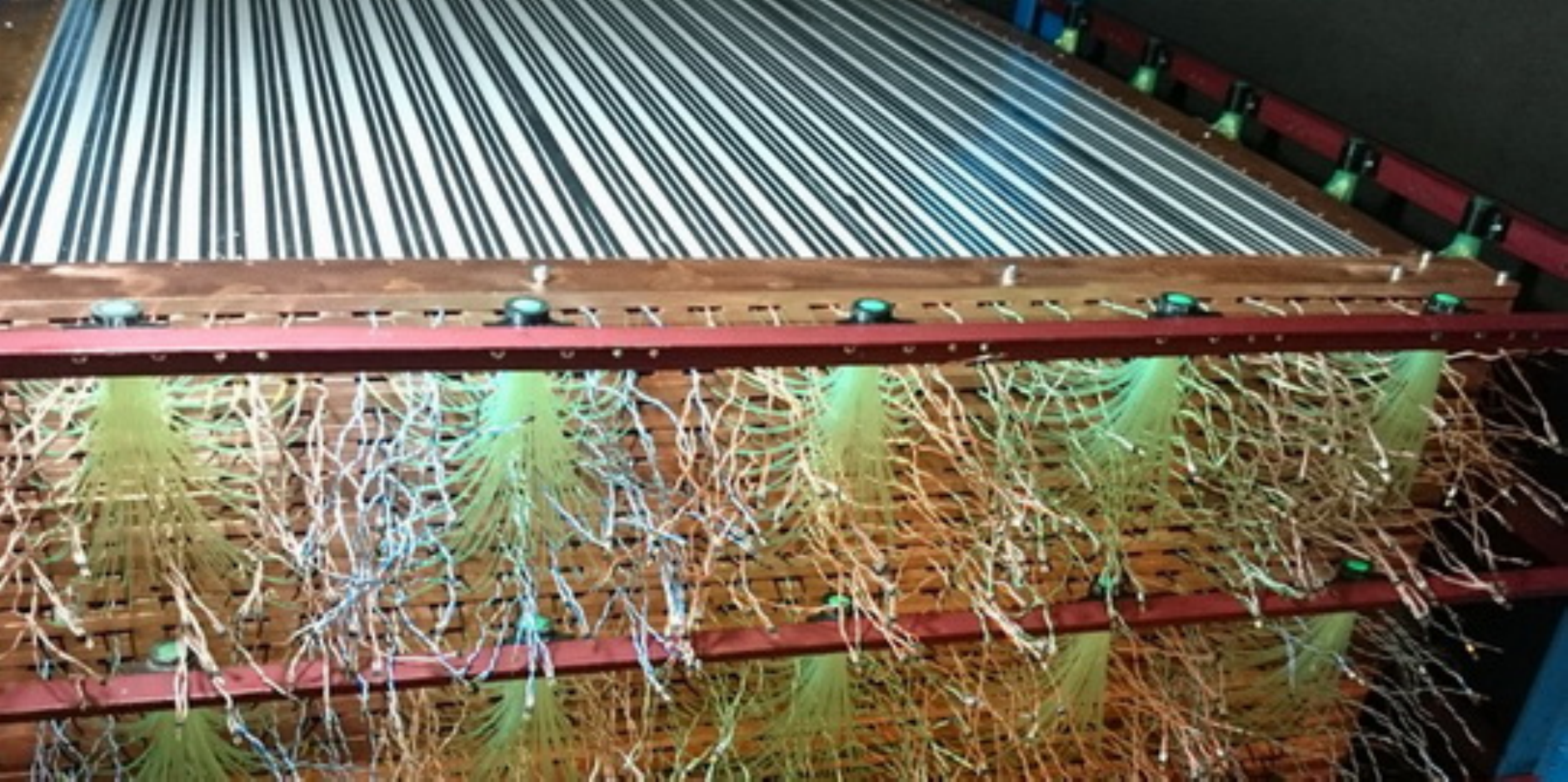}}
  \put(80,1.8){\includegraphics[height=42mm]{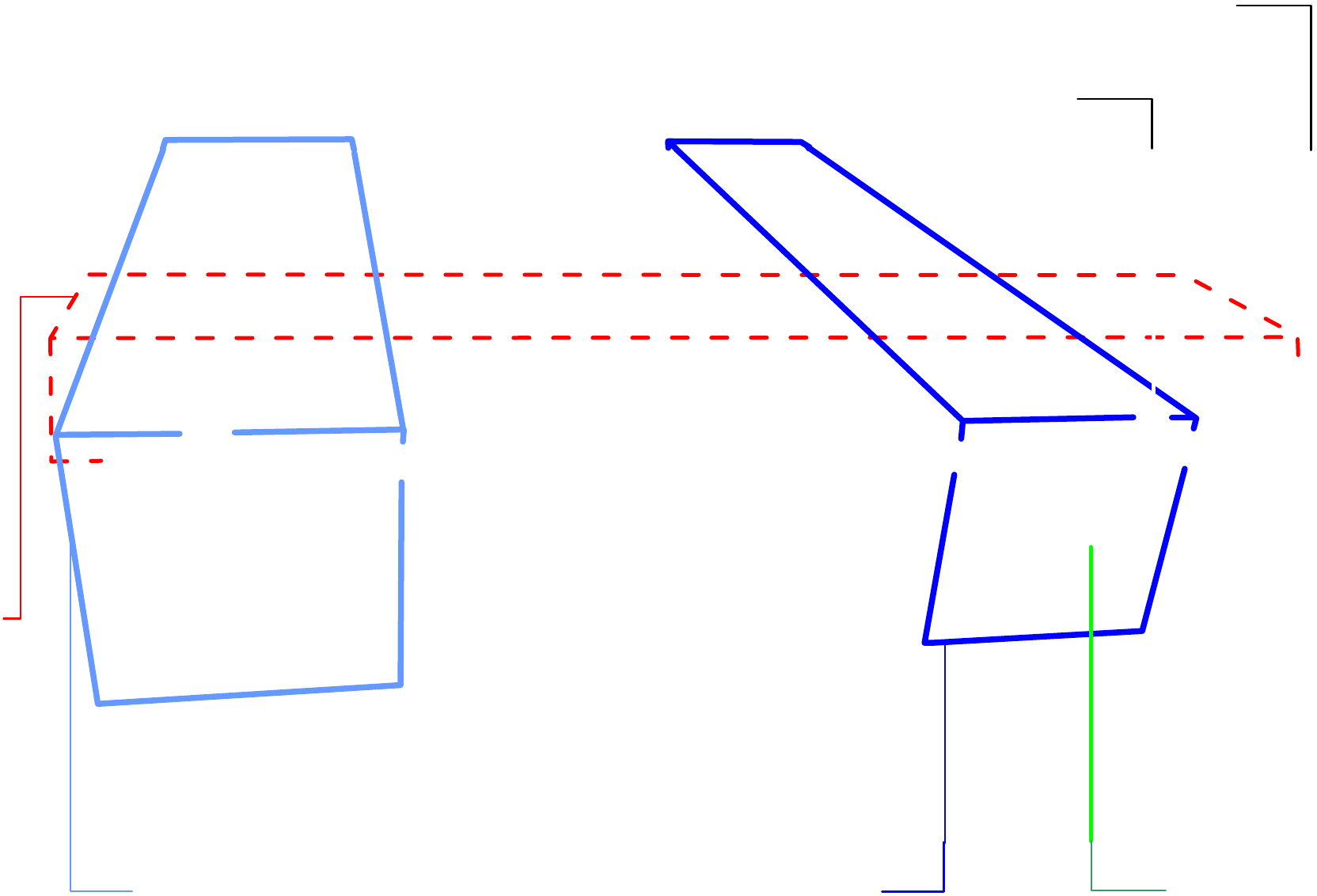}}

  \put(137.0,43.5){\makebox(0,0)[r]{\scriptsize\sf Adapter with fibers of the X2 module}}
  \put(129.5,39.5){\makebox(0,0)[r]{\scriptsize\sf Adapter with fibers of the Y2 module}}
  {\color{red}
  \put(78.5,15.0){\makebox(0,0)[r]{\scriptsize\sf module}}
  }
  \put(87.0,1.0){\makebox(0,0)[lb]{\scriptsize\sf Y5 module}}
  \put(120.5,1.0){\makebox(0,0)[rb]{\scriptsize\sf Y2 module}}
  \put(136.0,1.0){\makebox(0,0)[lb]{\scriptsize\sf WLS fibers}}
\end{picture}
\caption{One section consisting of 5 X and 5 Y modules.}
\label{Fig.Section}
\end{figure}

Ten X-Y planes stacked over each other form a Section (Fig.~\ref{Fig.Section}) which in fact can operate as an independent neutrino detector.
Each 50 parallel strips are combined logically to a module of a bar shape, so that the whole section (500 strips) is a structure of 10 intercrossing modules.
The module dimensions (20$\times$20$\times$100~cm$^3$) were optimized with numerous MC simulations showing that a typical neutrino signal is generated within a sphere with the diameter of about 30~cm. Each module is viewed by a compact PMT Hamamatsu R7600U-300 coupled with 3d-printed adapter to all 50 strips of the module via 100 wavelength shifting (WLS) fibers, two per strip.

The modular structure provides the following advantages with respect to a conventional single-volume detector:\\
--- It is very easy to realise coincidence operation mode (for instance, when the PMT signal is used as a short strobe for 50 MPPCs thus counteracting their random dark-current noise);\\
--- Coincident signal from an X-Y pair of intercrossing modules provides immediate rough position of the event (in space and in time) and could be used as a hard trigger (see subsection~\ref{Section.ACQ});\\
--- Each X-Y pair of intercrossing modules could operate separately as an independent detector with individual calibration;\\
--- Events in different parts of the detector could be analysed separately, depending on the desired parameter: upper/lower position (distance to the reactor core), inner/outer detector part (edge effects), etc.

%\begin{itemize}
% \item It is very easy to realise coincidence operation mode (for instance, when the PMT signal is used as a short strobe for 50 MPPCs thus counteracting their random dark-current noise);
% \item Coincident signal from an X-Y pair of intercrossing modules provides immediate rough position of the event (in space and in time) and could be used as a hard trigger (see subsection~\ref{Section.ACQ});
% \item Each X-Y pair of intercrossing modules could operate separately as an independent detector with individual calibration;
% \item Events in different parts of the detector could be analysed separately, depending on the desired parameter: upper/lower position (distance to the reactor core), inner/outer detector part (edge effects), etc.
%\end{itemize}

\subsection{Calibration system} \label{Section.Calibration}

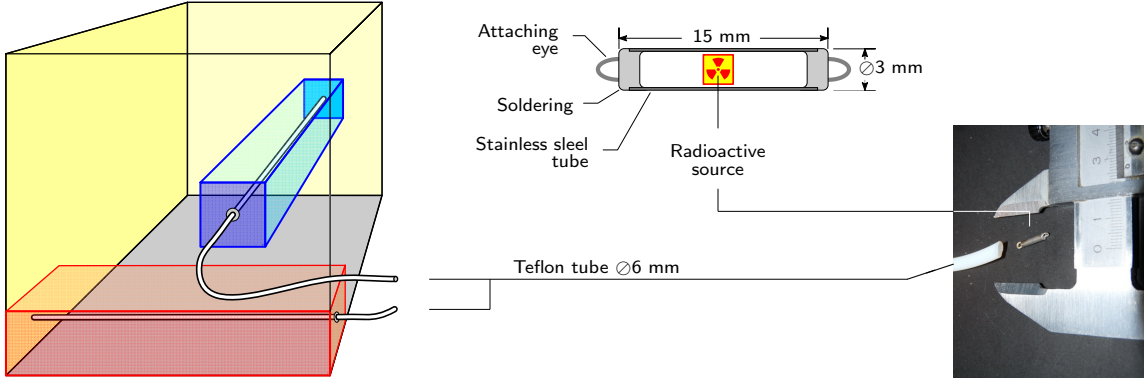
\begin{figure}[hbt]
 \setlength{\unitlength}{1mm}
 \begin{picture}(150,50)(0,0)
  \linethickness{0.005in}
  %\put(0,0){\framebox(150,50)[b]{}}
  \put(0,0){\includegraphics[height=50mm]{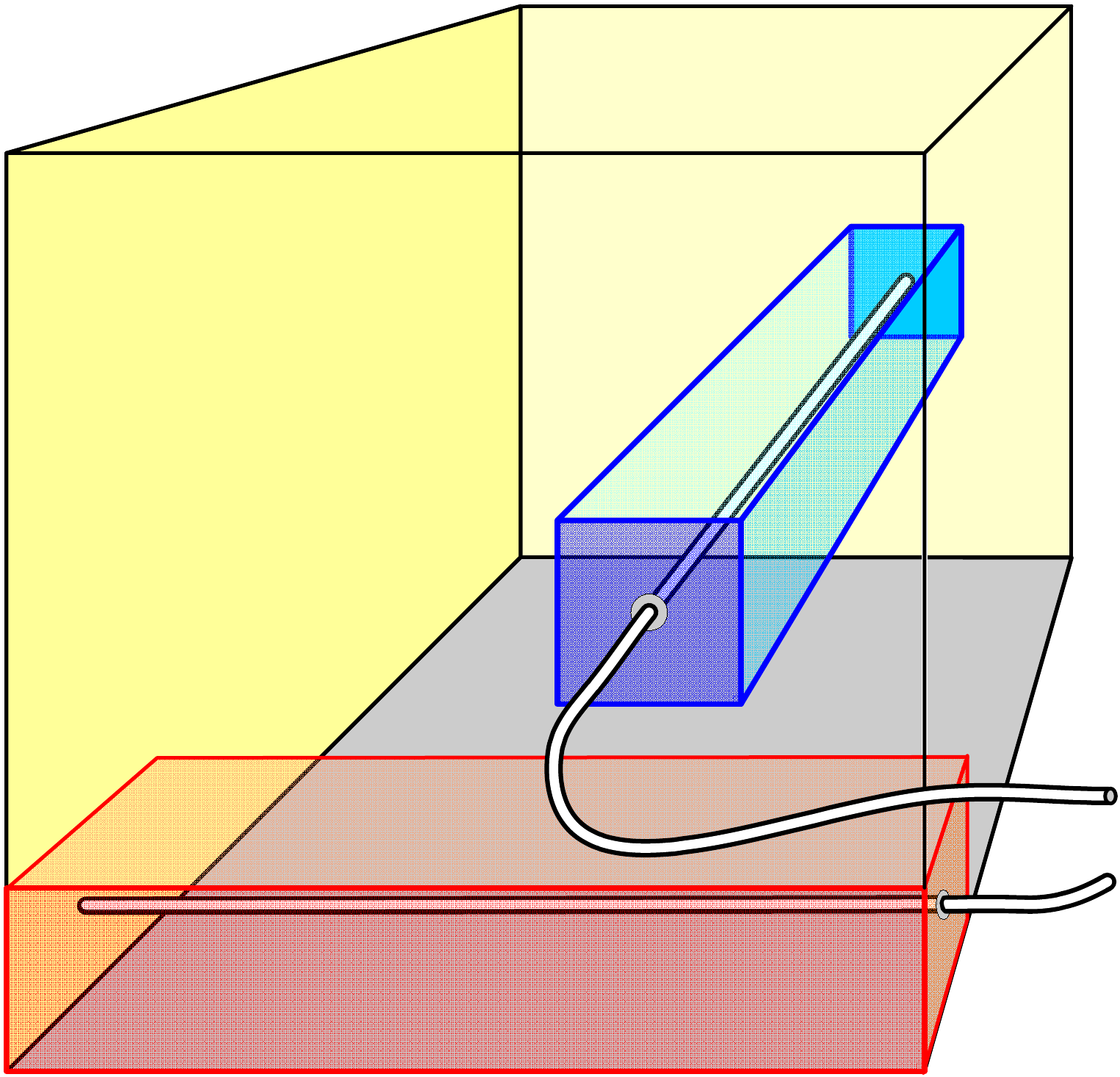}}
  \put(125,0){\includegraphics[width=25mm]{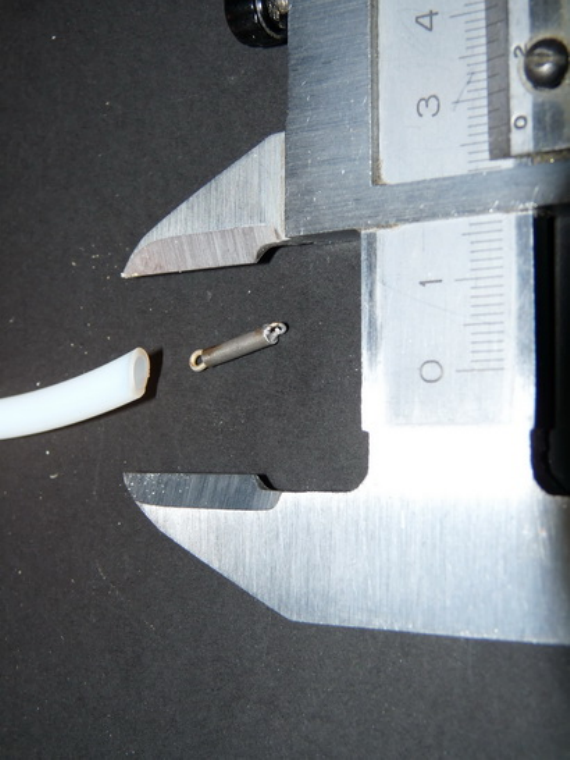}}
  \put(64.0,13.0){\line(0,-1){4}}
  \put(119.0,13.0){\line(-1,0){63}}
  \put(125,15.0){\line(-3,-1){6}}
  \put(64.0,9.0){\line(-1,0){8}}
\put(89,13.5){\makebox(0,0)[br]{\scriptsize\sf Teflon tube $\oslash$6 mm}}

  \put(74.8,30){\includegraphics[width=40mm]{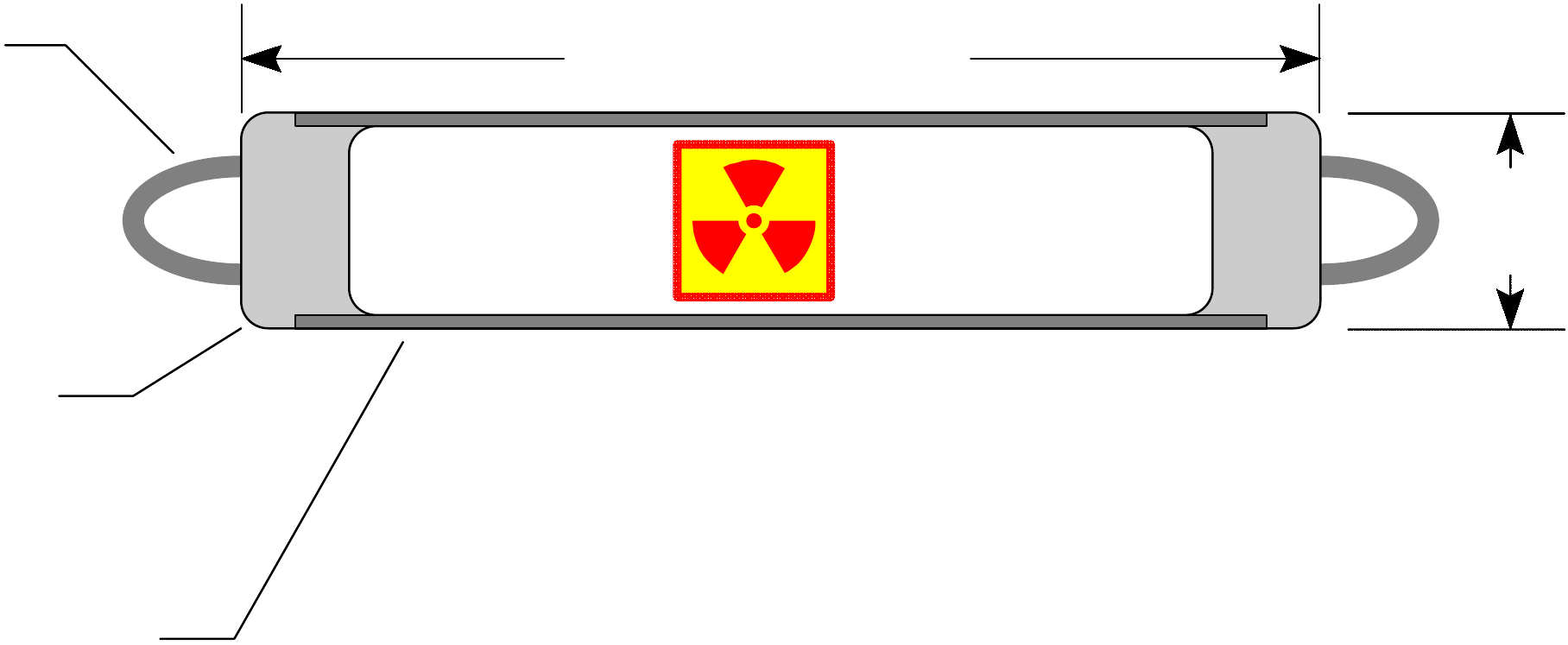}}
  \put(94.5,44.5){\makebox(0,0)[b]{\scriptsize\sf 15 mm}}
  \put(121,40.0){\makebox(0,0)[br]{\scriptsize\sf $\oslash$3 mm}}
  \put(58.0,42.0){\parbox{15mm}{\begin{flushright}\scriptsize\sf Attaching\\[-0.5mm]  eye\end{flushright}}}
  \put(75.0,37.0){\makebox(0,0)[rt]{\scriptsize\sf Soldering}}
  \put(57.0,29.0){\parbox{20mm}{\begin{flushright}\scriptsize\sf Stainless sleel\\[-0.5mm]  tube\end{flushright}}}
  \put(85.0,28.0){\parbox{18mm}{\begin{center}\scriptsize\sf Radioactive\\[-0.5mm]  source\end{center}}}
  \put(94.0,33.0){\line(0,1){7}}
  \put(94.0,22.0){\line(0,1){3}}
  \put(94.0,22.0){\line(1,0){31}}
  {\color{white}
   \put(125,22){\line(1,0){10}}
   \put(135,22){\line(0,-1){2}}
  }
\end{picture}
 \caption{Teflon tubes permeate the detector body and allow to introduce a compact calibration source.}
 \label{Fig.Calibration_Source}
\end{figure}

To perform energy calibration, some of DANSS modules are equipped with a teflon tube along their axes (Fig.~\ref{Fig.Calibration_Source}), so that a tiny radioactive source can be inserted into different position inside the detector by means of a thin flexible string. For this purpose several long-lived gamma and neutron sources ($^{137}$Cs, $^{60}$Co, $^{22}$Na, $^{248}$Cm) with activity of few Bq were produced and soldered hermetically in ampoules.

\begin{figure}[htb]
\setlength{\unitlength}{1mm}
\begin{minipage}[t]{70mm}
\centering
\begin{picture}(70,35)(0,0)
  %\put(0,0){\framebox(70,35)[b]{}}
\put(0,0){\includegraphics[height=35mm]{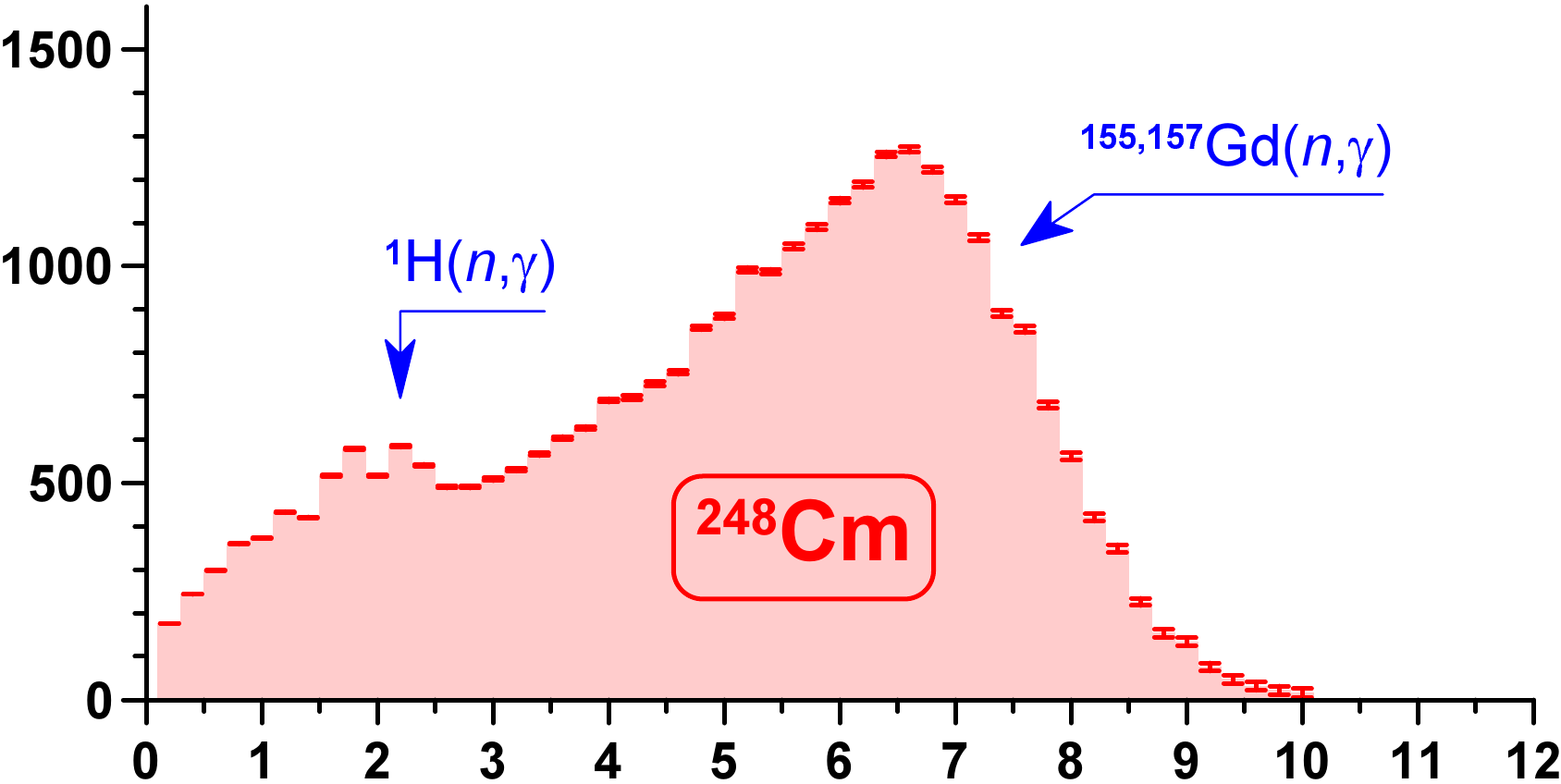}}
\put(9.0,34.0){\makebox(0,0)[lt]{\scriptsize\sf Count Rate, a.u.}}
\put(50.0,5.5){\parbox{20mm}{\begin{flushright}\scriptsize\sf Energy deposit\\[-0.5mm]  in the detector,\\[-0.5mm] MeV\end{flushright}}}
\end{picture}
\caption{Energy spectrum of the Delayed signal measured with $^{248}$Cm calibration source.}
\label{Fig.248Cm_neutron_spectrum}
\end{minipage}\hfill{ }
\begin{minipage}[t]{70mm}
\centering
\begin{picture}(70,35)(0,0)
  %\put(0,0){\framebox(70,35)[b]{}}
\put(0,0){\includegraphics[height=35mm]{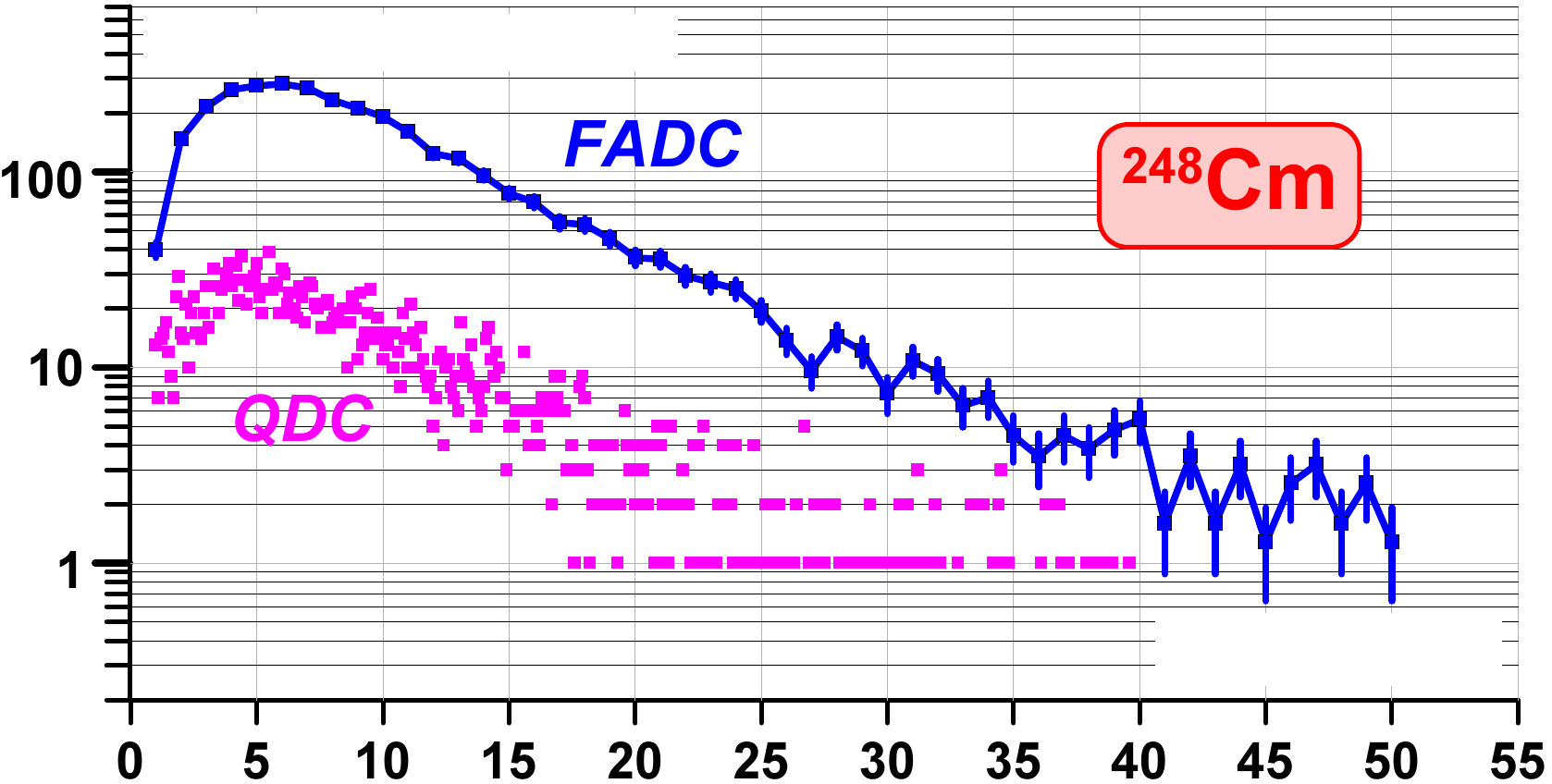}}
\put(9.0,34.0){\makebox(0,0)[lt]{\scriptsize\sf Count Rate, a.u.}}
\put(65.0,4.5){\makebox(0,0)[rb]{\scriptsize\sf $T_{\rm P-D}$, $\mu$s}}
\end{picture}
\caption{$T_{\rm P-D}$ distribution measured for $^{248}$Cm with different equipment and expositions.}
\label{Fig.TPD_248Cm}
\end{minipage}
\end{figure}

Fig.~\ref{Fig.248Cm_neutron_spectrum} demonstrates energy spectrum of the Delayed signal measured with the $^{248}$Cm neutron source installed inside the detector. Two peaks correspond to the neutron capture by Gd and hydrogen. Position of these peaks is a bit lower than expected because some part of gamma-rays escape the detector volume without detection. Time distribution of these delayed signals is shown in Fig.~\ref{Fig.TPD_248Cm}. It was measured with two independent acquisition systems (see subection~\ref{Section.ACQ}) but is in a very good agreement with each other and with the previous DANSSino results.

Regular monitoring of all optic sensors is performed with 50 test fibers bringing short light pulses from external LED to all modules. Using the method described above in the Section~\ref{Section.Strips.Yield}, these pulses allow to measure the number of photoelectrons produced at the PMT photocathode when a cosmic muon crosses the detector at known zenith angle (for instance, vertically).

\begin{figure}[bht]
\setlength{\unitlength}{1mm}
\begin{minipage}[c]{70mm}
\centering
\begin{picture}(70,27)(0,0)
%\put(0,0){\framebox(70,27)[b]{}}
\put(0,0){\includegraphics[height=27mm]{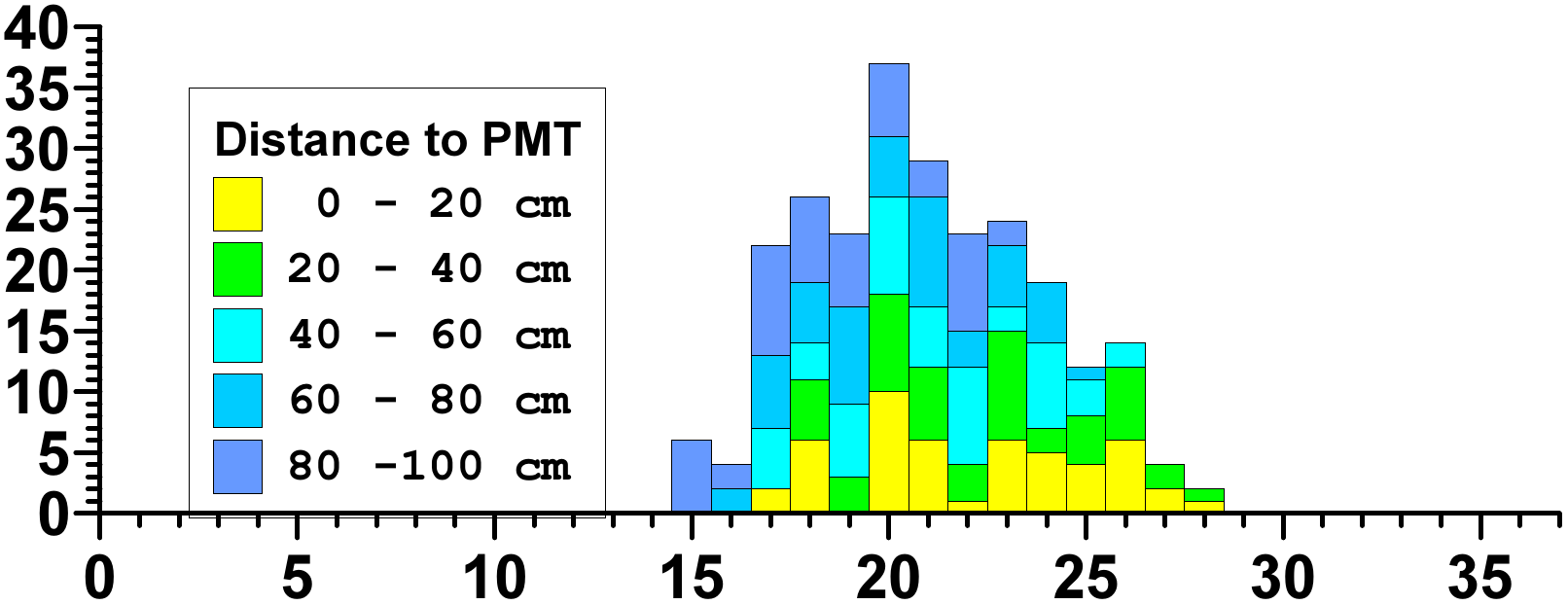}}
\put(5.0,27.0){\makebox(0,0)[lt]{\scriptsize\sf Rate}}
\put(70.0,4.5){\makebox(0,0)[rb]{\scriptsize\sf $n$, p.e./MeV}}
\end{picture}
\caption{Distribution of the signal yield from 50~PMTs, 5 distant fragments per each.}
\label{Fig.PMT_Sensitivity}
\end{minipage}\hfill{ }
\begin{minipage}[c]{73mm}
\centering
\begin{picture}(73,27)(0,0)
 %\put(0,0){\framebox(73,27)[b]{}}
\put(0,0){\includegraphics[height=27mm]{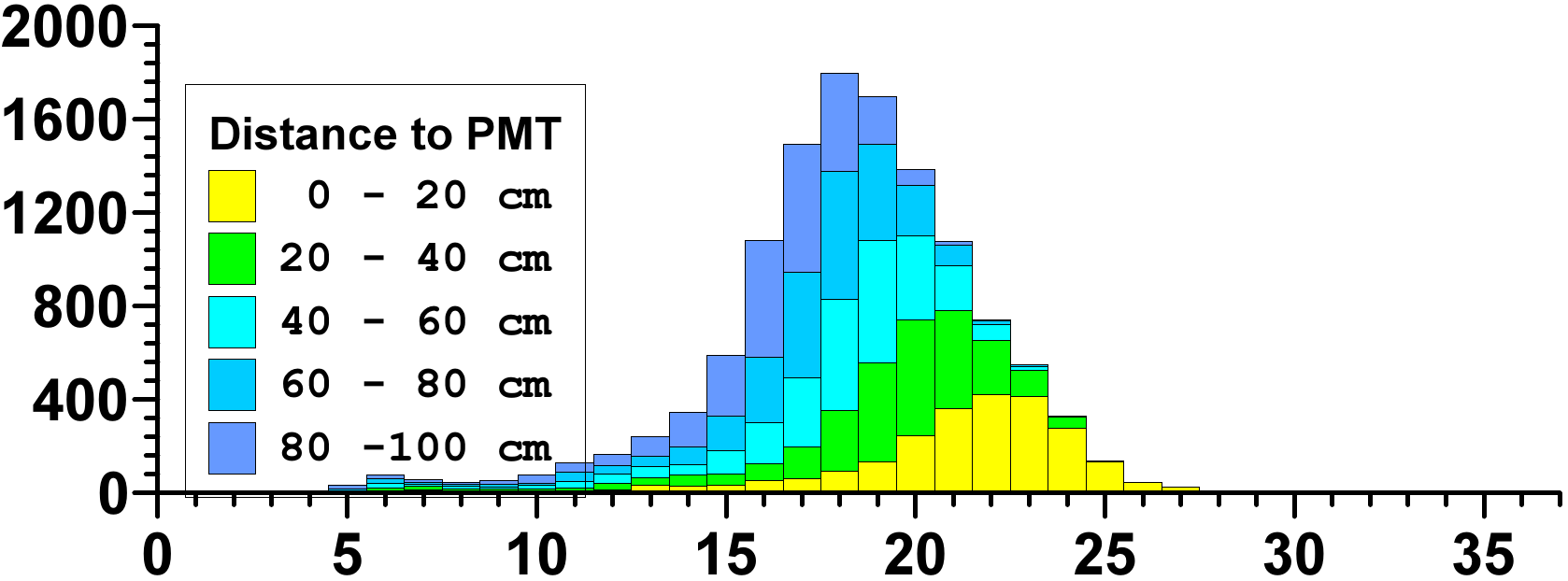}}
\put(8.0,27.0){\makebox(0,0)[lt]{\scriptsize\sf Rate}}
\put(73.0,4.5){\makebox(0,0)[rb]{\scriptsize\sf $n$, p.e./MeV}}
\end{picture}
\caption{Distribution of the signal yield from 2500~MPPCs, 5 distant fragments per each.}
\label{Fig.MPPC_Sensitivity}
\end{minipage}
\end{figure}

Figs.~\ref{Fig.PMT_Sensitivity} and \ref{Fig.MPPC_Sensitivity} show results of such measurement. It is seen that with PMTs and MPPCs we get in total about $n$=35 photoelectrons per 1~MeV energy deposit in the detector. At the mean value of the prompt signal of $E_P\simeq 3$~MeV it corresponds to the energy resolution $\Delta E_P/E_P = 1/\sqrt{n\cdot E_P}\simeq 10\%$ (i.e., 23\% {\sl FWHM}).

\subsection{Shielding} \label{Section.Shielding}

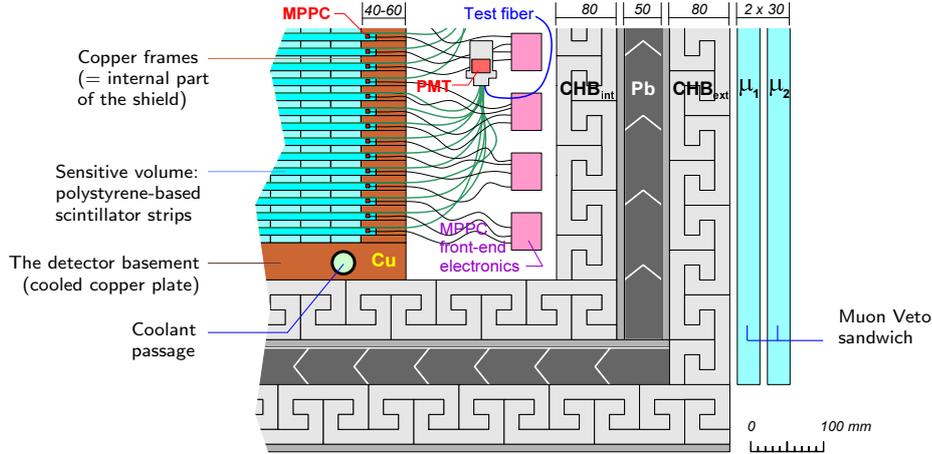
\begin{figure}[htb]
 \setlength{\unitlength}{1mm}
 \begin{picture}(150,60.0)(0,0)  %75
  %\put(0,0){\framebox(150,60.0)[b]{}}
  \put(40.0,0.0){\includegraphics[height=60mm]{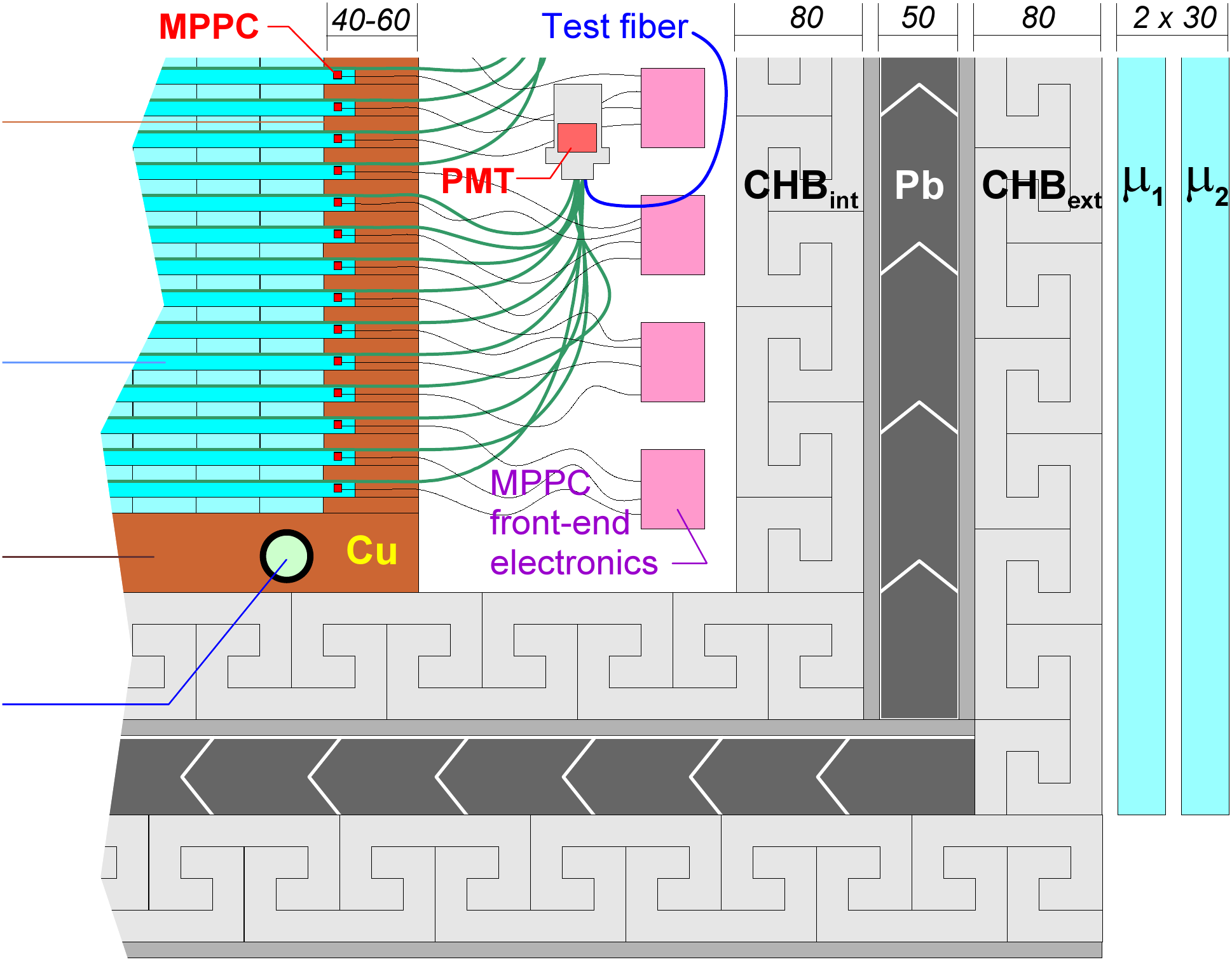}}
  \put(111.0,0.0){\includegraphics[width=16.5mm]{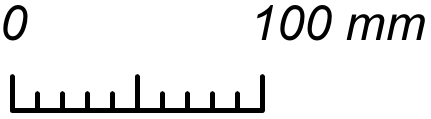}}
  \linewidth=1pt
   \put(20.0,33.5){\parbox{35mm}{\scriptsize\sf Sensitive volume:\\ polystyrene-based\\ scintillator strips}}
   \put(22.9,48.7){\parbox{30mm}{\scriptsize\sf Copper frames\\(= internal part\\ of the shield)}}
   \put( 4.0,21.0){\parbox{35mm}{\begin{flushright}\scriptsize\sf The detector basement \\(cooled copper plate)\\[-0.5mm]\end{flushright}}}
   \put(30.0,14.0){\parbox{10mm}{\scriptsize\sf Coolant \\ passage}}
   {\color{blue}
    \put(111,15){\line(1,0){10}}
    \put(111,15){\line(0,1){2}}
    \put(115,15){\line(0,1){2}}
   }
   \put(123.0,16.0){\parbox{20mm}{\scriptsize\sf Muon Veto \\ sandwich}}
 \end{picture}
 \caption{Composition of the detector shielding.}
 \label{Fig.Shielding}
\end{figure}

All scintillator strips and calibration tubes are carried by copper frames (Cu) which at the same time act as internal part of gamma-shielding (Fig.~\ref{Fig.Shielding}). An outer surface of frames is used as a heatsink for the PMT and MPPC front-end electronics (HV power supply, preamplifiers and discriminators). Then detector is surrounded with a combined passive shield of lead (Pb) and borated polyethylene (CHB). Lead protects the detector against $\gamma$-rays (mainly, 1.461~MeV of $^{40}$K from concrete walls), whereas the external layer of borated polyethylene moderates and captures thermal and epithermal neutrons which could penetrate to the room through the reactor shield holes. The second (internal) CHB layer is used to suppress secondary neutrons induced by cosmic muons in massive lead shield.  As this CHB-Pb-CHB sandwich completely prevents heat exchange, copper basement and roof of the frames are cooled\footnote{Total heat release of the front-end electronics is not high (350-400~W), but the MPPC operation temperature should not be higher than 20-25$^\circ$C.} with an external water chiller TAEevo M03.

\begin{figure}[ht]
 \setlength{\unitlength}{1mm}
\begin{minipage}[t]{90mm}
\centering
 \begin{picture}(90,17)(0,1)
  %\put(0,0){\framebox(90,18)[b]{}}
  \put(5.0,0){\includegraphics[width=83mm]{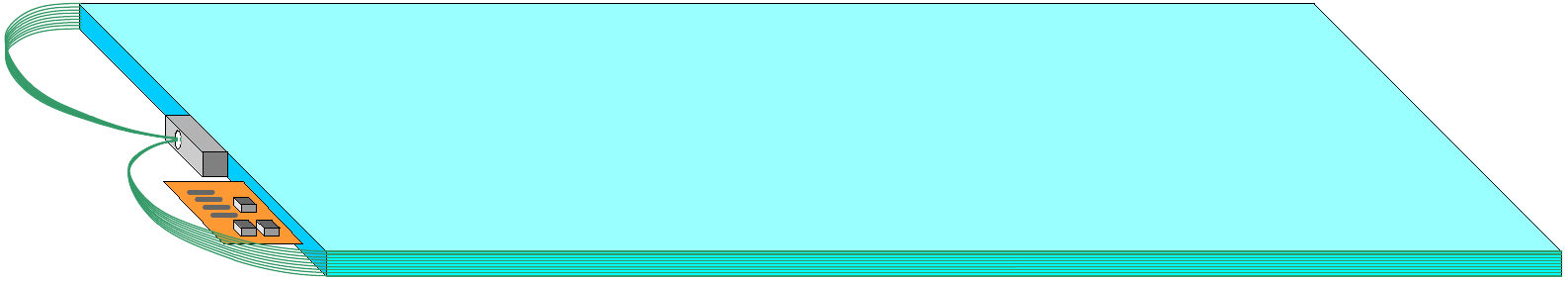}}

  \put(23,8){\line(-1,0){5}}
   \put(23.5,8){\makebox(0,0)[l]{\scriptsize\sf Photosensor H10720-20}}

  \put(26,5){\line(-1,0){5}}
   \put(26.5,5){\makebox(0,0)[l]{\scriptsize\sf Front-end electronics}}

  \put(0,1){\line(1,0){2}}
   \put(0,1){\line(0,1){12}}
   \put(0,13){\line(1,0){4.5}}
   \put(0,5){\line(1,0){11}}
   \put(2.5,1){\makebox(0,0)[l]{\scriptsize\sf WLS fibers}}

  \put(42,17){\makebox(0,0)[c]{\tiny L = 1300, 1500, 2000 mm}}
   \put(09.5,15.0){\line(0,1){3}}
   \put(74.5,15.0){\line(0,1){3}}
   \put(09.5,17.0){\line(1,0){17}}
   \put(74.5,17.0){\line(-1,0){17}}

  \put(87.5,2.0){\line(0,1){3}}
   \put(74.5,17.0){\rotatebox{-45}{\line(1,0){4}}}
   \put(85.0,6.5){\rotatebox{-45}{\line(1,0){3.5}}}
   \put(78.5,12.5){\rotatebox{-45}{\tiny 500 mm}}

  \put(90.0,1.9){\line(-1,0){2}}
  \put(90.0,0.4){\line(-1,0){2}}
  \put(90.0,17){\line(0,-1){17}}
  \put(90.0,17){\line(-1,0){2}}
  \put(87,17){\makebox(0,0)[r]{\tiny 30 mm}}

 \end{picture}
 \caption{Muon-veto detector -- polystyrene-based scintillator plate.}
 \label{Fig.Mu_Plate}
 \end{minipage}\hfill{ }
\begin{minipage}[t]{50mm}
\centering
\begin{picture}(50,17)(0,1)
  %\put(0,0){\framebox(50,18)[b]{}}
  \put(0,0){\includegraphics[width=50mm]{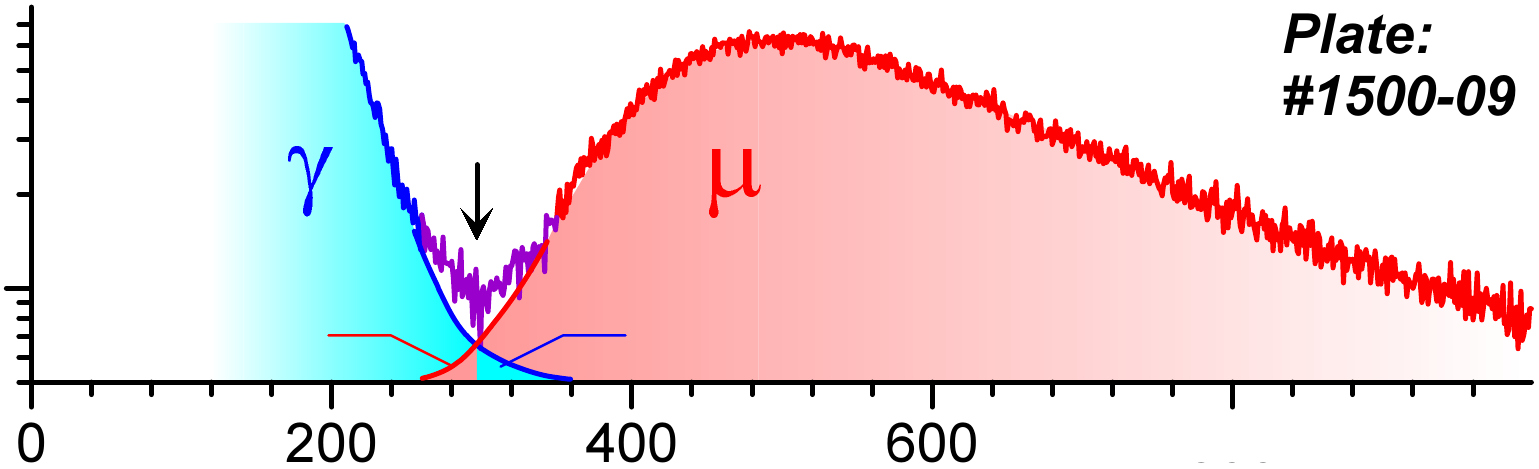}}
    \put(0.0,17.5){\makebox(0,0)[lt]{\tiny\sf Relative Count Rate}}
    \put(50.0,0.2){\makebox(0,0)[rb]{\tiny\sf ADC channels}}
  {\color{red}
  \put(3.0,4.5){\parbox{10mm}{\scriptsize\sf Lost \\[-1.5mm]muons}}
   \color{blue}
  \put(20.5,4.5){\makebox(0,0)[l]{\scriptsize\sf False veto}}
   }
  \put(16.0,10.5){\makebox(0,0)[b]{\scriptsize\sf $E_{\rm thr}$}}
 \end{picture}
 \caption{Energy spectrum measured with the $\mu$-veto plate under laboratory conditions.}
 \label{Fig.MuVeto_Spectrum_lab}
 \end{minipage}

\begin{minipage}[t]{50mm}
\centering
\begin{picture}(50,25)(0,1)
 % \put(0,0){\framebox(50,25)[b]{}}
  \put(0,0){\includegraphics[width=50mm]{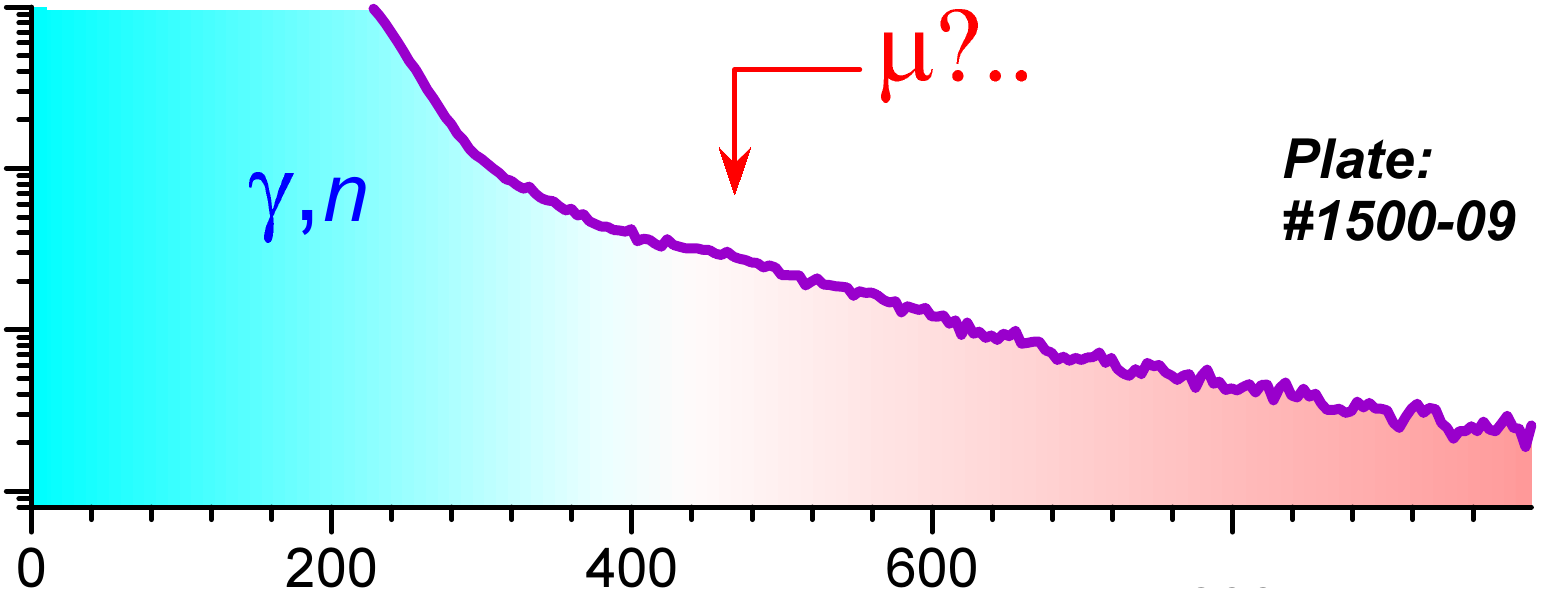}}
    \put(0.0,21.5){\makebox(0,0)[lt]{\tiny\sf Relative Count Rate}}
    \put(50.0,0.2){\makebox(0,0)[rb]{\tiny\sf ADC channels}}
 \end{picture}
 \caption{Energy spectrum measured with a plate under operating reactor.}
 \label{Fig.MuVeto_Spectra_site}
 \end{minipage}\hfill{ }
\begin{minipage}[t]{90mm}
\centering
\begin{picture}(90,25)(0,1)
 % \put(0,0){\framebox(90,25)[b]{}}
  \put(0,0){\includegraphics[height=25mm]{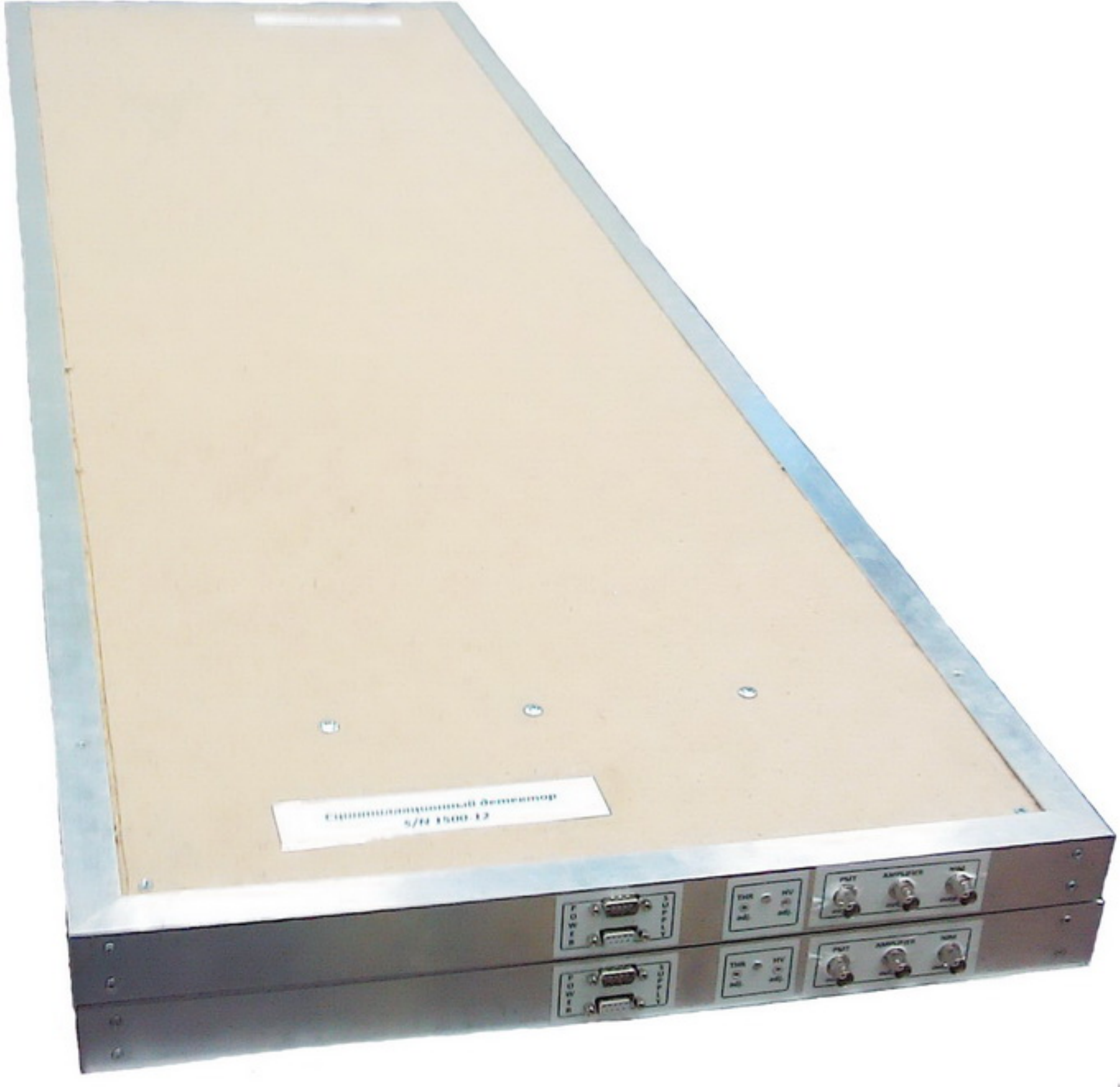}}
  \put(40,0){\includegraphics[width=50mm]{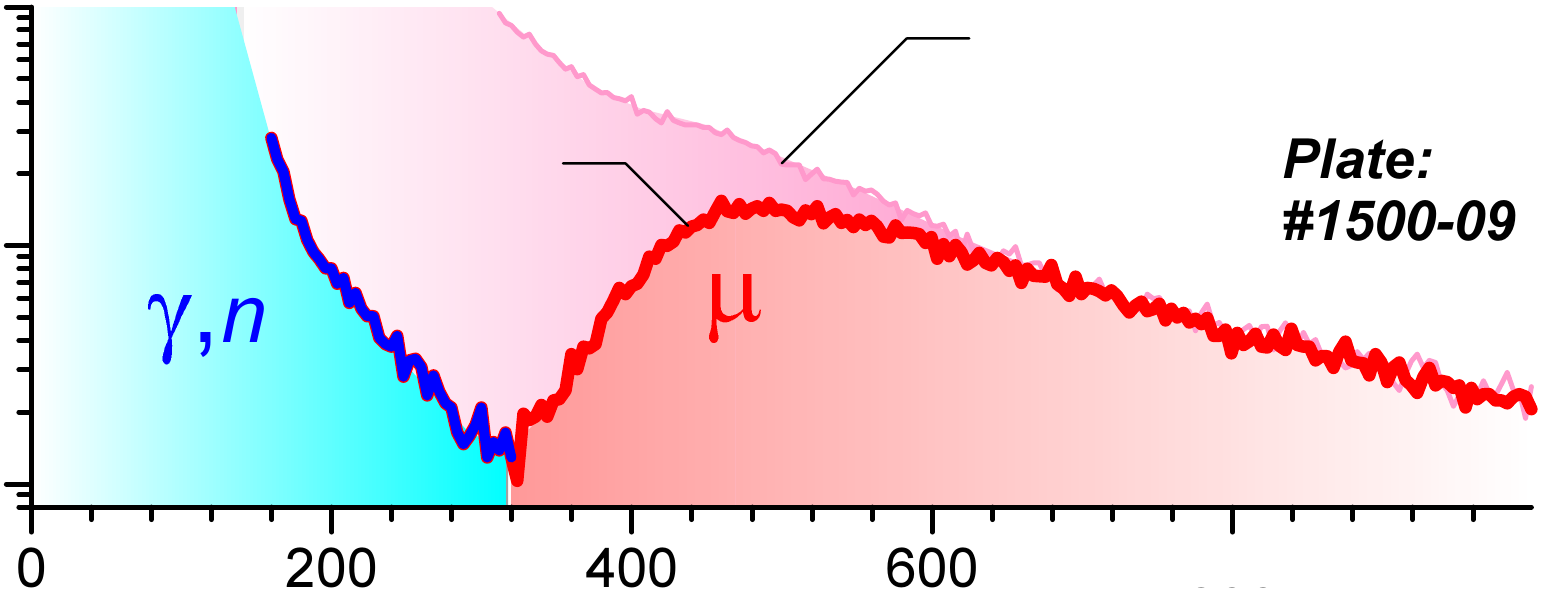}}
    \put(40.0,21.5){\makebox(0,0)[lt]{\tiny\sf Relative Count Rate}}
    \put(90.0,0.2){\makebox(0,0)[rb]{\tiny\sf ADC channels}}
    \put(71.5,17.2){\parbox{20mm}{\color{magenta}\tiny\sf Single\\[-0.5mm]mode}}
    \put(52.5,13.7){\parbox{20mm}{\color{red}\tiny\sf Coinc.\\[-0.5mm]mode}}
 \end{picture}
 \caption{Sandwich of two plates and the spectrum measured with one of them in coincident mode.}
 \label{Fig.MuVeto_Sandwich}
 \end{minipage}
\end{figure}

A set of big scintillator plates (Fig.~\ref{Fig.Mu_Plate}) form an active veto system which is used to tag the events associated with cosmic muons. Operating under normal laboratory conditions a plate detects not only muons but gamma background also (Fig.~\ref{Fig.MuVeto_Spectrum_lab}). With appropriate quality of the scintillator and photosensor it is possible to fix an energy threshold $E_{\rm thr}$ low enough to ensure both efficiency and selectivity of the muon veto. In other words, both percentage of undetected muons and a number of false veto signals caused by gamma-background are negligible.

Unfortunately, it is not the case of the room \#A336. Relatively high neutron flux there is detected by the plate rather efficiently, thus increasing non-muon background and percentage of false veto signals (and dead time, resp.). Fig.~\ref{Fig.MuVeto_Spectra_site} shows the spectrum measured with the same plate at the DANSS site.

In order to increase the veto selectivity, each two plates compose a sandwich (Fig.~\ref{Fig.MuVeto_Sandwich}) operating in coincidence mode. Then it becomes possible to separate true muon signals (about 110~Hz in total) from gamma and neutron background. It must be mentioned that the operation mode of each two plates (coincident or independent) is not hardware-fixed but can be chosen in a posteriori analysis.

\subsection{Lifting system}
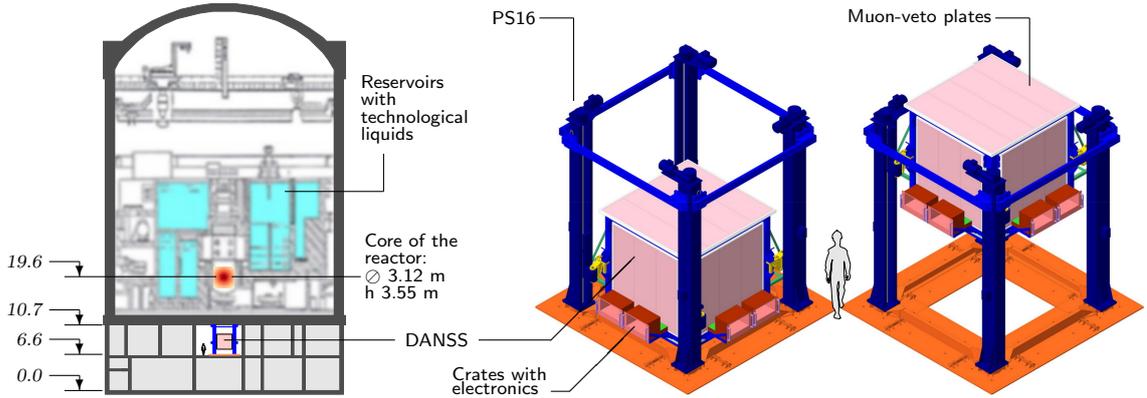
\begin{figure}[bht]
 \setlength{\unitlength}{1mm}
 \begin{picture}(150,52.0)(0,0)
  %\put(0,0){\framebox(150,52.0)[b]{}}
  \put(14.0,10.0){\includegraphics[width=30mm]{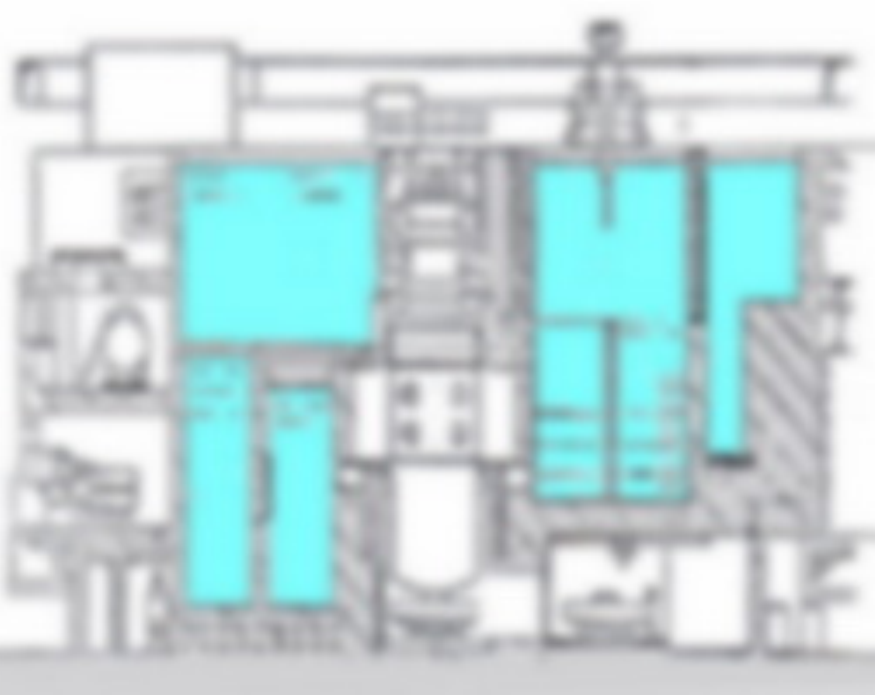}}%[width=0.4\textwidth]
  \put(14.0,35.0){\includegraphics[width=30mm]{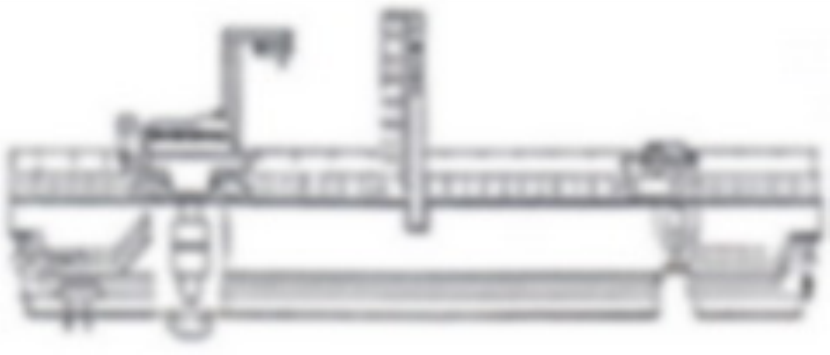}}%[width=0.4\textwidth]
\put(13.0,0.0){\includegraphics[height=52mm]{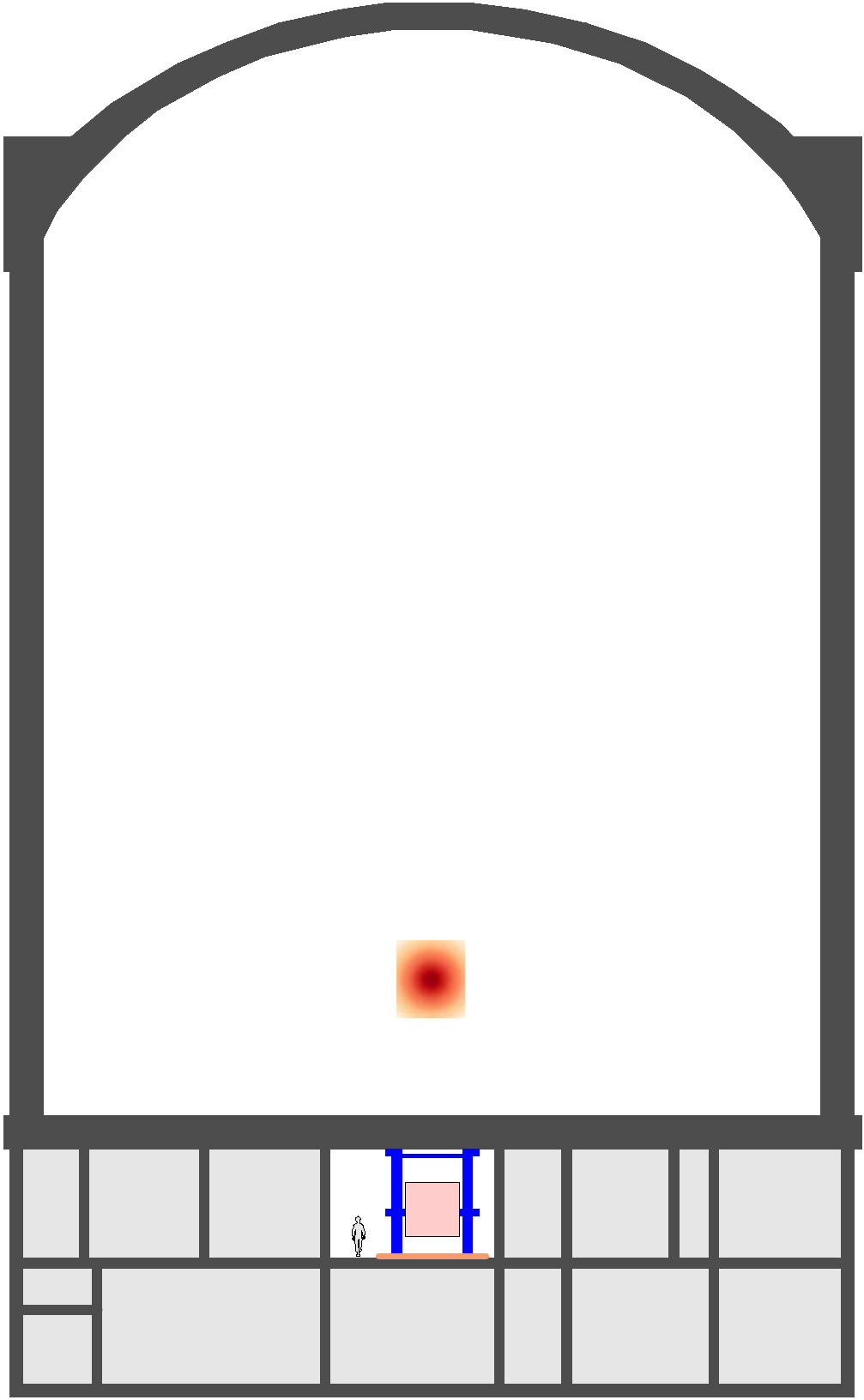}}%[width=0.4\textwidth]

  \put( 70.0,0.0){\includegraphics[width=40mm]{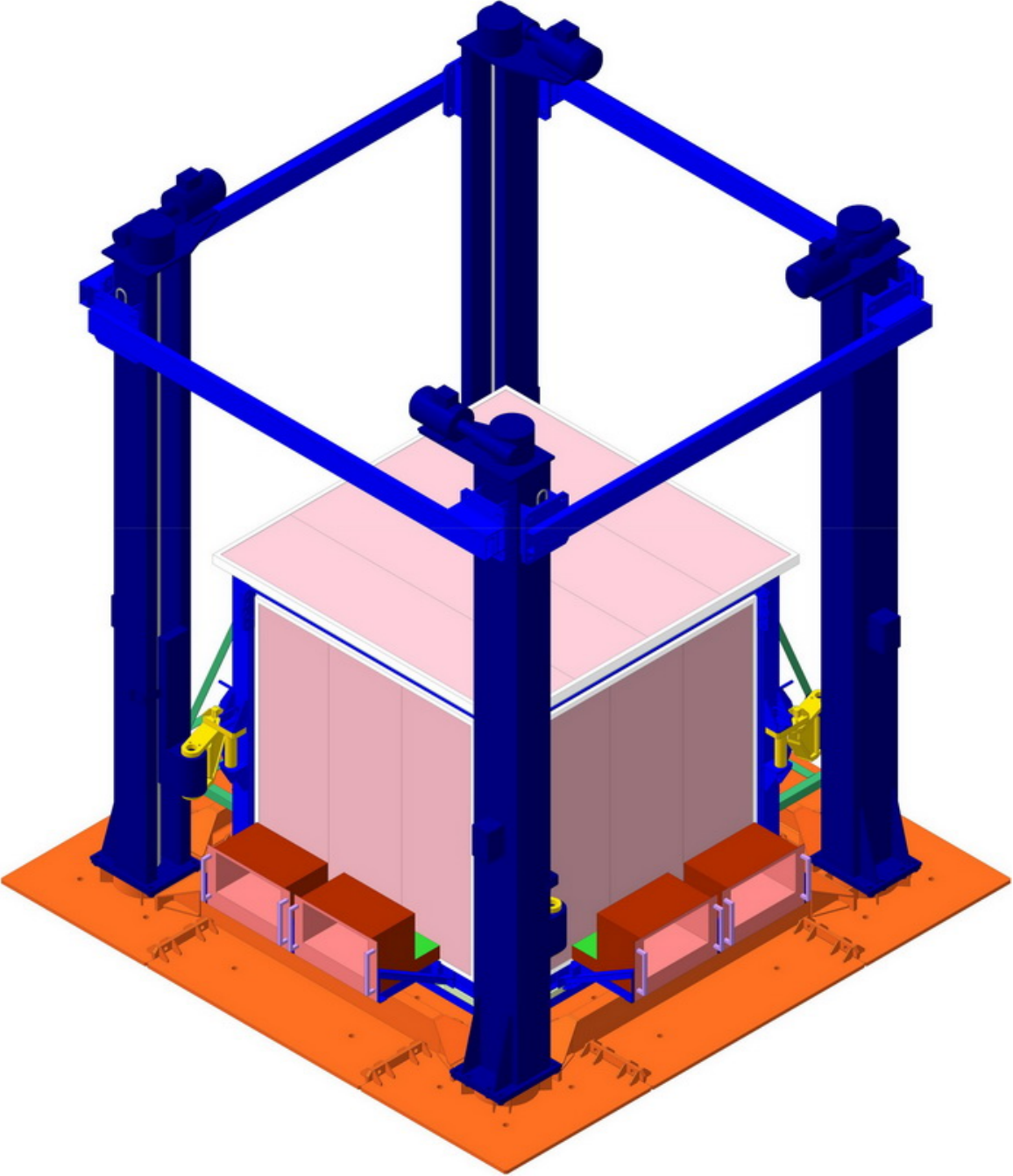}}%[width=0.4\textwidth]
  \put(110.0,0.0){\includegraphics[width=40mm]{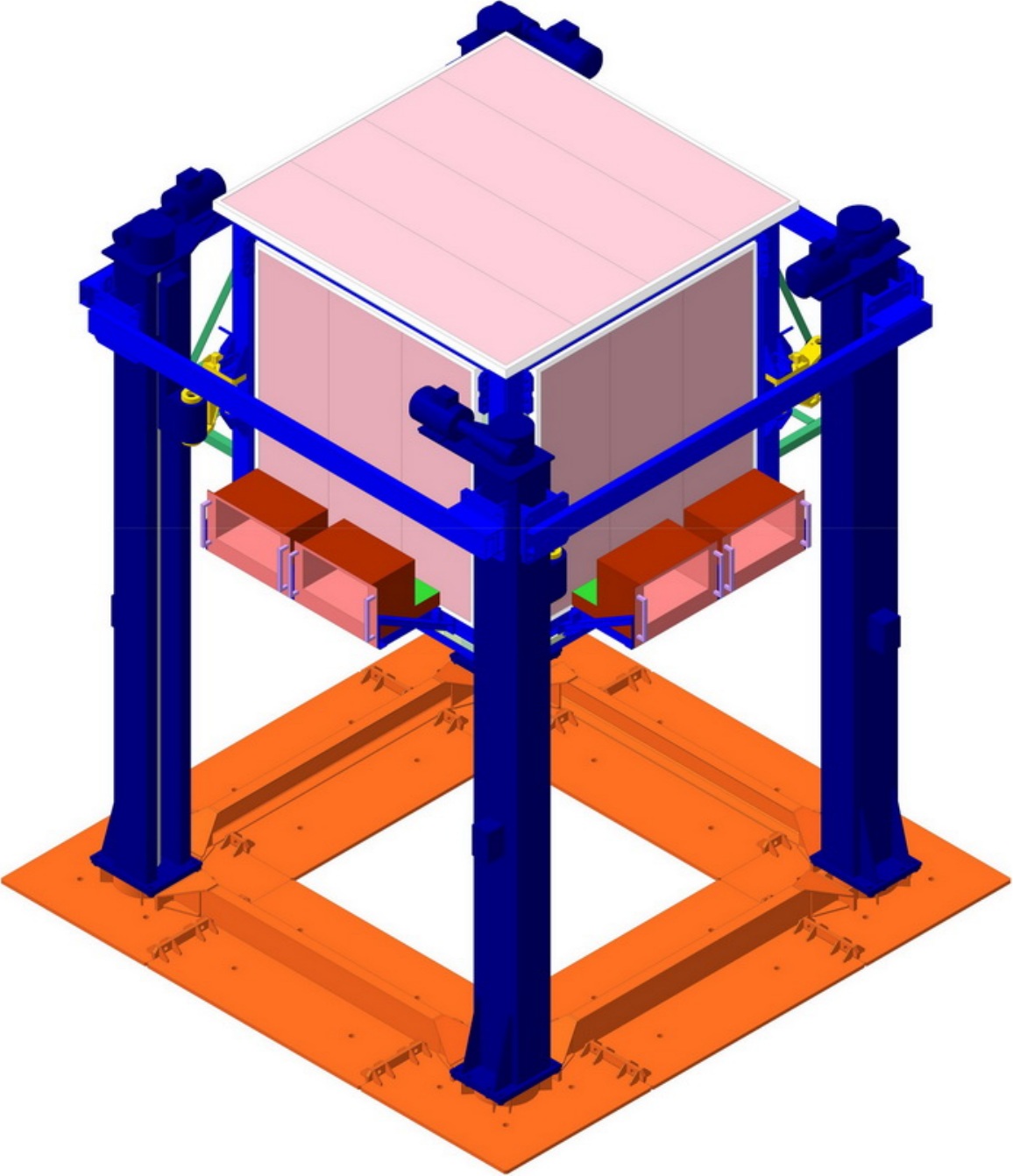}}%[width=0.4\textwidth]
  \put(108.0,10.0){\includegraphics[height=12mm]{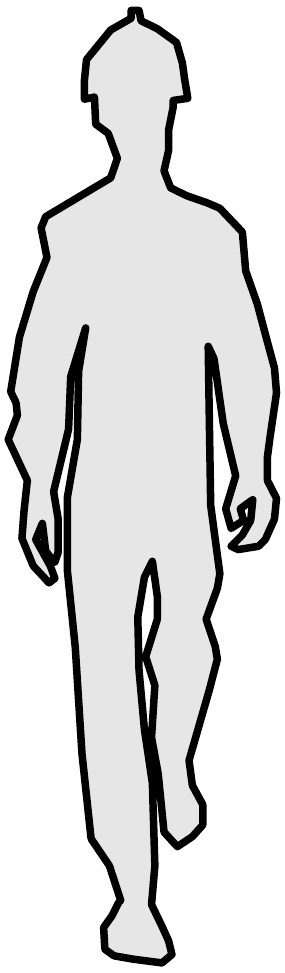}}%[width=0.4\textwidth]
  \linewidth=1pt

  \put(08.0,15.7){\line(1,0){20.0}}
   \put(10.0,17.7){\vector(0,-1){2}}
   \put(10.0,17.7){\line(-1,0){4}}
   \put( 5.0,17.7){\makebox(0,0)[r]{\scriptsize\sl 19.6}}

  \put (08.0, 9.3){\line(1,0){5.0}}
   \put(10.0,11.3){\vector(0,-1){2}}
   \put(10.0,11.3){\line(-1,0){4}}
   \put( 5.0,11.3){\makebox(0,0)[r]{\scriptsize\sl 10.7}}

  \put (08.0,5.4){\line(1,0){5.0}}
   \put(10.0,7.4){\vector(0,-1){2}}
   \put(10.0,7.4){\line(-1,0){4}}
   \put( 5.0,7.4){\makebox(0,0)[r]{\scriptsize\sl 6.6}}

  \put (08.0, 0.6){\line(1,0){5.0}}
   \put(10.0, 2.6){\vector(0,-1){2.0}}
   \put(10.0, 2.6){\line(-1,0){4}}
   \put( 5.0, 2.6){\makebox(0,0)[r]{\scriptsize\sl 0.0}}

  \put(51.5,7.3){\line(-1,0){22.5}}
  \put(57.0,7.3){\makebox(0,0)[c]{\scriptsize\sf DANSS}}
  \put(72.0,7.3){\line(-1,0){9.5}}
   \put(72.0,7.3){\line(1,1){11}}

  \put(47.0,15.7){\line(-1,0){17.0}}
   \put(47.5,14.5){\parbox{20mm}{\begin{flushleft}\scriptsize\sf Core of the\\[-0.5mm]
    reactor:\\[-0.5mm] $\oslash$ 3.12 m \\[-0.5mm] h 3.55 m\end{flushleft}}}

  \put(75.0,50.0){\line(-1,0){4.5}}
  \put(75.0,50.0){\line(0,-1){11.0}}
   \put(70.0,50.0){\makebox(0,0)[r]{\scriptsize\sf PS16}}

  \put(50,27){\line(-1,0){13}}
  \put(50,27){\line(0,1){5}}
  \put(47,37){\parbox{20mm}{\begin{flushleft}\scriptsize\sf Reservoirs\\[-0.5mm] with\\[-0.5mm] technological\\[-0.5mm] liquids\end{flushleft}}}

 \put(59.0, 1.0){\parbox{20mm}{\scriptsize\sf Crates with\\[-1.0mm]electronics}}
  \put(74.5,0.5){\line(-1,0){2.5}}
   \put(74.5,0.5){\line(1,1){8.5}}

\put(130.0,50.0){\makebox(0,0)[r]{\scriptsize\sf Muon-veto plates}}
  \put(130.5,50.0){\line(1,0){4.5}}
   \put(135.0,50.0){\line(0,-1){9}}

 \end{picture}
 \caption{The movable neutrino detector under the industrial reactor WWER-1000.}
 \label{Fig.Lift}
\end{figure}

To make the detector movable, a special lifting system (Fig.~\ref{Fig.Lift}) was designed on the basis of commercial hoisting gear PS16 which is commonly used in auto-repair centers to lift heavy tracks. After modification, it is able to move the DANSS detector with shielding ($m\simeq15$~ton) to the height up to 2.5~m. Finally, taking into account thickness of the shielding and other mechanical details, a distance between the centers of the detector and the reactor core varies from 10 to 12~m.

The PS16 gear is equipped with four standard motors AIR90L4, 2.2~kW each. Such three phase induction motor with short-circuited rotor generates rotational torque which is proportional to so-called ``slip'' -- the difference between synchronous speed and operating speed. In our case it means that rotation speed of four motors could be slightly different because of their non-equal loads.

In order to prevent any significant warp or non-horizontality of the movable platform and ensure comfortable sparkless switching\cite{CRYDOM} a special stabilization electronic system is used. It includes four solid state three phase reversing relays %MCT-311P
controlled by ultrasonic sensors and providing appropriate motion characteristics.

\subsection{Data acquisition system} \label{Section.ACQ}
It is useless to register {\sl all} the information from numerous photo sensors of the spectrometer -- 50 PMTs of 25 X and 25 Y modules and 2500 individual MPPCs of all strips. In order to perform some preliminary selection at the very first stage of the data acquisition, so-called ``Hardware Trigger'' ({\sl HT}) is required. The {\sl HT} is a signal produced by the detector elements at a hardware level, and only appearance of {\sl HT} causes start of an acquisition procedure for each event.

\begin{figure}[ht]
 \setlength{\unitlength}{1mm}
 \begin{minipage}[t]{72mm}
  \centering
  \begin{picture}(70,41)(0,0)
   %\put(0,0){\framebox(70,41)[b]{}}
   \put(6,25){\includegraphics[width=60mm]{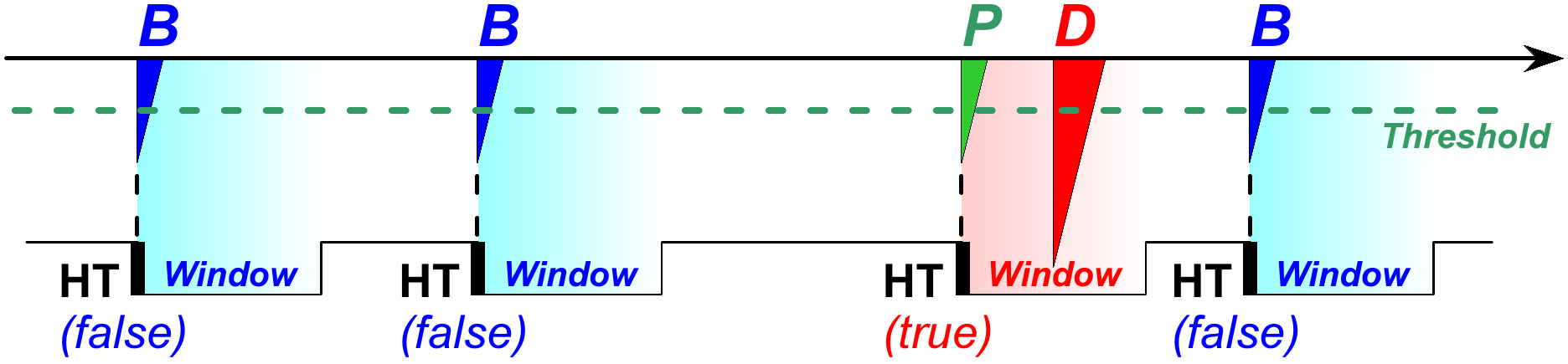}}
   \put(70.0,38.3){\makebox(0,0)[r]{\tiny\sf Time}}
   \put(3.2,38.0){\circle{6}}
   \put(3.2,38.0){\makebox(0,0)[c]{\LARGE\sf a}}

   \put(6, 0){\includegraphics[width=60mm]{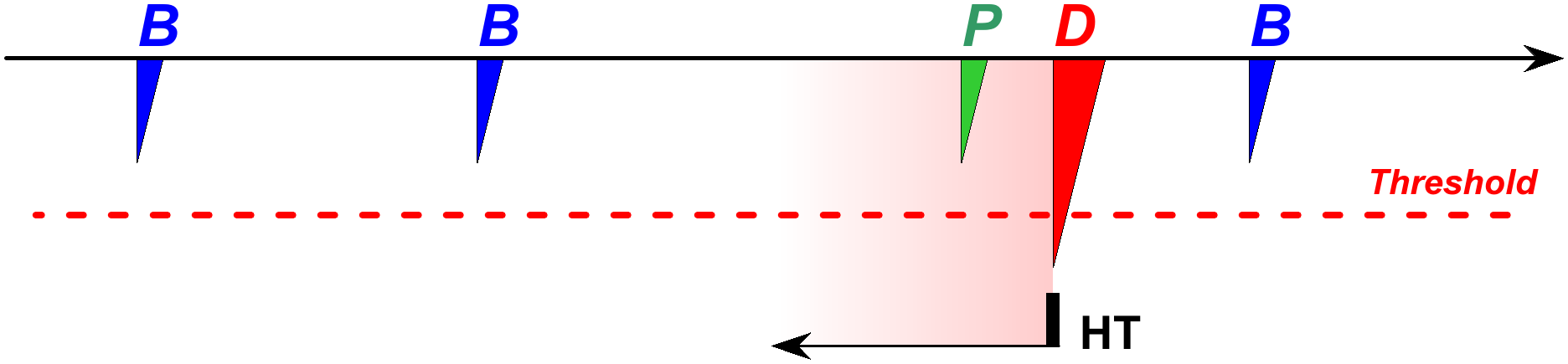}}
   \put(70.0,13.3){\makebox(0,0)[r]{\tiny\sf Time}}
   \put(3.2,13.0){\circle{6}}
   \put(3.2,13.0){\makebox(0,0)[c]{\LARGE\sf b}}
   \put(24.5,0.0){\parbox{20mm}{\tiny\sl \begin{flushright}Retrospective\\[-0.7mm] analysis\end{flushright}}}
  \end{picture}
  \caption{Two alternative types of {\sl HT} for the true IBD event consisting of the Prompt ($P$) and Delayed ($D$) signals in presence of background pulses ($B$).}
  \label{Fig.TwoHT}
 \end{minipage}\hfill{ }
 \begin{minipage}[t]{70mm}
  \centering
  \begin{picture}(70,41)(0,0)
   %\put(0,0){\framebox(70,41)[b]{}}
   \put(0, 0){\includegraphics[width=70mm]{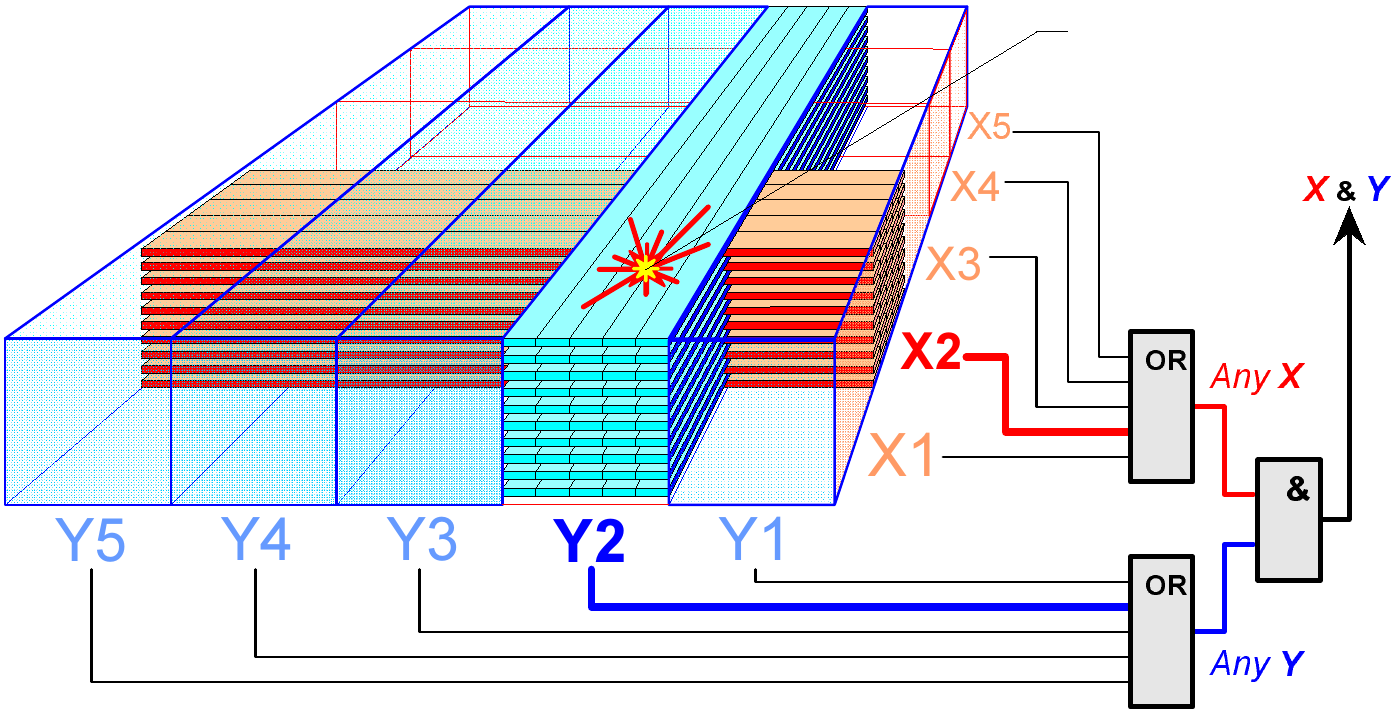}}
   \put(53.5,35.0){\makebox(0,0)[l]{\normalsize\sf $^{157}$Gd($n$,$\gamma$)}}
  \end{picture}
  \caption{Detection of the neutron-induced $\gamma$-cascade by two intercrossing modules of the same section as a possible Hardware Trigger.}
  \label{Fig.HardTrigger}
 \end{minipage}
 \end{figure}

In our case there are two obvious types of the {\sl HT}.
The first of them is the most simple one. Within this method (Fig.~\ref{Fig.TwoHT}a), the {\sl HT} is produced by any Prompt signal, and then the system waits for the Delayed signal during some fixed time window. The energy of both Prompt and Delayed signals ($E_P$ and $E_D$) detected by all X and Y fired modules are measured with a number of Charge-to-Digital Converters (QDC$_P$ and QDC$_D$), which are gated separately by $S_P$ and $S_D$ strobes (Fig.~4c in~\cite{DANSSino2}). Finally, each collected event contains two energies ($E_{P}$ and $E_{D}$) with their specific space patterns, time between the P and D pulses ($T_{PD}$),  and information about the muon veto (which of the plates were fired and when).

This scheme was successfully tested with a prototype detector DANSSino \cite{DANSSino2} in the room 3-A336 but can successfully function only under a relatively low background count rate. As was mentioned in Sec.~\ref{Section.Neutron_BG}, the neutron background in the room 4-A336 is four times higher. Nevertheless, the first preliminary tests performed with DANSS show that with the energy threshold of each PMT at a level of 500~keV, the total count rate from all 50 modules does not exceed 900 Hz with 2.5\% of dead time.

An alternative (and more reliable) type of {\sl HT} is detection of the Delayed neutron capture (Fig.~\ref{Fig.HardTrigger}), as the amplitude and multiplicity of the neutron signal are much higher than those of the natural $\gamma$-background producing absolute majority of signal pulses.

Unfortunately, the neutron is detected later than the positron. That is why a method based on this type of {\sl HT} (Fig.~\ref{Fig.TwoHT}b) requires permanent digitization of the total data stream with flash ADCs and subsequent retrospective analysis of the preceding signals (in this way one can spot the Prompt signal which could happen few tens of a microsecond before the {\sl HT} and have relatively low energy).

Data acquisition system of the DANSS spectrometer includes both methods described above. The QDC-based universal subsystem\cite{Hons2015,Hons2016} provides results immediately and therefore can be easily used for on-line monitoring and hardware tuning.

On the other hand, the main subsystem \cite{Svirida} is more informative and has no dead time, but requires more sophisticated software and much higher disk space.
Being capable of gathering data from all PMTs and MPPCs, it is based on the detection of \emph {all} events that may represent either positron production or neutron capture. In this case the correlation of the events by time, space and energy is performed completely in software. Such approach requires very fast digitization technique and assumes large data flows. To achieve this high performance 64-channel flash ADC modules were designed and manufactured in ITEP specially for DANSS detector. 46 FADC modules are placed in 4 VME crates and react to the HT with recording a certain portion of the photo-detector waveform digitized at 125 MHz. All PMT and MPPC signals are simultaneously analyzed on-the-flight and only those exceeding zero suppression threshold are buffered for further data transfer. Two dedicated FADC module are  additionally programmed to produce the HT itself. One module produces HT based on the digital sum of all 50 PMT signals given this sum exceeds a threshold of about 0.5 MeV and the other does the same for muon-veto counters. Extra low electronic noise together with 12 bit FADC dynamic range allows to record single pixel MPPC signal as well as 15 MeV energy deposit in a single strip. Such soft trigger is perfect to produce the most unbiased data sample, but results in the event rate of about 1 kHz. Yet the FADC modules are capable of fast VME data transfers and can easily handle the data flow produced by the above event rate.

\subsection{Slow control system}
Slow control system performs permanent monitoring of the slow-varying conditions such as internal and external background, air temperature, etc. Special attention is paid to MPPC operating conditions.
The MPPC power supply system consists of 170 hv-boards each serving a group of 15 MMPC. The HV-board produces three types of output voltages on the basis of input 15 V level: (i) $-4$~V to feed MMPC preamplifiers, (ii) base MMPC power, which is common for the group and could be regulated in the range $0\ldots65$~V, and (iii) 15 individual MMPC powers, which are necessary to tune MMPC gains on separate basis and are adjustable in the range $-10\ldots+10$~V via 16-bit DAC.

 \begin{figure}[hbt]
 \setlength{\unitlength}{1mm}
 \begin{minipage}[t]{45mm}
  \centering
  \begin{picture}(45,32)(0,0)
  % \put(0,0){\framebox(45,32)[b]{}}
\put(3,0){\includegraphics[height=32mm]{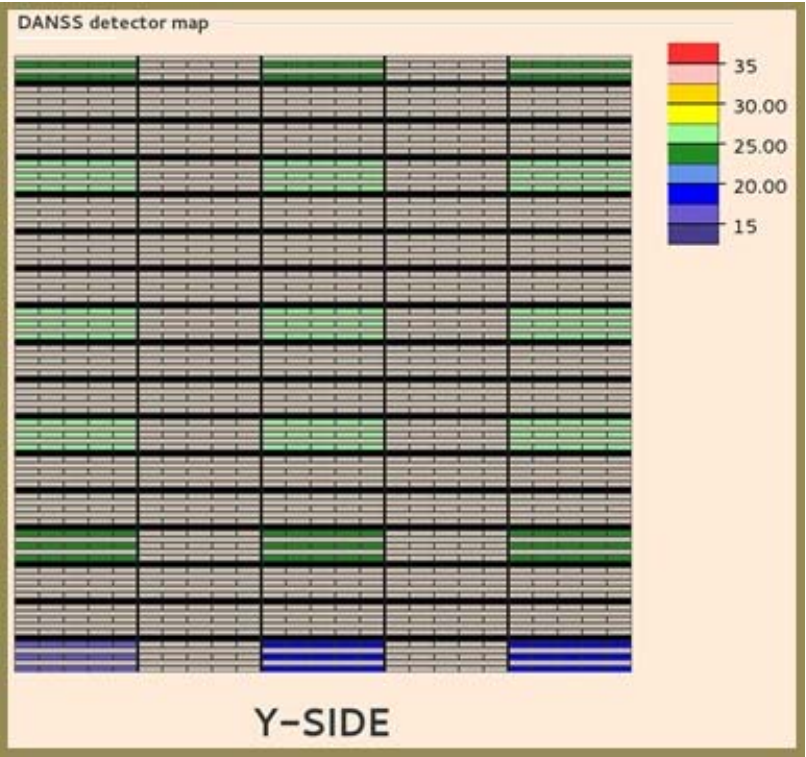}}
\end{picture}
  \caption{Typical temperature map for the Y-side.}
  \label{Fig.Screenshot}
 \end{minipage}\hfill{ }
 \begin{minipage}[t]{95mm}
  \centering
  \begin{picture}(95,32)(0,0)
   %\put(0,0){\framebox(95,32)[b]{}}
   \put(0, 0){\includegraphics[width=95mm]{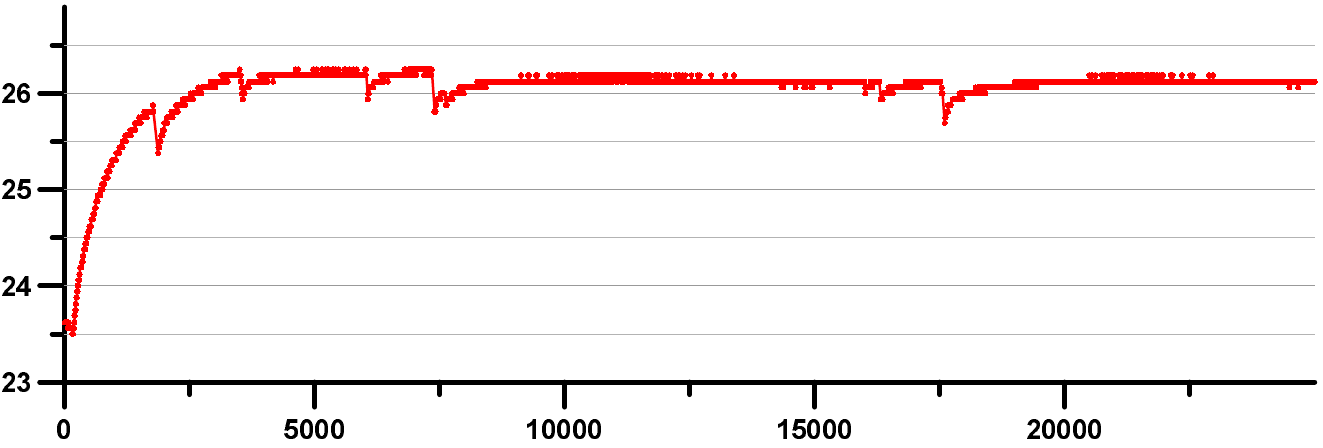}}
   \put(6.0,31.3){\makebox(0,0)[lt]{\tiny\sf MPPC temperature, $^\circ$C}}
   \put(94.5,0.2){\makebox(0,0)[rb]{\tiny\sf Time,  minutes}}
  \end{picture}
  \caption{MMPC temperature trend for the 19-days period.}
  \label{Fig.MPPC_temperature}
 \end{minipage}
 \end{figure}

Each HV-board is interfaced with microprocessor Silicon Labs C8051F353 with built-in 16-bit
ADC allowing measuring the board's control parameters, namely the values of voltages, current consumption and temperatures of microprocessor and DAC chips. Apart from that, 36 out of 170 HV-boards are coupled with external temperature sensors distributed uniformly on readout sides of the detector cube to control MMPC operating conditions.

The MMPC power supply system is governed by slow control software running constant automatic cycles of monitoring and storage of control parameters. The GUI in the form of two-dimensional detector map allows the manual mode of detector operation. It gives the possibility to set the desirable values of voltages for the arbitrary groups of MMPC channels as well as to readout a custom pattern of control parameters and visualize them in the form of colored maps or trends. As an example, Fig~\ref{Fig.Screenshot} shows the partial screenshot of GUI main window demonstrating typical temperature map of DANSS Y-side. The increase of temperatures in the central sections of the detector is clearly seen, explained by the specifics of the DANSS cooling, which is maintained via the top and the bottom sides of the cube. Fig~\ref{Fig.MPPC_temperature} shows typical MMPC temperature trend for the pilot DANSS data taking, which took place over the period from February 20th to March 9th of 2016. The initial increase corresponds to the stabilization process immediately after turning on the cooling and MMPC power supply systems. Then no temperature variations of any significance are registered which bodes stable MMPC behavior over long periods of future regular data taking.

 \begin{figure}[hbt]
 \setlength{\unitlength}{1mm}
 \begin{picture}(150,30)(0,0)
  %\put(0,0){\framebox(150,30)[b]{}}
  \put(  0.0,0.0){\includegraphics[height=30mm]{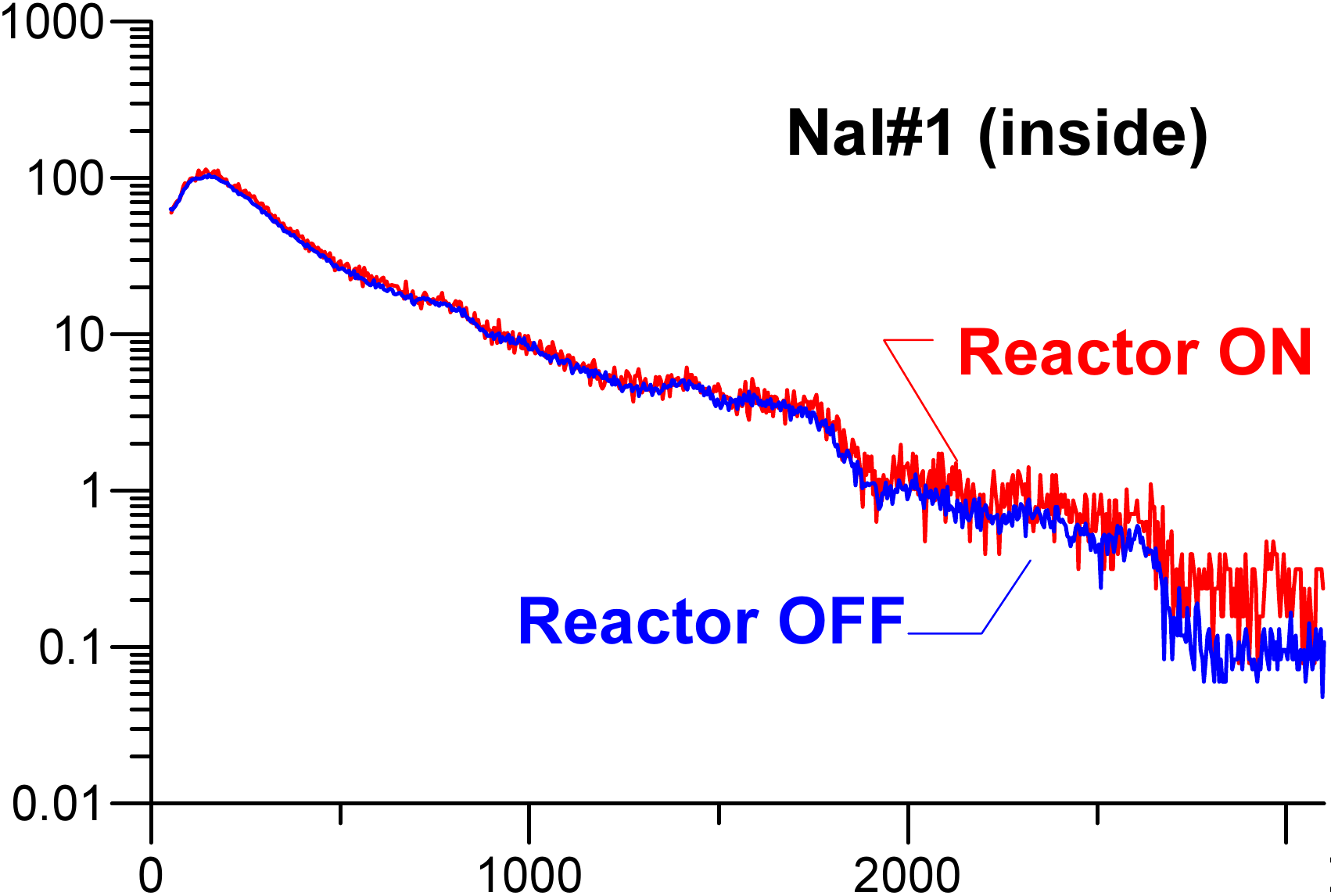}}
  \put( 52.0,0.0){\includegraphics[height=30mm]{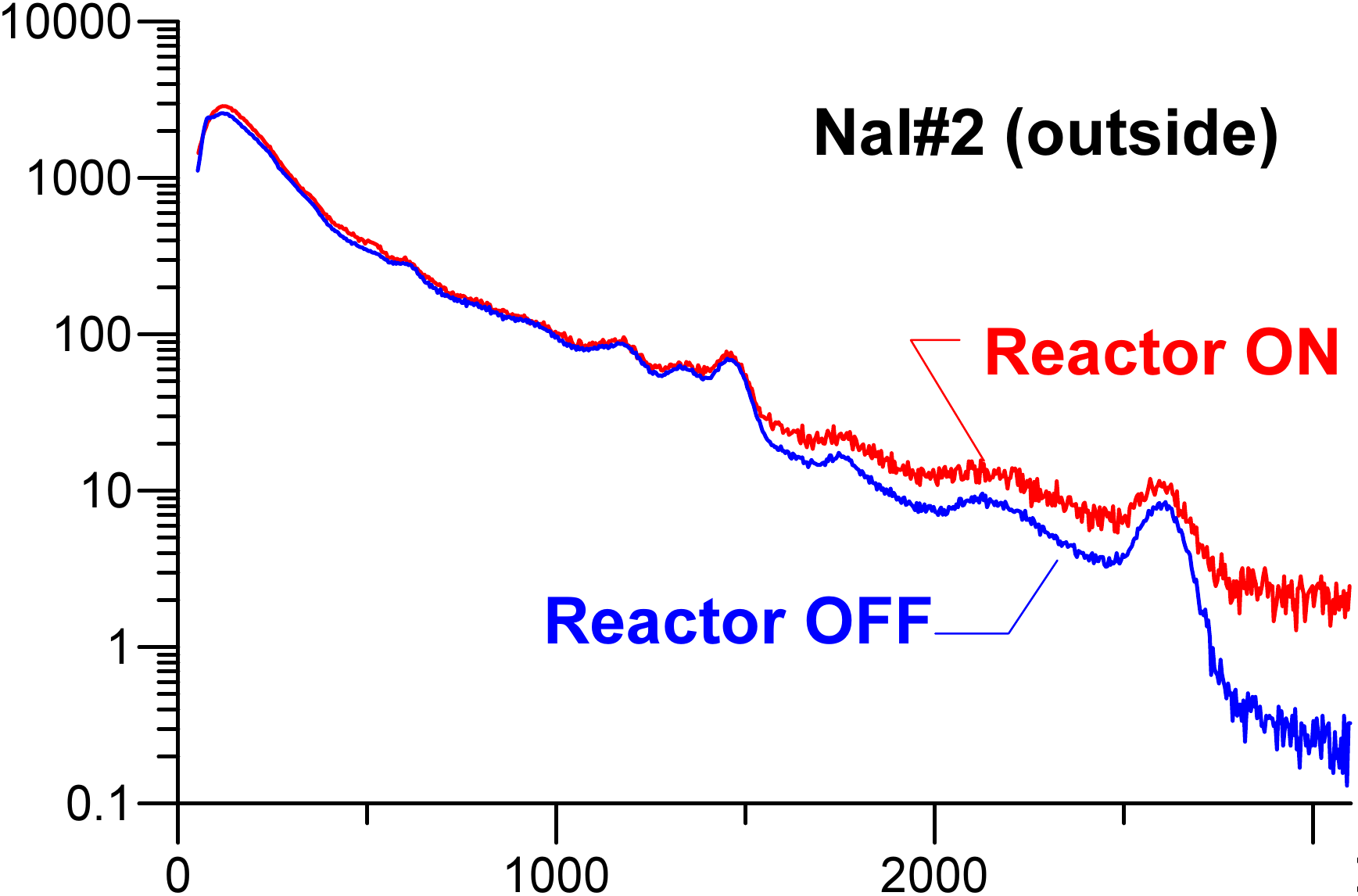}}
  \put(104.0,0.0){\includegraphics[height=30mm]{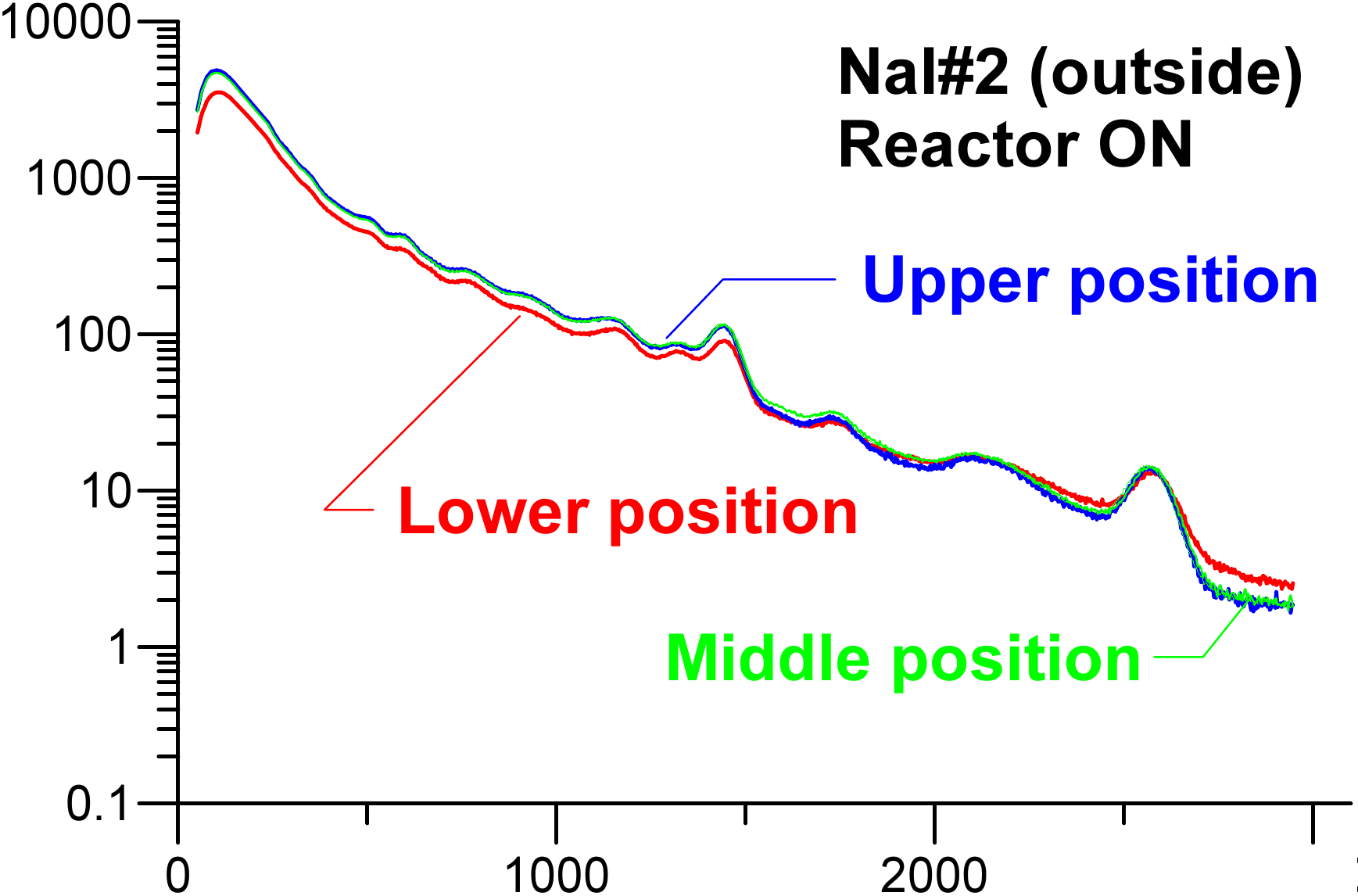}}
  \put(  6,30){\makebox(0,0)[lt]{\tiny\sf Counts/keV/hour}}
  \put( 59,30){\makebox(0,0)[lt]{\tiny\sf Counts/keV/hour}}
  \put(111,30){\makebox(0,0)[lt]{\tiny\sf Counts/keV/hour}}
  \put( 45,0){\makebox(0,0)[rb]{\tiny\sf $E$, keV}}
  \put( 98,0){\makebox(0,0)[rb]{\tiny\sf $E$, keV}}
  \put(150,0){\makebox(0,0)[rb]{\tiny\sf $E$, keV}}

 \end{picture}
 \caption{Examples of gamma energy spectra measured with detectors of the slow control system.}
 \label{Fig.SlowControlSpectra}
 \end{figure}
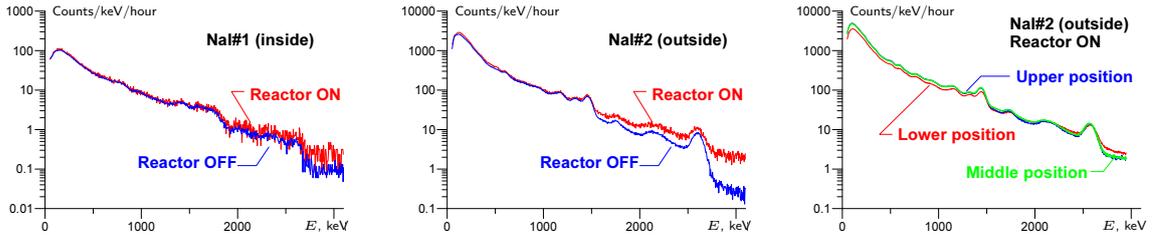

Gamma background is measured with four NaI crystals. One of them (NaI\#1$_{\rm in}$)is installed inside the DANSS shielding, in a gap between copper frames and CHB-Pb-CHB ``sandwich''. Two others (NaI\#2$_{\rm out}$ and NaI\#3$_{\rm out}$) are placed outside the shielding at opposite sides of the DANSS. The fourth detector (NaI\#4$_{\rm fix}$) is fixed close to the room ceiling; contrary to the previous three, it is not moved with the lifting gear.

Neutron flux in the room (about 590~n/m$^2$/s when the reactor is ON and 1.2~n/m$^2$/s when it is OFF) is measured permanently with the four-fold detector 4n$_{\rm out}$  (Fig.~\ref{Fig.He3}) which is placed on the movable platform near the NaI\#2$_{\rm out}$. The flux inside the shielding is expected to be negligible. Nevertheless, it is monitored with a single $^3$He tube without moderator. This detector (1n$_{\rm in}$) is installed in the same gap as the NaI\#1$_{\rm in}$ crystal and indicates the flux of thermal neutrons at the level of 0.05~n/m$^2$/s, irrespective of the reactor status.

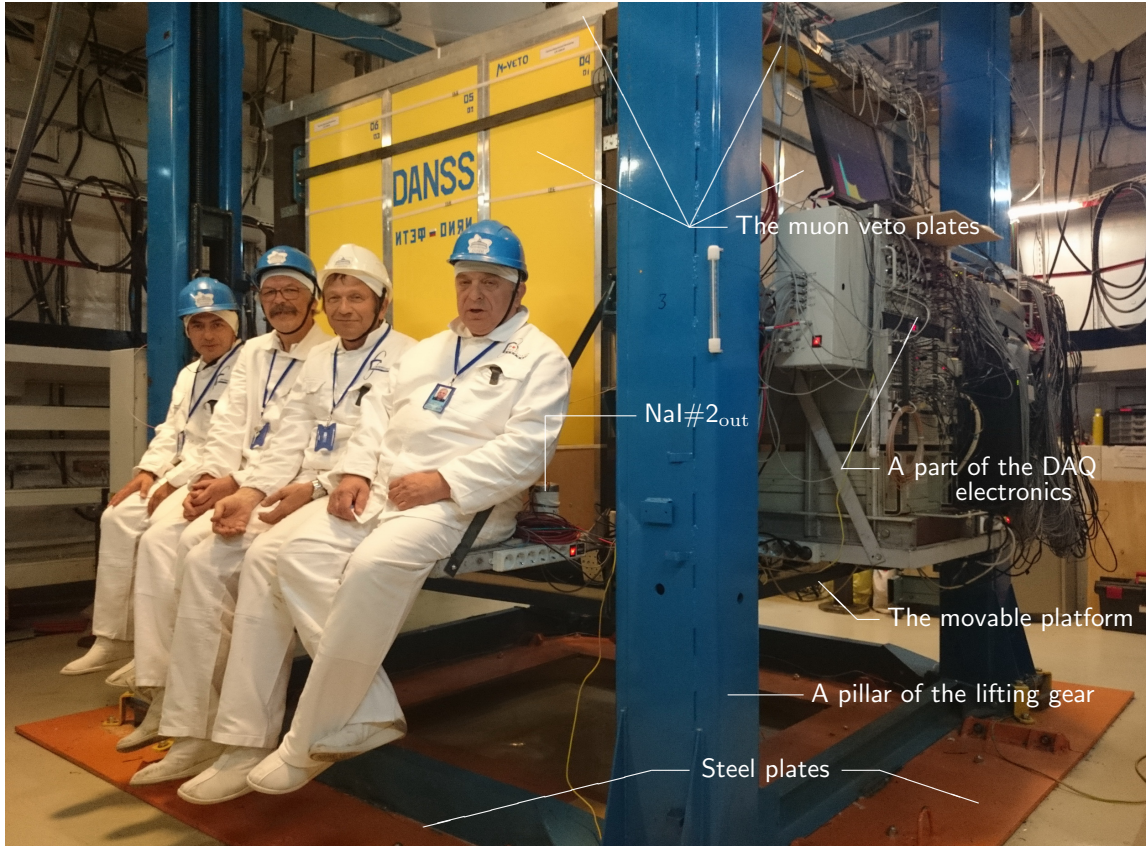
\begin{figure}[hbt]
 \setlength{\unitlength}{1mm}
 \begin{picture}(150,111)(0,1)
  %\put(0,0){\framebox(150,112)[b]{}}
 \put(0,0){\includegraphics[width=150mm]{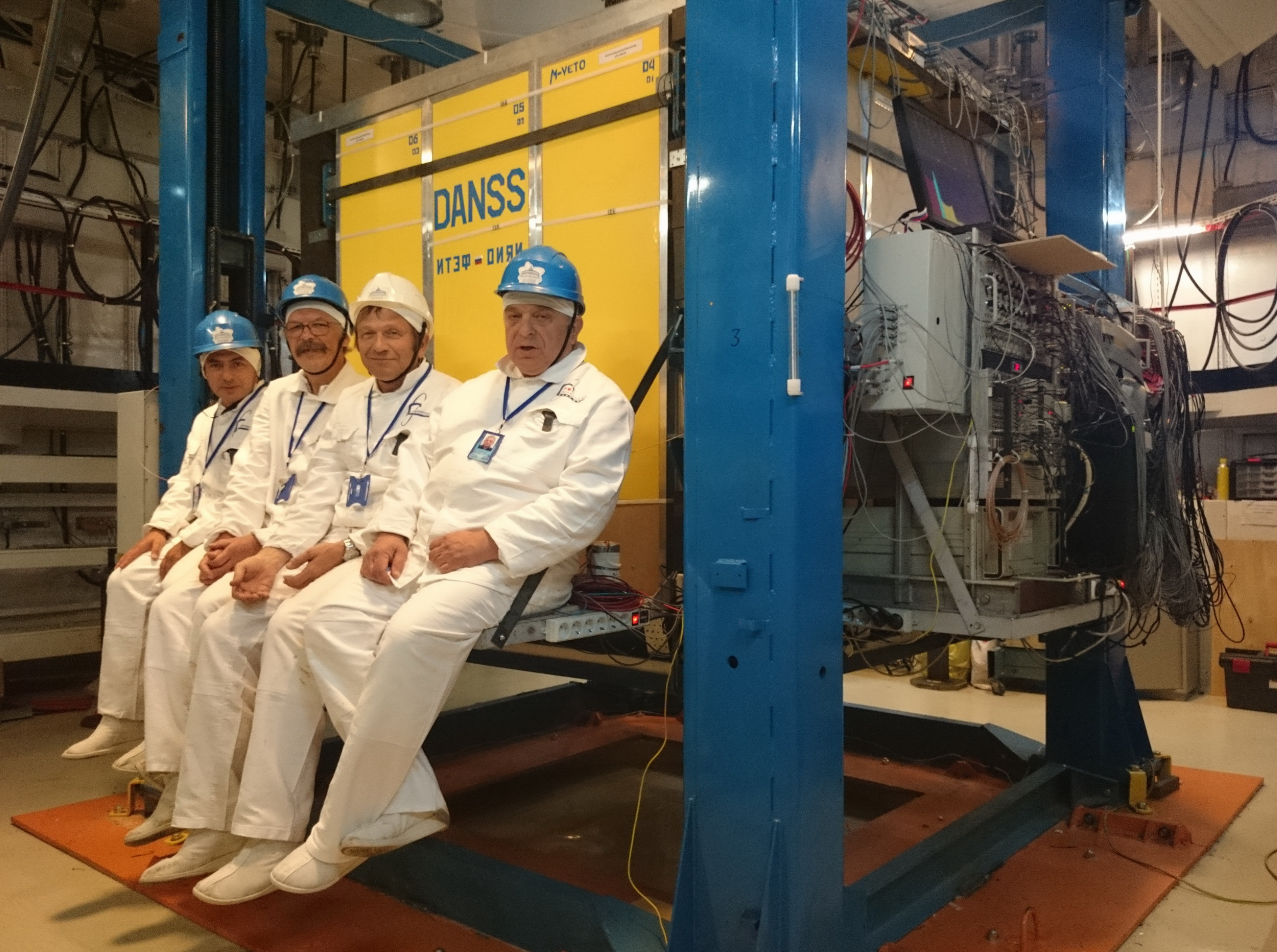}}
  {\color{white}
   \put(116.0,30.0){\makebox(0,0)[l]{\normalsize\sf The movable platform}}
   \put(115.0,30.0){\line(-1,0){3}}
   \put(112.0,30.0){\line(-1,1){5}}
   \put(106.0,20.0){\makebox(0,0)[l]{\normalsize\sf A pillar of the lifting gear}}
   \put(105.0,20.0){\line(-1,0){10}}
   \put(84.0,57.0){\makebox(0,0)[l]{\normalsize\sf NaI\#2$_{\rm out}$}}
   \put(83.0,57.0){\line(-1,0){12}}
   \put(71.0,57.0){\line(0,-1){10}}
   \put(116.0,50.0){\makebox(0,0)[l]{\normalsize\sf A part of the DAQ}}
   \put(125.0,47.0){\makebox(0,0)[l]{\normalsize\sf electronics}}
   \put(115.0,50.0){\line(-1,0){5}}
   \put(110.0,50.0){\line(1,2){10}}
   \put(96.0,82.0){\makebox(0,0)[l]{\normalsize\sf The muon veto plates}}
   \put(95.0,82.0){\line(-1,0){5}}
   \put(90.0,82.0){\line(-1,2){14}}
   \put(90.0,82.0){\line(1,2){12}}
   \put(90.0,82.0){\line(-2,1){20}}
   \put(90.0,82.0){\line(2,1){15}}
   \put(100.0,10.0){\makebox(0,0)[c]{\normalsize\sf Steel plates}}
   \put(85.0,10.0){\line(1,0){5}}
   \put(85.0,10.0){\line(-4,-1){30}}
   \put(115.0,10.0){\line(-1,0){5}}
   \put(115.0,10.0){\line(3,-1){13}}
  }
 \end{picture}
 \caption{I.~Alexeev, D.~Svirida, V~Egorov and V~Brudanin on board the DANSS operating in some intermediate position. (Photo by I.~Zhitnikov).}
 \label{Fig.Team_on_board}
 \end{figure}

Energy spectra from the above gamma and neutron detectors are measured with conventional spectrometric channels and written once per hour.  Similar spectrum from one of muon-veto plates is registered as well. Examples of the gamma spectra are shown in Fig.~\ref{Fig.SlowControlSpectra}. It is seen that external gamma background in the room does not depend significantly on the reactor operation or the detector position. Slight difference is caused mainly by thick steel plates of the floor (Fig.~\ref{Fig.Team_on_board}). At lower detector position it serves as an additional shield against $^{40}$K radiation, suppressing low energy part of the spectrum. On the other hand, neutron capture by iron increases high energy part. Nevertheless, gamma background becomes almost constant inside passive shielding, being suppressed with it by a factor of $\simeq$20.

Additional files contain information about intensities of selected regions-of-interest (ROI), individual count rates of all muon-veto plates, air temperature inside and outside the shielding and position of the movable platform.

\section{Estimated ``sterile'' sensitivity of DANSS}
\begin{figure}[htb]
 \setlength{\unitlength}{1mm}
 \begin{minipage}[t]{70mm}
  \centering
 \begin{picture}(70,42)(0,0)
 %\put(0,0){\framebox(70,42)[b]{}}
 \put(0,0){\includegraphics[height=42mm]{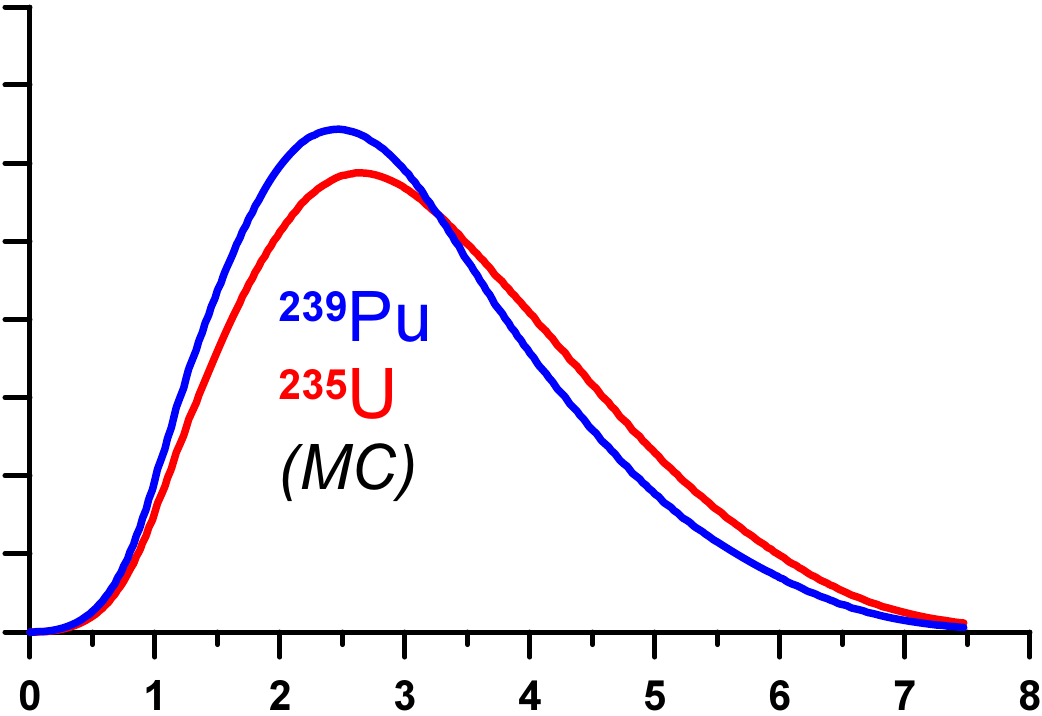}}
  \put(4.0,42.0){\makebox(0,0)[tl]{\footnotesize\sf Spectral density $S_{Z}$, a.u.}}
  \put(68.0, 6.0){\makebox(0,0)[r]{\footnotesize\sf $E_P$}}
  \put(69.0, 0.2){\makebox(0,0)[rb]{\footnotesize\sf MeV}}
 \end{picture}
 \caption{Energy Prompt spectra simulated for fission of $^{235}$U and $^{239}$Pu without oscillation effect.}
 \label{Fig.Spectra(U-Pu)}
 \end{minipage}
 \hfill{ }
 \begin{minipage}[t]{70mm}
  \centering
 \begin{picture}(70,42)(0,0)
 %\put(0,0){\framebox(70,49)[b]{}}
 \put(0,0){\includegraphics[height=42mm]{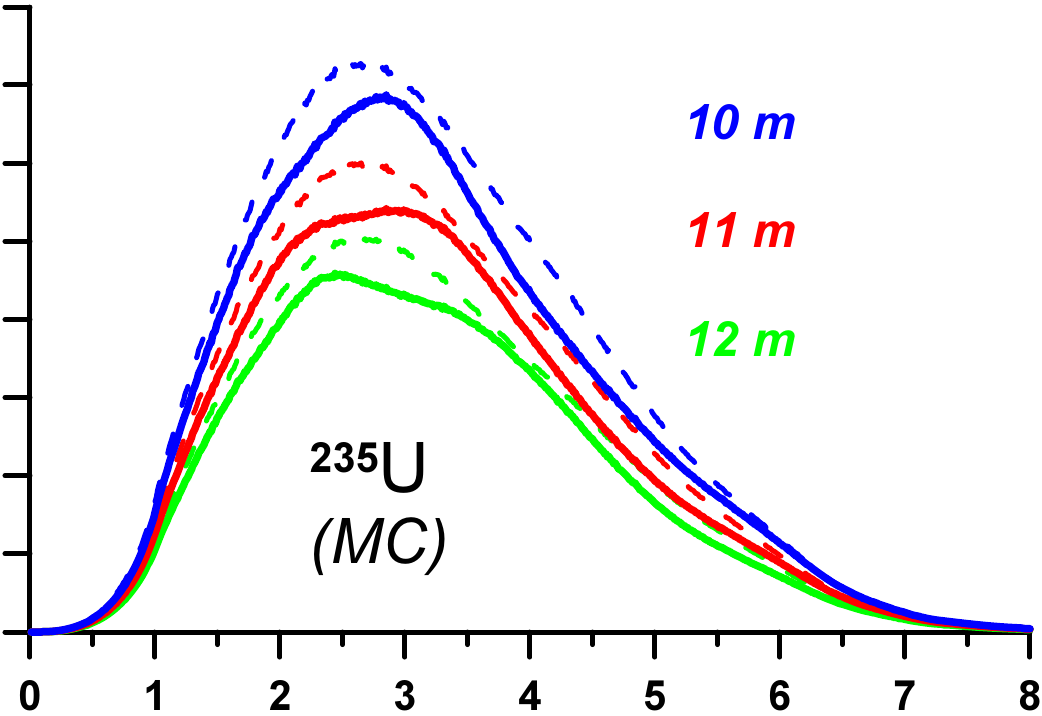}}
  \put(4.0,42.0){\makebox(0,0)[tl]{\footnotesize\sf Spectral density $S_{Z,F}$, a.u.}}
  \put(68.0, 6.0){\makebox(0,0)[r]{\footnotesize\sf $E_P$}}
  \put(69.0, 0.2){\makebox(0,0)[rb]{\footnotesize\sf MeV}}
 \end{picture}
 \caption{$E_P$ spectra simulated for fission of $^{235}$U without (dashed curves) and with (solid curves) oscillation under assumption (\ref{Eq.French}) at different distance.}
 \label{Fig.Spectra(L)}
 \end{minipage}
\end{figure}

As it was shown more than once\footnote{See, e.g., proceedings of AAP Int. Workshops -- AAP2015, AAP2013, AAP2011, AAP2009, etc.}, neutrino energy spectrum $S(E_\nu)$ depends on the fuel composition (which changes during the reactor campaign) and therefore could be used for an on-line reactor monitoring (Fig.~\ref{Fig.Spectra(U-Pu)}).

DANSS sensitivity to the amount of bomb-grade $^{239}$Pu produced in the reactor core is estimated as 1 SQ (Significant Quantity\footnote{The SQ-unit equals to amount of fissile material enough to produce nuclear warhead.}) per 2 weeks of measurement. It should be mentioned that this value is rather rough and depends very much on the real conditions (final energy resolution, long-term stability, background level, etc.) which are not known yet at this stage of the project.

In addition to the above {\bf applied goal}, the main {\bf fundamental goal} of the project is searching for short-range oscillation of the reactor neutrino to a sterile state. As it was recently claimed by our French colleagues \cite{Mention2011}, neutrino oscillates to a new 4th type with the following oscillation parameters:
\begin{equation}
\left.
\begin{array}{lcl}
\sin^2(2\theta)&=&0.17\;,\\
\Delta m^2&=&2.0\; {\rm eV}^2\;.
\end{array}
\right\}\;(F)
\label{Eq.French}
\end{equation}
Neutrino survival probability is expressed as\vspace{-1mm}
\begin{equation}
 P_{\rm osc}(\nu_e\rightarrow\nu_e)\; =\; 1-\sin^2 (2\theta)\cdot \sin^2\left(1.267\; \frac{\Delta m^2 \;L}{E_\nu}\right)\;,
 \label{Eq.SurvivalProbability}
\end{equation}
where the source-detector distance $L$ is given in metres and the neutrino energy $E_\nu$ in units of MeV. With finite non-zero dimensions of the source and detector, as well as finite energy resolution, one has to integrate the formula (\ref{Eq.SurvivalProbability}) over some $\delta L$ and $\delta E$ intervals.

Taking into account size of the reactor core and space distribution of the fission probability, realistic energy resolution (see Section~\ref{Section.Strips} above) and other detector parameters, we have estimated the neutrino spectral density $S(E)$ which could be measured by DANSS in two cases -- when the above oscillation with ``French'' parameters really exists ($S_F$) and when the phenomenon probability is zero ($S_Z$).

For the typical $E_\nu$ range of 2--6 MeV the oscillations manifest themselves mainly at short distances, transforming both the spectrum shape and integral count rate (Fig.~\ref{Fig.Spectra(L)}).

\begin{figure}[htb]
 \setlength{\unitlength}{1mm}
 \begin{minipage}[t]{70mm}
  \centering
 \begin{picture}(70,46)(0,0)
 %\put(0,0){\framebox(70,46)[b]{}}
 \put(0,0){\includegraphics[width=70mm]{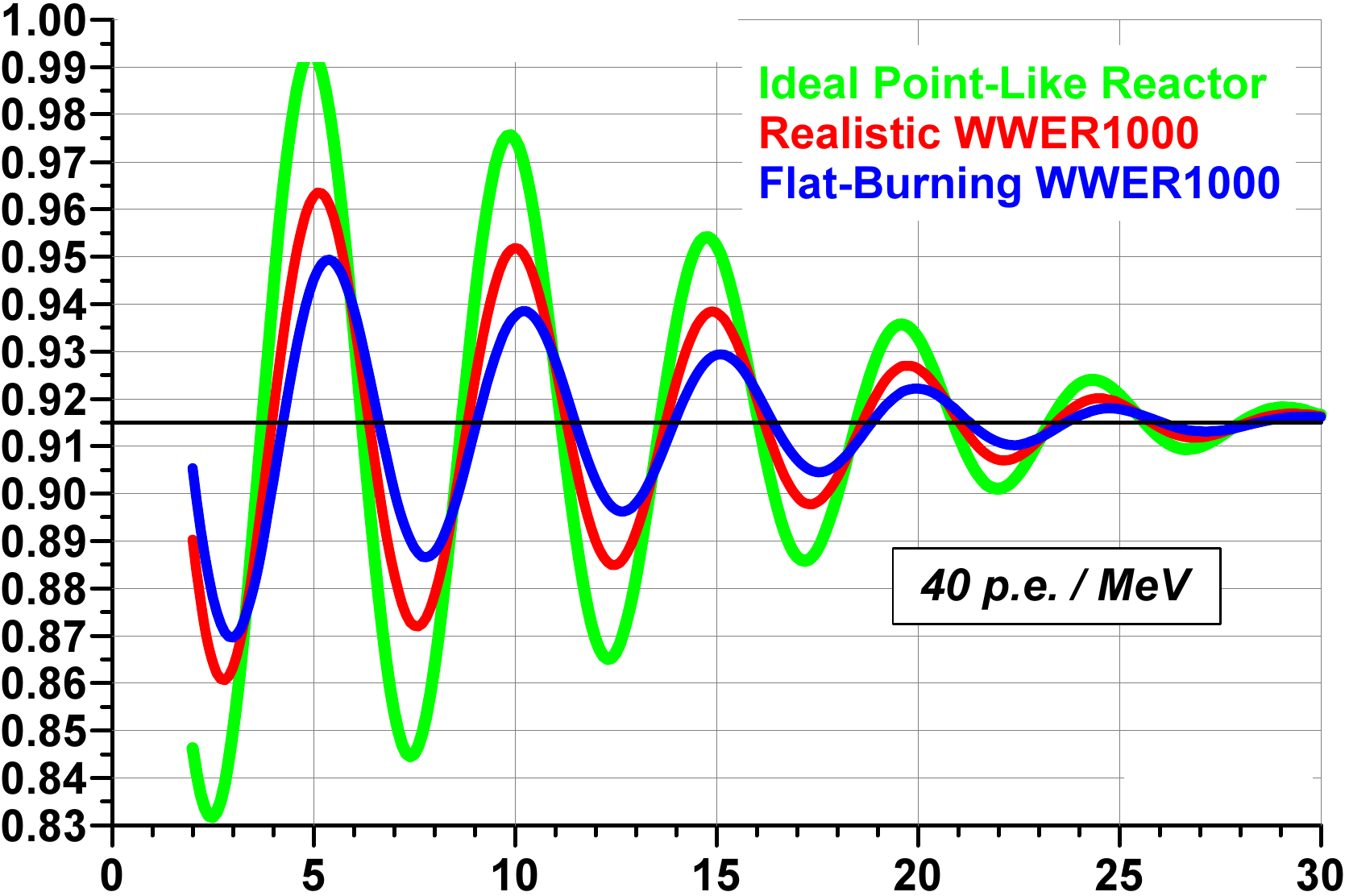}}
 \put(7.0,46.0){\makebox(0,0)[tl]{\tiny\sf SF / SZ @ $E_P$=(3.00$\pm$0.28) MeV}}
 \put(66.0,5.0){\makebox(0,0)[rb]{\tiny\sf $L$, m}}
 \end{picture}
 \caption{Oscillation curves simulated for realistic DANSS energy resolution but different reactor sizes.}
 \label{Fig.Oscillation_Curve(dimensions)}
 \end{minipage}
 \hfill{ }
 \begin{minipage}[t]{70mm}
  \centering
 \begin{picture}(70,46)(0,0)
 %\put(0,0){\framebox(70,46)[b]{}}
 \put(0,0){\includegraphics[width=70mm]{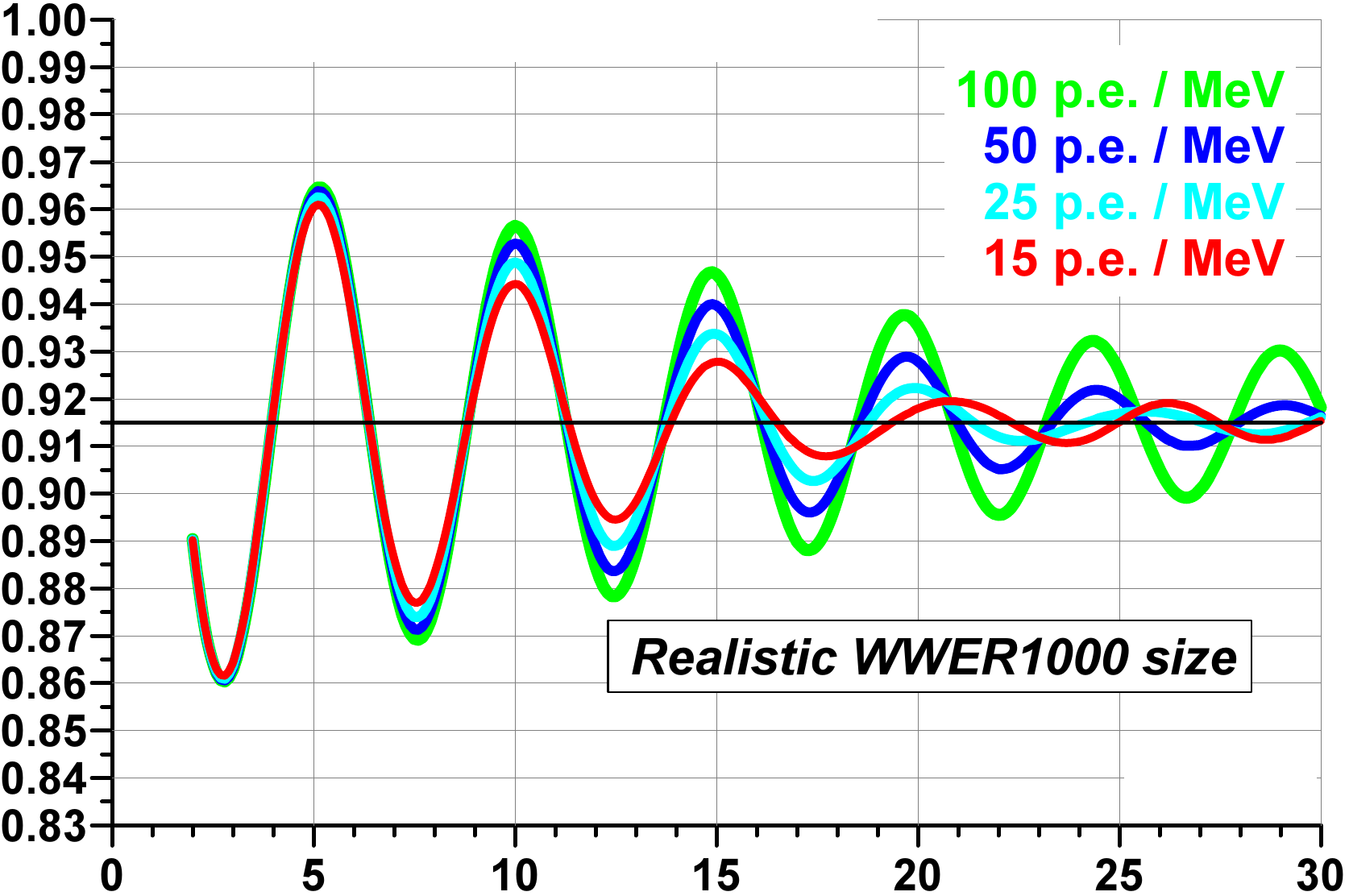}}
 \put(7.0,46.0){\makebox(0,0)[tl]{\tiny\sf SF / SZ @ $E_P$= (3.00$\pm \delta E$) MeV}}
 \put(66.0,5.0){\makebox(0,0)[rb]{\tiny\sf $L$, m}}
 \end{picture}
 \caption{Oscillation curves simulated for realistic WWER1000 size but different energy resolution $\delta E$.}
 \label{Fig.Oscillation_Curve(resolution)}
 \end{minipage}
 \end{figure}

At higher distances oscillations are partially washed-out because of the integration, so that only some decrease of the absolute count rate can be observed. Relative role of the dimensions and energy resolution in the washing-out is shown in Figs.~\ref{Fig.Oscillation_Curve(dimensions)} and \ref{Fig.Oscillation_Curve(resolution)}, respectively. It can be seen that big dimensions of an industrial reactor do not kill oscillation waves completely, but only reduce their amplitude by 30-40\% with respect to an ideal point-like hypothetical source. On the other hand, energy resolution is rather important and should not be poor (when the signal yield is less than 25 photo electrons per MeV energy deposit), especially at higher distance ($L>$15-20~m).

\begin{figure}[hbt]
 \setlength{\unitlength}{1mm}
 \begin{minipage}[t]{70mm}
  \centering
 \begin{picture}(70,65)(0,0)
 %\put(0,0){\framebox(70,65)[b]{}}
 \put(1,0){\includegraphics[height=65mm]{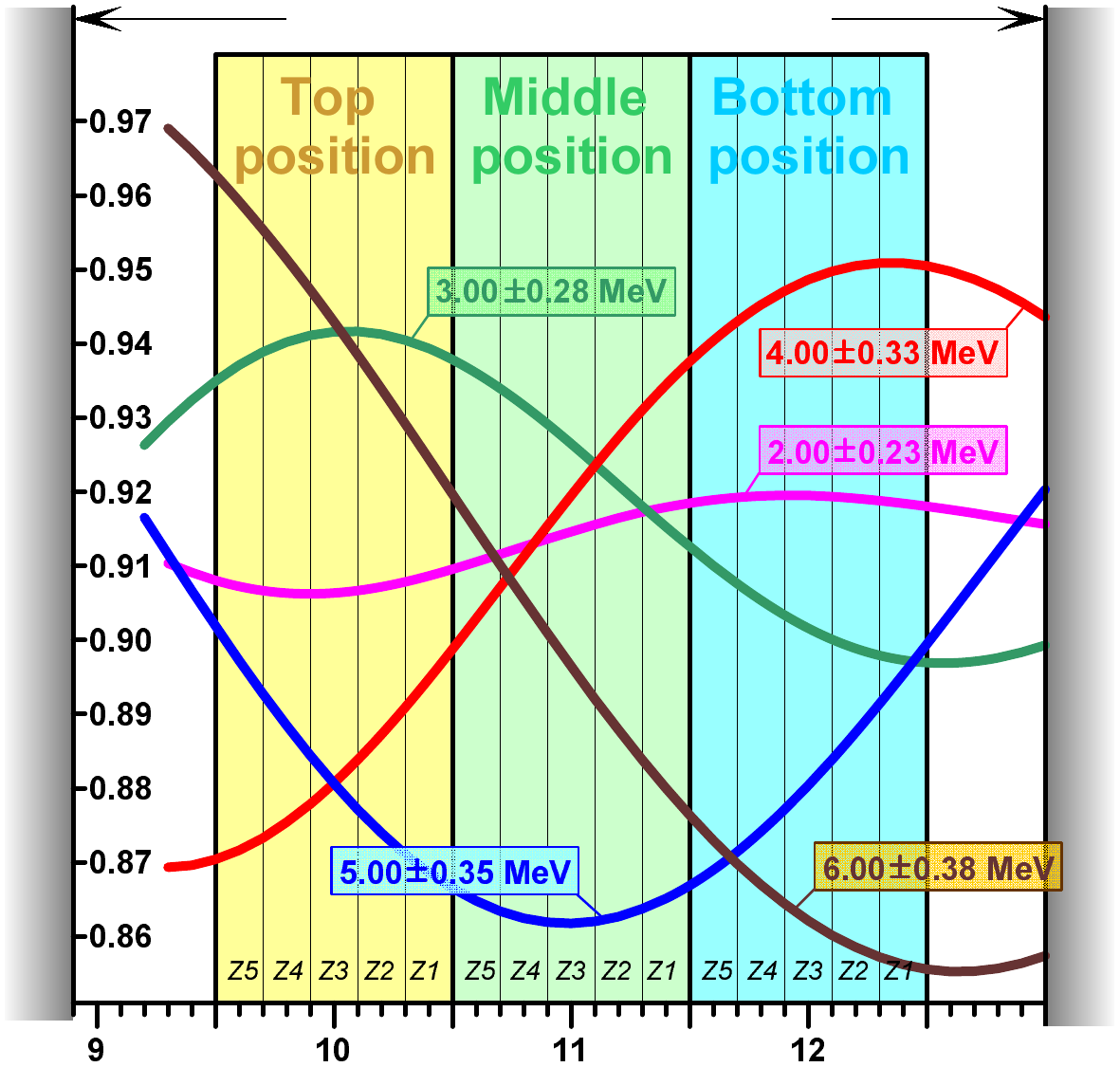}}
 \put(0.5,30){\rotatebox{90}{\small\sl Ceiling}}
 \put(67,30){\rotatebox{90}{\small\sl Floor}}
 \put(35.0,65.0){\makebox(0,0)[t]{\footnotesize\sl Height available: 4.1 m}}
 \put(7.0,61.0){\makebox(0,0)[l]{\scriptsize\sf $\frac{\bf SF}{\bf SZ}$}}
 \put(63.0,0.0){\makebox(0,0)[rb]{\scriptsize\sf $L$, m}}
 \end{picture}
 \caption{Oscillation curves simulated for realistic DANSS conditions under assumption (\ref{Eq.French}).}
 \label{Fig.Oscillation_Curve(zoom)}
 \end{minipage}\hfill{ }
 \begin{minipage}[t]{70mm}
  \centering
 \begin{picture}(70,65)(0,0)
 %\put(0,0){\framebox(70,65)[b]{}}
 \put(3.0,0){\includegraphics[height=65mm]{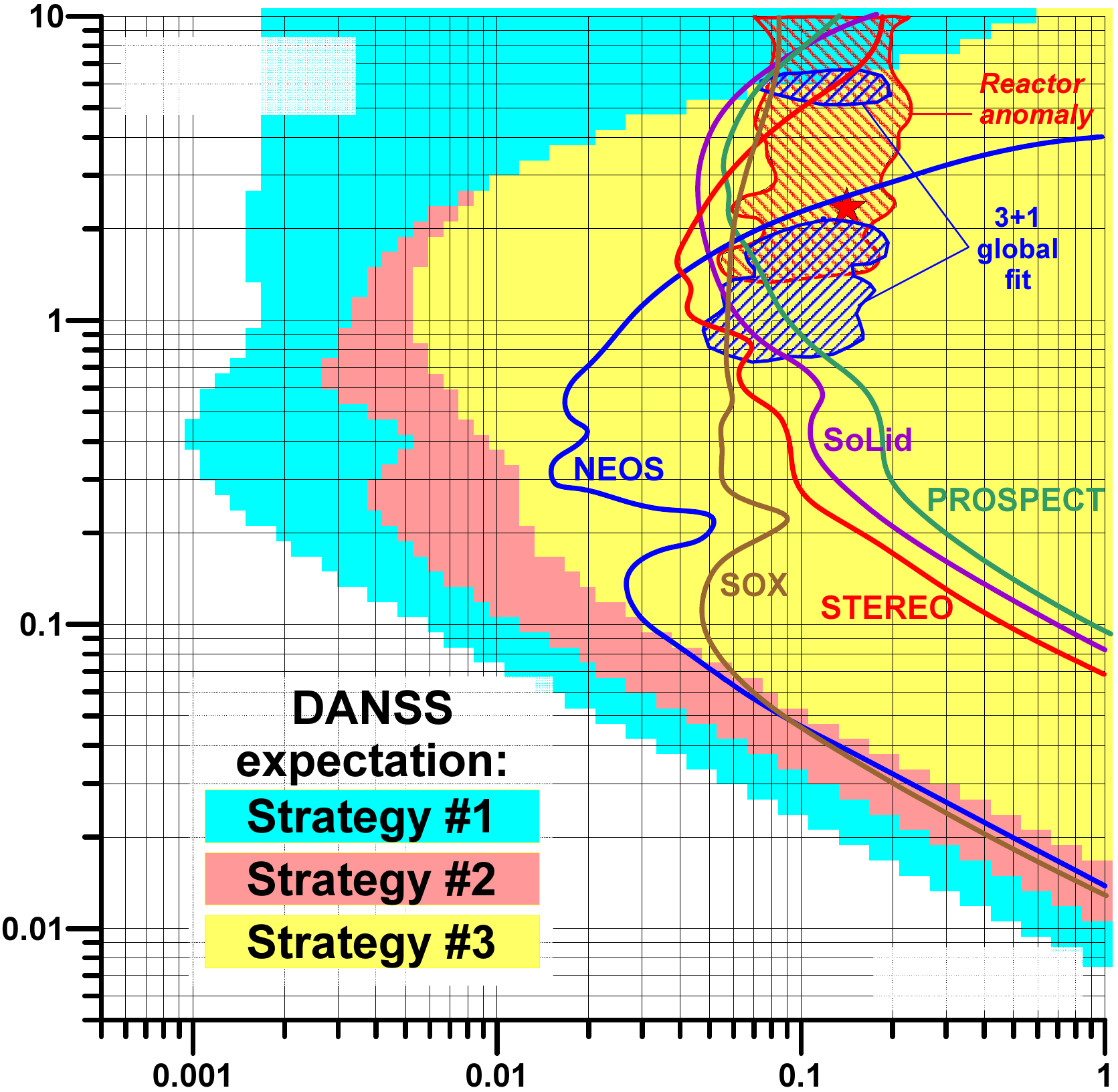}}
 \put( 11.0,62.5){\makebox(0,0)[lt]{\footnotesize $\Delta m^2$,  eV$^2$}}
 \put( 66.5,5.5){\makebox(0,0)[rb]{\footnotesize $\sin^2(2\theta)$}}
 \end{picture}
 \caption{Estimation of the DANSS sensitivity at 95\%CL to the oscillation parameters in case of one-year measurement.}
 \label{Fig.Sensitivity_Chart}
 \end{minipage}

\end{figure}

Figure~\ref{Fig.Oscillation_Curve(zoom)} shows the $S_F/S_Z$ ratio for different energy spectrum fragments (with the full width  $2\delta E$ corresponding to 40 photo electrons per MeV) as a function of the distance $L$. As previously, each curve reflects oscillation of neutrinos with the given energy and represents relative deviation of the detector counting rate from the $1/L^2$ rule. Moving the detector to a top, middle or bottom position by means of the lifting gear, we expect to observe the shown deviation of few percent within a week.

But what if the oscillation takes place, but its parameters are different from (\ref{Eq.French})? In order to estimate DANSS sensitivity, numerous MC simulations have been performed and several methods of approaching tested. As a result, two measurement policies and three strategies of the data analysis can be considered.

The measurement can be done with immovable detector or, alternatively, with the detector located sequentially in bottom, middle and top position. In the first case the 1~m detector body is considered as 5 independent sections (Z1...Z5) 20 cm each, so that one could compare five spectra measured with these sections. In the second case the statistics taken in each position is 3 times lower, but the total scanned $L$-region is 3 times longer (3 metres instead of 1).

According to the first (worldwide used) strategy of the data analysis, energy spectra measured at each detector position are compared channel by channel with the calculated ones. This strategy requires exact knowing of the absolute initial neutrino spectrum and flux, as well as the absolute detector efficiency. This method is the most sensitive, but seems to be the most susceptible of systematic errors.

Using the second strategy, one compares {\bf relative} spectra shapes instead of their {\bf absolute} values. Here the value of initial neutrino flux and detector efficiency are not used and therefore do not introduce an error.

With the third strategy one observes evolution of energy spectral intervals with distance, as it is shown in Fig.~\ref{Fig.Oscillation_Curve(zoom)}. This strategy is the most free of systematic errors because it does not require theoretical calculation of the spectrum shape (which has a quite questionable precision) and does not depend on the fuel composition.

Figure \ref{Fig.Sensitivity_Chart} shows sensitivity (at 95\%CL) of the movable DANSS estimated for one-year measurements in three positions. Two lashed areas correspond to the claim \cite{Mention2011} and the 3+1 global fit to all relevant accelerator, source, and reactor data given in \cite{Giunti_global_fit}.

We have also investigated the influence on the sensitivity of systematic uncertainties in the energy scale, background level, as well as the non-uniformity of the scintillator strip response \cite{Danilov_EPS2013,Skorobova}. The reduction of the sensitivity due to these effects is small.

Starting from the pioneer Russian experiments of L.~Mikaelyan group with RONS facility\cite{RONS}, physicists of several countries apply their efforts to build a compact detector of reactor neutrino. Some of them are aimed at monitoring of an industrial reactor (SONGS\cite{SONGS}, Nucifer\cite{Nucifer}, ANGRA\cite{ANGRA}, CORMORAD\cite{CORMORAD}, PANDA\cite{PANDA}, WATCHMAN\cite{WATCHMAN}), whereas others are intended for short-range neutrino oscillations only (NEUTRINO4\cite{NEUTRINO4}, STEREO\cite{Stereo}, POSEI\-DON\cite{Poseidon}) or combine the above goals (HANARO/NEOS\cite{NEOS}, NuLat\cite{NuLat}, SoLid\cite{SoLid}, PROSPECT\cite{PROSPECT}, ASDC\cite{ASDC}). Big underground neutrino detectors are planned to be equipped with artificial radioactive source (SOX\cite{SOX}, CeLAND\cite{CeLAND}).

Most of the above projects are reviewed in \cite{Lasserre_review,Giunti_review}. Colored curves in Fig.~\ref{Fig.Sensitivity_Chart} represent 95\%CL sensitivity expected for one-year measurements with some of them. It can be seen that our project is quite competitive with others.

\section{Data taking}
Regular data taking with DANSS has been started in April 2016. The measurements are performed in 3 positions of the lifting platform: on the floor, 1 and 2 metres above the floor. Change of the position is done once per day and takes 4-5 minutes. The relatively short one-day expositions are necessary in order to avoid probable systematic errors caused by instability of the reactor power, fuel composition, etc.

\section{Acknowledgments}

{The authors are grateful to the directorates of ITEP and JINR for constant support of this work. The authors appreciate the administration of the KNPP and the staff
of the KNPP Radiation Safety Department for permanent assistance in the
experiment.
This work is supported in part by the Russian State Corporation ROSATOM (state contracts H.4x.44.90.13.1119 and H.4x.44.9B.16.1006)
and by the Russian Foundation for Basic Research, pro\-ject 09-02-00449. Young JINR physicists were supported by JINR grants 14-202-(07,08), 15-203-(02,03,07,10), 16-202-(03,04), 16-203-(02,03) and Czech Ministry of Education, Youth and Sports INGO II-LG14004.}

\end{document}